\documentclass[aps,10pt,prd,superscriptaddress,preprintnumbers,eqsecnum,%
showkeys,nofootinbib,showpacs]{revtex4-1}
\usepackage{bm} 
\usepackage{graphicx}
\newcommand{\be}{\begin{eqnarray}}
\newcommand{\ee}{\end{eqnarray}}
\newcommand{\beq}{\begin{equation}}
\newcommand{\eeq}{\end{equation}}
\begin{document}
\title{Electron-deuteron deep-inelastic scattering with spectator nucleon tagging \\
and final-state interactions at intermediate $x$} 
\author{M.~Strikman}
\email[ E-mail: ]{strikman@phys.psu.edu}
\affiliation{Department of Physics, Pennsylvania State University,
University Park, PA 16802, USA}
\author{C.~Weiss}
\email[ E-mail: ]{weiss@jlab.org}
\affiliation{Theory Center, Jefferson Lab, Newport News, VA 23606, USA}
\begin{abstract}
We consider electron-deuteron deep-inelastic scattering (DIS) with detection of a
proton in the nuclear fragmentation region (``spectator tagging'') as a method 
for extracting the free neutron structure functions and studying their nuclear modifications.
Such measurements could be performed at a future Electron-Ion Collider (EIC) with suitable
forward detectors. The measured proton recoil momentum ($\lesssim$ 100 MeV in the deuteron
rest frame) specifies the deuteron configuration during the high-energy process and permits 
a controlled theoretical treatment of nuclear effects. Nuclear and nucleonic structure are separated 
using methods of light-front quantum mechanics. The impulse approximation (IA) to the tagged DIS cross
section contains the free neutron pole, which can be reached by on-shell
extrapolation in the recoil momentum. Final--state interactions (FSI) distort
the recoil momentum distribution away from the pole. In the intermediate--$x$ region
$0.1 < x < 0.5$ FSI arise predominantly from interactions of the spectator proton with 
slow hadrons produced in the DIS process on the neutron (rest frame momenta 
$\lesssim 1\, \textrm{GeV}$, target fragmentation region). We construct a schematic model 
describing this effect, using final--state hadron distributions measured in nucleon DIS
experiments and low-energy hadron scattering amplitudes. We investigate the magnitude of 
FSI, their dependence on the recoil momentum (angular 
dependence, forward/backward regions), their analytic properties, and their effect on 
the on-shell extrapolation. We comment on the prospects for neutron structure extraction 
in tagged DIS with EIC. We discuss possible extensions of the FSI model to other kinematic 
regions (large/small $x$). In tagged DIS at $x \ll 0.1$ FSI resulting from diffractive scattering 
on the nucleons become important and require separate treatment.
\end{abstract}
\preprint{JLAB-THY-17-2488}
\maketitle
\tableofcontents
\newpage
\section{Introduction}
\label{sec:introduction}
Measurements of deep-inelastic lepton scattering (DIS) from nuclei with mass number
$A > 1$ address several key topics in short-range nuclear structure
and quantum chromodynamics (QCD). One is the partonic structure of the neutron,
which is needed for the flavor decomposition of the nucleon's valence and sea
quark densities and for the separation of singlet and non-singlet nucleon
structure functions in studies of the scale dependence (QCD evolution,
higher-twist effects).  Another topic are the modifications of the nucleon's
partonic structure in the nucleus and their dependence on the scaling variable
$x$ (EMC effect at $x > 0.3$, antishadowing at $x \sim 0.1$), which attest to
the presence of non-nucleonic degrees of freedom in nuclei and reveal the QCD
structure of nucleon-nucleon interactions
\cite{Frankfurt:1988nt,Arneodo:1992wf}. Yet another topic are coherence
phenomena at $x \ll 0.1$ such as nuclear shadowing, which arise from the participation of multiple
nucleons in the DIS process and govern the approach to the unitarity limit at
high energies \cite{Frankfurt:2011cs}.  Experiments in nuclear DIS have been
carried out in fixed-target $eA/\mu A$ scattering at several facilities (SLAC,
HERMES, CERN EMC and COMPASS, FNAL E665, JLab; see Ref.~\cite{Malace:2014uea}
for a review) and will be extended further with the Jefferson Lab 12 GeV
Upgrade.  A much wider kinematic range would become accessible in
colliding--beam experiments with a future Electron--Ion Collider (EIC)
\cite{Boer:2011fh,Accardi:2012qut,2015NSAC}.  A medium--energy EIC with a squared electron--nucleon 
center-of-mass energy $s_{eN} \equiv s_{eA}/A \sim$ 200--2000 GeV$^2$
would be ideally suited for nuclear DIS measurements in the region 
$x \gtrsim 10^{-3}$ and enable detailed studies of sea quarks and
gluons in the nucleon and their nuclear modifications \cite{Accardi:2011mz,Abeyratne:2012ah}. 
Complementary information is provided by
measurements of hard processes in high-energy hadron and photon scattering on
nuclei (RHIC, LHC) \cite{Baltz:2007kq}.

The main challenge in the analysis of nuclear DIS experiments is to account for
the multitude of nuclear configurations that can be present in the initial state
of the high-energy scattering process and affect its outcome
\cite{Frankfurt:1981mk}.  The scattering can take place on any of the
constituent protons and neutrons ($p$ and $n$), in different states of their
quantum--mechanical motion in the nucleus (momentum, spin). In addition,
non-nucleonic degrees of freedom such as $\Delta$ isobars are excited by the
nuclear binding.  In the extraction of neutron structure one needs to isolate
the DIS cross section arising from scattering on the neutrons and eliminate the
effects of nuclear binding (Fermi motion, non-nucleonic degrees of freedom). For
neutron spin structure one must also infer the effective polarization of the
neutron in the polarized nucleus and account for the polarization of
non-nucleonic degrees of freedoms, particularly intrinsic $\Delta$'s in
polarized ${}^3$He \cite{Frankfurt:1996nf,Bissey:2001cw,Boros:2000af}.  In the study of nuclear
modifications at $x \gtrsim 0.1$ (EMC effect, antishadowing) one wants to relate
the modifications to the nucleon interactions taking place in particular nuclear
configurations (short-range correlations, exchange mechanisms). In traditional
inclusive nuclear DIS measurements $e + A \rightarrow e' + X$ these issues
are addressed by modeling the nuclear effects for typical nuclear configurations
and averaging over all possible configurations. The resulting theoretical
uncertainty usually represents the dominant systematic error in neutron
structure extraction. Likewise, this method provides limited possibilities for
unraveling the dynamical origin of nuclear modifications.  Major progress could
come from experiments that provide information on the nuclear configurations
present during the high--energy process through measurements of the nuclear
final state.

Deep-inelastic scattering on the deuteron ($d, A = 2$) with detection of a nucleon ($N = p$ or $n$) 
in the nuclear fragmentation region, $e + d \rightarrow e' + N + X$, represents a unique method for 
performing DIS measurements in controlled nuclear configurations (``spectator tagging''). 
The nucleon emerges with a typical recoil momentum
$|\bm{p}_N| \sim$ few 10--100 MeV in the deuteron rest frame.\footnote{We use units in which
$\hbar = c = 1$ and quote momenta in MeV/GeV.} At such momenta the deuteron's non-relativistic
$pn$ wave function is well known from low--energy measurements and can be used to construct 
the $pn$ light-front (LF) wave function entering in high--energy processes (see below). 
Because the deuteron has isospin $I = 0$,
$\Delta$ isobars in the wave function are strongly suppressed (they can occur
only in $\Delta\Delta$ configurations), so that the deuteron can be treated as a
$pn$ system for most of the configurations relevant to DIS
\cite{Frankfurt:1981mk}. Under these conditions the detection of the recoil nucleon and the 
measurement of its momentum positively identify the active nucleon and control its momentum during the 
DIS process. By measuring DIS with a tagged proton and extrapolating the measured
recoil momentum dependence to the on-shell point near $|\bm{p}_p| = 0$ (in the deuteron rest frame) 
one can eliminate nuclear binding effects and extract free neutron structure in a
model-independent manner \cite{Sargsian:2005rm}. DIS on the deuteron with proton tagging
was measured in the JLab CLAS BONuS experiment at 6 GeV beam 
energy \cite{Tkachenko:2014byy,Griffioen:2015hxa} and will be explored further 
at 11 GeV \cite{bonus12}. This setup covers recoil momenta $|\bm{p}_p| \gtrsim$ 70 MeV, 
which are larger than the typical nucleon momenta in the deuteron (the median of
the nonrelativistic momentum distribution is $\sim 70$ MeV).
In such fixed--target experiments it is difficult to get slow protons (or neutrons) out of the target 
and measure their momenta with sufficient resolution, which restricts the measurements to large
recoil momenta and prevents on-shell extrapolation.

Much more suitable for tagged DIS measurements are colliding--beam experiments,
where the spectator nucleon moves on with approximately half the deuteron beam
momentum and can be detected using forward detectors. Both EIC designs presently
discussed include capabilities for forward nucleon detection
\cite{Abeyratne:2012ah,Aschenauer:2014cki,JLEIC_webpage,eRHIC_webpage}. 
The JLab EIC detector is designed to provide 
full coverage for spectator protons down to zero transverse momentum, and with a momentum
resolution corresponding to $|\bm{p}_p| \sim$ 20 MeV in the rest frame, as well as
forward neutron detection. This setup would enable measurements of
deuteron DIS with spectator tagging over the entire $(x, Q^2)$ range covered by
the collider and thus permit extraction of neutron structure and study of 
nuclear modifications with control of the nuclear configuration. It would also
allow for tagged measurements on the polarized deuteron, which is potentially
the most precise method for determining neutron spin structure.

The theoretical analysis of tagged DIS on the deuteron relies essentially on the
analytic properties of the scattering amplitude (and cross section) in the
recoil proton momentum.  As a function of the invariant 4--momentum transfer
between the deuteron and the recoiling proton, $t \equiv (p_d - p_p)^2$, the
cross section has a pole at $t = M_N^2$ (we assume isospin symmetry and
denote the common nucleon mass by $M_N \equiv M_{p, n}$). The pole is contained 
in the impulse approximation (IA) amplitude and corresponds to the scattering from 
an on-shell neutron in the deuteron in unphysical kinematics. According to the general
principles of scattering theory, the residue at the pole is given by the
structure function of the \textit{free neutron.} Nuclear binding and final-state
interactions (FSI) affect only the tagged deuteron structure functions away from the
pole, at $t - M_N^2 \neq 0$, not the residue at the pole. This makes it possible to
extract the free neutron structure function, by measuring the
proton-tagged DIS cross section as a function of $t$ and extrapolating to the
on-shell point $t \rightarrow M_N^2$. In terms of the recoil momentum in the
rest frame $t - M_N^2 = -2 |\bm{p}_p|^2 + t'_{\rm min}$, where $t'_{\rm min} = -M_d
\epsilon_d = -0.0041 \, \textrm{GeV}^2$ ($\epsilon_d$ = 2.2 MeV is the deuteron binding energy), 
so that the on-shell point corresponds to unphysical values of the recoil momentum
extremely close to zero, $|\bm{p}_p|^2 = -t'_{\rm min}/2$. The method is
model-independent and relies only on general properties of the tagged DIS cross
section (analyticity, position of singularities).  It has considerable
theoretical appeal and can be turned into a practical tool, given sufficiently precise
data at small recoil momenta.

Away from the nucleon pole, at $t - M_N^2 \neq 0$, the recoil momentum
dependence of the tagged DIS cross section is modified by FSI.
They result from amplitudes in which the final state
produced in the DIS process on the active nucleon rescatters from the spectator
nucleon and changes its momentum. They exhibit a complex dependence on the
recoil momentum angle and magnitude, dictated by the kinematics of the
rescattering process, and on $x$, because the character of the nucleon DIS final
state changes as a function of the latter. The FSI effects in the tagged cross
section need to be estimated quantitatively in order to assess the feasibility
of neutron structure extraction through on-shell extrapolation. The same is
needed in order to explore the possibility of separating initial-state nuclear
modifications from final-state interactions in tagged DIS. Such an estimate
requires a theoretical model of FSI appropriate to the region of $x$ explored in
the tagged DIS experiments.

In this article we develop the theoretical framework for tagged DIS measurements
on the deuteron in the kinematic region explored with a medium-energy EIC.
We use methods of LF quantum mechanics to separate nuclear and nucleonic 
structure in the high-energy process and enable dynamical calculations of the 
deuteron structure elements. We derive the tagged structure functions in the IA
and study their symmetries and analytic properties in the recoil momentum. We then develop a 
dynamical model of nuclear FSI at ``intermediate'' $x$, defined as the region between 
the extreme valence quark regime at $x \gtrsim 0.5$ and the coherent regime at $x \ll 0.1$
(the role of coherent phenomena in FSI at $x \ll 0.1$ will be considered in a separate 
study \cite{Guzey17}). This intermediate region is 
of prime interest for the study of sea quarks and gluons and their nuclear modifications.
We use our model to estimate the magnitude and kinematic dependence of the FSI effects, 
demonstrate their analytic properties, and study the implications for neutron 
structure extraction through on-shell extrapolation.

\begin{figure}[t]
\includegraphics[width=.28\textwidth]{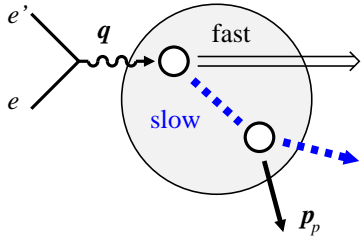}
\caption[]{Physical picture of FSI in electron--deuteron DIS with proton tagging,
$e + d \rightarrow e' + p(p_p) + X$, in the deuteron rest frame. A slow hadron in the 
final state produced by DIS on the active neutron scatters from the spectator proton, 
changing its momentum compared to the IA. The fast component of the DIS final state 
does not interact strongly with the spectator.}
\label{fig:fsi_phys}
\end{figure}

Our treatment is based on a definite physical picture of FSI at intermediate $x$.
The DIS process on the nucleon with momentum transfers
$|\bm{q}| \gg 1\, \textrm{GeV}$ (in the deuteron rest frame) produces a broad spectrum of
hadrons, ranging in momenta from $|\bm{p}_h| \sim |\bm{q}|$ to $|\bm{p}_h|
\lesssim 1\, \textrm{GeV}$. The ``fast'' part of the nucleon DIS final state
does not interact strongly with the spectator nucleons in the nucleus. This
assertion is supported by empirical and theoretical arguments. Nuclear DIS
data show that fast hadrons with $|\bm{p}_h| \gg 1\, \textrm{GeV}$ are not
substantially attenuated in nuclei \cite{Arvidson:1984fz,Ashman:1991cx,%
Adams:1993wv,Adams:1995nu,Airapetian:2000ks,Airapetian:2007vu}; the soft neutron spectra produced 
by nuclear breakup in DIS likewise indicate the absence of 
strong FSI \cite{Adams:1995nu,Strikman:1998cc}. 
Theoretical estimates of the hadron formation time show that such fast hadrons 
form mainly outside of the nucleus and cannot
interact with hadronic cross sections; see Section 5.10 of Ref.~\cite{Boer:2011fh}
for an overview. The dominant FSI in tagged DIS on the
deuteron therefore comes from ``slow'' hadrons with $|\bm{p}_h| \lesssim 1\,
\textrm{GeV}$ in the nucleon DIS final state. Such hadrons are formed inside the nucleus
and can interact with the spectator with hadronic cross sections (see
Fig.~\ref{fig:fsi_phys}). In the terminology of DIS, the ``fast'' and ``slow''
part of the DIS final state on the nucleon correspond to the current and target
fragmentation regions (see Sec.~\ref{sec:hadron}). We note that the physical picture of 
FSI proposed here is consistent with the general QCD factorization theorem for target 
fragmentation in DIS, which is a rigorous asymptotic result and holds irrespective 
of the type of target (nucleon or nucleus) \cite{Trentadue:1993ka,Collins:1997sr}. 
FSI of the ``fast'' DIS hadrons with the nuclear remnant would amount to a violation 
of factorization for the nuclear target in the asymptotic regime. In contrast, FSI of 
the ``slow'' DIS hadrons with the nuclear remnant represent a particular soft-interaction 
contribution to the nuclear fracture function that is allowed by the factorization theorem.
%
%

We express this physical picture of FSI in a schematic model. We calculate
the tagged DIS cross section on the deuteron using LF quantum mechanics, including 
the IA and FSI amplitudes. We use empirical hadron distributions,
measured in $ep/ed$ DIS, to describe the slow part of the hadronic final state
produced on the nucleon in the deuteron. The interactions of the slow hadrons
with the spectator are treated as on-shell scattering with an effective cross
section. Off-shell effects can be absorbed into the effective cross section and
the slow hadron distribution; they are physically indistinguishable from effects
of the finite hadron formation time and can consistently be accounted for in this way.
This model amounts to a minimal description of FSI based on the space-time evolution 
of the DIS process and empirical hadron distributions.

In the present study we use the apparatus of LF quantum mechanics to describe the
initial-state nuclear structure and final-state interactions in tagged DIS.
High-energy processes such as DIS effectively probe a strongly interacting system 
at fixed LF time $x^+ = x^0 + x^3$, along the direction defined by the 
reaction axis. In LF quantization one follows the time evolution of the system in 
$x^+$ and describes its structure by wave functions and densities at $x^+ =$ const.
\cite{Dirac:1949cp,Leutwyler:1977vy,Coester:1992cg,Brodsky:1997de,Heinzl:1998kz}.
The scheme is unique in that it permits a composite description in which effects of
the off-shellness of the constituents remain finite as the scattering 
energy becomes large \cite{Frankfurt:1981mk}. 
It makes possible a composite description of nuclear structure in DIS in terms of nucleon degrees 
of freedom, which exhibits a close correspondence with non-relativistic nuclear structure ($NN$ 
interactions, wave functions), satisfies sum rules (baryon number, LF momentum), 
and enables a smooth matching with nucleon structure (parton picture, QCD degrees of 
freedom) \cite{Frankfurt:1981mk,Frankfurt:1988nt}.
It is important to understand that the structure thus described is ``low-energy'' nuclear structure,
governed by interactions and degrees of freedom on the nuclear scale; it is only presented in a 
way that is appropriate for the initial state of high-energy processes. The application of LF 
quantum mechanics to nuclear high-energy processes is described in detail in 
Refs.~\cite{Frankfurt:1981mk,Frankfurt:1988nt}; the elements used in the present calculation
are summarized below.

Conservation of baryon number and LF momentum is an important consideration 
in describing nuclear DIS.  The LF IA for the inclusive
DIS structure functions correctly implements the baryon number and momentum sum rules 
for the deuteron, i.e., the baryon numbers of the $p$ and $n$ add up to the
total baryon number of the deuteron when integrating over all configurations 
in the wave function, and the LF ``plus'' momenta of the $p$ and $n$ add up to the total 
momentum of the deuteron \cite{Frankfurt:1981mk}. It means that at this level there are 
no non-nucleonic degrees of freedom and ensures that, when the nucleons are resolved into
partons, the partonic sum rules for the deuteron are satisfied. In the tagged
DIS structure functions one recovers these sum rules when integrating over the 
spectator recoil momentum. FSI in tagged DIS may distort the recoil momentum distribution
but should not modify the sum rules for the recoil momentum--integrated structure functions.
In our picture this can be accomplished by modeling the slow hadron--nucleon
rescattering process as elastic scattering (no additional hadrons are produced)
and implementing unitarity of the rescattering amplitude. In this sense our model
of FSI in tagged DIS preserves the baryon number and momentum sum rules and is consistent 
with the standard LF treatment of the inclusive DIS structure functions.

The plan of the article is as follows. In Sec.~\ref{sec:cross} we present the
kinematic variables and invariant structure functions in tagged DIS, introduce
the collinear frame used in the LF description, and discuss the recoil
momentum variables. In Sec.~\ref{sec:lightfront} we summarize the elements of
LF quantum mechanics used in our calculations --- the single-nucleon
states, the deuteron LF wave function, and its rotationally symmetric
representation. In Sec.~\ref{sec:ia} we calculate the tagged DIS cross section
in the IA and study its properties. We discuss the LF current components, 
compute the IA current and structure functions, introduce the LF spectral function, 
and discuss the non-relativistic limit and the analytic properties in $t$.  
In Sec.~\ref{sec:hadron} we discuss the slow hadron
distribution in DIS on the nucleon --- the kinematic variables, structure
functions, and the features of empirical distributions.  In Sec.~\ref{sec:fsi}
we calculate the FSI effects in tagged DIS in LF quantum mechanics and
study their properties. The FSI effects are expressed in terms of a distorted
spectral function.  We formulate a factorized approximation using the fact
that the range of the rescattering interaction is small compared to the deuteron
size. We demonstrate the positivity of the cross section, investigate the recoil
momentum dependence of the FSI effects, and discuss the analytic properties of
the distorted spectral function. We also comment on the role of unitarity of the 
rescattering process and the implementation of sum rules for the LF spectral
function with FSI. In Sec.~\ref{sec:neutron}
we discuss the strategy for neutron structure measurements in tagged DIS with
EIC. A summary and outlook are given in Sec.~\ref{sec:summary}.

Technical material needed to reproduce the calculations is summarized in
appendices.  Appendix~\ref{app:deuteron} describes a simple two-pole
parametrization of the deuteron wave function with correct analytic properties,
which we use in the numerical calculations.
Appendix~\ref{app:projection} contains the projection formulas for extracting
the tagged structure functions from the deuteron tensor in the IA and
with FSI. Appendix~\ref{app:rescattering_amplitude} summarizes our
parametrization of the nucleon-nucleon cross section used in the numerical
estimates of FSI effects. Appendix~\ref{app:integral} describes the evaluation 
of the rescattering phase-space integral in LF coordinates. 

The physical picture of FSI in tagged DIS proposed here is specific to the region 
of intermediate $x$ as defined above, and our model should be used in this context only.  
In tagged DIS in the limit $x \rightarrow 1$ the minimal rest-frame momentum of the ``slow'' 
hadrons produced on the nucleon becomes large, as their LF momentum fractions are bounded 
by $1 - x$, and the interactions are suppressed by the hadron formation time 
(see Fig.~\ref{fig:fsi_phys}). The framework developed in the present work could be extended 
to this region when supplied with empirical information about the formation time effects. 
(The work of Refs.~\cite{Cosyn:2010ux,Cosyn:2013uoa,Cosyn:2017ekf} considers tagged DIS 
at large $x$ in a subasymptotic domain of fixed energy and momentum transfer, where the 
limit $x \rightarrow 1$ corresponds to a small inelasticity $W^2 - M_N^2$, and the 
DIS final state is modeled as a superposition of baryonic resonances; this formulation is 
different from the asymptotic one presented here and not appropriate for collider energies.) 
In tagged DIS at small $x$ ($\ll 0.01$), diffractive
scattering on the nucleon becomes significant and gives rise to a new mechanism
of FSI. In diffractive scattering the nucleon appears intact in the final state
and recoils with a momentum $p_p \sim\textrm{few 100 MeV}$.  Because the
diffractive nucleon retains its quantum numbers (``vacuum exchange'') and
emerges with a small recoil momentum, there is a significant amplitude for the
final-state $pn$ system to revert back to a deuteron bound state. In tagged DIS
in this kinematics the outgoing $pn$ scattering state must be properly
orthogonalized to the deuteron bound state, which results in significant
distortion. This new mechanism of FSI in tagged DIS on the deuteron at small $x$
is closely related to shadowing in inclusive DIS and will be discussed elsewhere
\cite{Guzey17}.

In the present work we consider unpolarized electron--deuteron DIS and calculate
the FSI effects in the tagged cross section integrated over the azimuthal angle
of the recoil momentum, as relevant for the extraction of the neutron structure
functions $F_{2n}$ and $F_{Ln}$. The extension to polarized electron-deuteron
DIS with spectator tagging and azimuthal angle--dependent response functions
will be left to a future study, as the number of structures in the cross section
becomes very considerable \cite{Cosyn17}. A proper treatment of FSI in the polarized deuteron
would require also empirical information on the spin dependence of the slow DIS 
hadron distributions, which is not available at present. We note that the schematic 
model of FSI proposed here could be applied also to the time--reversal--odd ($T$--odd) 
response functions in tagged DIS, which are zero in the IA and can be used for 
sensitive tests of the FSI dynamics.

To simplify the presentation we suppress the internal spin structure of the deuteron 
and present the IA and FSI expressions for an S--wave bound state. This allows us to
leave aside for the moment the complications resulting from the treatment of spin
in LF quantum mechanics (Melosh rotations, angular conditions) and focus on the 
aspects essential to FSI. The resulting expressions are a good approximation at 
recoil proton momenta $|\bm{p}_p| < 200\, \textrm{MeV}$, where the $S$--wave dominates 
in the deuteron's nonrelativistic momentum density (see below), and which are of
prime interest for neutron structure extraction. The expressions for the IA
and FSI cross sections derived here can easily be generalized to account for the deuteron 
spin structure, by including the summation over LF helicity components \cite{Frankfurt:1983qs}.

FSI effects in DIS from nuclei and their kinematic dependence were studied in
Refs.~\cite{CiofidegliAtti:2002as,CiofidegliAtti:2003pb,Palli:2009it,Atti:2010yf} using a 
detailed microscopic model of hadron production in DIS (string breaking, gluon
radiation). In contrast to these studies we use a simple generic description of
hadron production and consider specifically the region of intermediate $x$.  FSI
effects have also been studied extensively in quasi-elastic scattering from
nuclei, including deuteron electrodisintegration $e + d \rightarrow e' + p + n$
\cite{Frankfurt:1994kt}; see Ref.~\cite{Boeglin:2015cha} for a review. 
There is an interesting formal analogy between FSI in
quasi-elastic deuteron breakup at $\sim$1--2 GeV incident momenta and our
picture of slow-hadron rescattering in DIS, and one can establish the correspondence 
between the formulas.
\section{Tagged DIS kinematics}
\label{sec:cross}
\subsection{Kinematic variables}
\begin{figure}
\includegraphics[width=.2\textwidth]{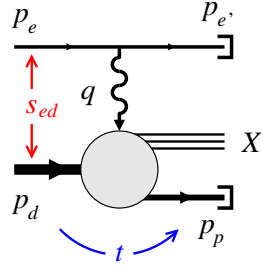}
\caption[]{Inclusive electron scattering from the deuteron with 
an identified proton in the deuteron fragmentation region, $e + d \rightarrow e' + p + X$
(``tagged DIS'').}
\label{fig:deut_tagged}
\end{figure}

We begin by summarizing the kinematic variables and cross section formulas for
inclusive electron scattering on the deuteron with an identified nucleon in the 
final state (``tagged DIS''). The kinematic factors are given in their exact 
form (no simplifications are made using the DIS limit) and expressed in terms of 
relativistic invariants, as suitable for collider experiments. The cross section
formulas given in this section are general and make no assumption regarding composite 
nuclear structure; particular results based on such approximations will be presented 
in Secs.~\ref{sec:ia} and \ref{sec:fsi}. To be specific we consider the case that the 
identified nucleon is a proton; equivalent formulas can be written for the case 
of an identified neutron. Thus, we consider the scattering process (see Fig.~\ref{fig:deut_tagged})
\be
e(p_e) + d(p_d) &\rightarrow & e'(p_{e'}) + p(p_p) + X ,
\label{dis_tagged}
\ee
where $X$ denotes an unresolved hadronic final state. 
Here $p_e$ and $p_{e'}$ are the 4--momenta of the initial and final electron,
$p_d$ is the 4--momentum of the deuteron, and $p_p$ is the 4--momentum of the 
identified proton. The 4--momentum transfer to the nuclear 
system, calculated from the initial and final electron 4--momenta, is
\beq
q \; \equiv \; p_e - p_{e'}, \hspace{2em} Q^2 \; \equiv \; - q^2 \; > \; 0.
\eeq
Invariants formed from the electron and deuteron 4--momenta are
\be
s_{ed} &=& (p_e + p_d)^2 ,
\\[1ex]
W_d^2 &=& (q + p_d)^2 ,
\ee
which describe, respectively, the electron--deuteron and the virtual photon--deuteron 
squared CM energies. Useful scaling variables are
\be
x_d \; &\equiv& \; \displaystyle \frac{-q^2}{2 (p_d q)} 
\; = \; \frac{Q^2}{W_d^2 - M_d^2 + Q^2} ,
\\[3ex]
y \; &\equiv& \; \displaystyle \frac{(p_d q)}{(p_d k)} \; = \; 
\frac{Q^2}{x_d (s_{ed} - M_d^2)} .
\ee
The variable $x_d$ is the Bjorken variable for the nuclear target $(0 < x_d < 1)$ 
and will be used in the kinematic formulas for the cross section, to facilitate comparison 
with the standard expressions for electron--proton scattering. In the description of
composite deuteron structure we shall use the alternative variable 
\beq
x \; \equiv \; 2 x_d \hspace{2em} (0 < x < 2),
\label{x_def}
\eeq
which corresponds to the effective Bjorken variable for scattering from a 
nucleon in the unbound nucleus (deuteron). The variable $y$ describes the electron's 
fractional energy loss (or inelasticity) in the deuteron rest frame.
The invariants and scaling variables formed with the recoil nucleon momentum $p_p$
will be presented in Sec.~\ref{subsec:recoil} below.
%
%

\subsection{Cross section and structure functions}
\label{subsec:cross_section}
The invariant amplitude for the electroproduction of a final state $p + X$, including the 
detected proton $p$ and a specified set of hadrons $X$, is in leading order of 
the electromagnetic coupling given by
\beq
\mathcal{M}[ed \rightarrow e'pX]
\;\; = \;\; \frac{e^2}{q^2} \; \langle e', \bm{p}_{e'}| \hat{J}^\mu (0) 
| e, \bm{p}_e \rangle \; 
\langle p, \bm{p}_p; X | \hat{J}^\mu (0) 
| d, \bm{p}_d \rangle ,
\eeq
where $e$ is the elementary charge, and the brackets denote the transition matrix elements
of the electromagnetic 4--vector current $\hat{J}^\mu (0)$ 
between the initial and final electron and nuclear/hadronic states.
All particle states (electron, deuteron, nucleon) are normalized according to the relativistic 
convention
\be
\langle e, \bm{p}_{e2} | e, \bm{p}_{e1} \rangle
&=&  (2\pi)^3 \, 2 E_e(\bm{p}_{e1}) \; \delta^{(3)}(\bm{p}_{e2} - \bm{p}_{e1}) , 
\hspace{2em} \textrm{etc.}
\label{relativistic_normalization}
\ee
The spin quantum numbers of the states are suppressed for brevity and will be specified below. 
The differential cross section for production of the specified hadronic state is \cite{LLIV}
\beq
d\sigma [ed \rightarrow e'pX] \;\; = \;\; (2\pi)^4 
\delta^{(4)} (p_e + p_d - p_{e'} - p_p - p_h) \; \frac{|\mathcal{M}|^2}{4I} 
\; d\Gamma_{e'} \; d\Gamma_p \; d\Gamma_X .
\label{dsigma_h}
\eeq
The invariant incident particle current is defined as (we neglect the electron mass)
\be
I &\equiv & \sqrt{(p_e p_d)^2} 
\;\; \equiv \;\; {\textstyle\frac{1}{2}} (s_{ed} - M_d^2 ) .
\ee
The invariant phase space elements of the scattered electron and
the identified proton are
\be
d\Gamma_{e'} \; &\equiv& \; \frac{d^3 p_{e'}}{(2\pi)^3 2 E_{e'}} ,
\hspace{2em}
E_{e'} \; =\;  |\bm{p}_{e'}| ,
\\[2ex]
d\Gamma_p \; &\equiv& \; \frac{d^3 p_p}{(2\pi)^3 2 E_p} ,
\hspace{2em}
E_p \; =\;  \textstyle{\sqrt{|\bm{p}_p|^2 + M_N^2}} .
\ee
The phase space element of the multi--hadron state $X$ can be defined analogously
in terms of the hadron momenta, but its explicit form is not needed in the following.
The cross section for tagged inclusive scattering 
is then given by integrating Eq.~(\ref{dsigma_h}) over the phase space of the 
unidentified hadron state $X$ and summing over all such states (we 
denote both operations together symbolically by $\sum\limits_X$),
\be
d\sigma [ed \rightarrow e'pX] \; &=& \; \sum_X \; (2\pi)^4 
\delta^{(4)} (p_e + p_d - p_{e'} - p_p - p_X) \; \frac{|\mathcal{M}|^2}{4I} 
\; d\Gamma_{e'} \; d\Gamma_p
\\
\; &=& \; \frac{4\pi e^4}{4I (-q^2)^2} \; w_{\mu\nu} W_d^{\mu\nu} 
\; d\Gamma_{e'} \; d\Gamma_p .
\label{dsigma_leptonic_hadronic}
\ee
The leptonic tensor is defined as
\be
w^{\mu\nu} \; &\equiv& \; w^{\mu\nu}(p_{e'}, p_e) \;\; = \;\; 
\langle e', \bm{p}_{e'}| \hat{J}^\mu (0) 
| e, \bm{p}_e \rangle^\ast \;
\langle e', \bm{p}_{e'}| \hat{J}^\nu (0) 
| e, \bm{p}_e \rangle .
\ee
In the case of scattering of an unpolarized electron beam (average over initial helicities) 
and unspecified polarization of the final electron (sum over final helicities)
its explicit form is
\beq
w^{\mu\nu} \;\; = \;\; 4 p_{e}^\mu p_{e}^\nu + q^2 g^{\mu\nu} \; + \; 
\textrm{terms $\propto q^\mu, q^\nu$}
\eeq
The deuteron tensor is defined as (using $q = p_e - p_{e'}$)
\be
W_d^{\mu\nu} &\equiv& W_d^{\mu\nu} (p_d, q, p_p)
\nonumber
\\[2ex]
&=& (4\pi)^{-1} \sum_X \; (2\pi)^4 \; 
\delta^{(4)} (q + p_d - p_p - p_X) \; 
\langle p X| \hat J^\mu (0) |d \rangle^\ast \;
\langle p X| \hat J^{\nu}(0) |d \rangle  \; 
\\[1ex]
&=& (4\pi)^{-1} \sum_h \; (2\pi)^4 \; 
\delta^{(4)} (q + p_d - p_p - p_X) \; 
\langle d| \hat J^{\mu\dagger}(0) |pX \rangle \;
\langle pX| \hat J^{\nu}(0) |d \rangle ,
\label{w_mu_nu_current}
\ee
in which $J^{\mu\dagger} = J^\mu$ (hermiticity of the electromagnetic current operator).
It obeys the transversality conditions $q_\mu W_d^{\mu\nu} = 0$ and $W_d^{\mu\nu} q_\nu = 0$
and can be expanded in tensors constructed from the 4--vectors $p_d, q$ and $p_p$, and the
invariant tensor $g^{\mu\nu}$. It is convenient to introduce the auxiliary 4--vectors
\be
L^\mu &\equiv& p_d^\mu - \frac{(p_d q) q^\mu}{q^2}, 
\hspace{2em} (qL) \; = \; 0, \hspace{2em} L^2 \; > \; 0, 
\label{L_def}
\\[1ex]
T^\mu &\equiv& p_p^\mu - \frac{(p_p q) q^\mu}{q^2} 
- \frac{(p_p L) L^\mu}{L^2}, 
\hspace{2em} (qT) \; = \; 0,  \hspace{1em} (LT) \; = \; 0, 
\hspace{2em} T^2 \; < \; 0 .
\label{T_def}
\ee
Their particular meaning in a frame where $p_d$ and $q$ are collinear is
explained in Sec.~\ref{subsec:collinear}. We decompose the deuteron tensor as
\be
W_d^{\mu\nu} 
&=& \left( g^{\mu\nu} - \frac{q^\mu q^\nu}{q^2} \right) \frac{F_{Ld}}{2}
\; + \; \left( \frac{L^\mu L^\nu}{L^2} + \frac{q^\mu q^\nu}{q^2} 
- g^{\mu\nu} \right) \frac{F_{Td}}{2}
\nonumber
\\[1ex]
&+& \frac{T^\mu L^\nu + L^\mu T^\nu}{\sqrt{-T^2}\sqrt{L^2}} \; \frac{F_{LT, d}}{2}
\; + \; \left( g^{\mu\nu} - \frac{L^\mu L^\nu}{L^2} - \frac{q^\mu q^\nu}{q^2} 
- \frac{2 T^\mu T^\nu}{T^2} \right) \frac{F_{TT, d}}{2}
\label{W_decomposition}
\ee
Here $F_{Ld}, F_{Td}, F_{LT, d}$ and $F_{TT, d}$ are invariant structure functions,
depending on the kinematic invariants formed from the vectors
$q, p_d$ and $p_p$. The first and second tensor structures in
Eq.~(\ref{W_decomposition}) do not involve the identified proton momentum $p_p$
are are present also in untagged (fully inclusive) scattering. Our definition of the
longitudinal and transverse structure functions, $F_{Ld}$ and $F_{Td}$,
is identical to that of Ref.~\cite{Ellis:1982cd}. Their relation to the conventional 
functions $F_{1d}$ and $F_{2d}$ is 
\be
F_{Td} \; &=& \; \frac{(-q^2) L^2}{(p_d q)^2} \; \frac{F_{2d}}{x_d}
\; = \; \left( 1 + \frac{4 x_d^2 M_d^2}{Q^2} \right) \; \frac{F_{2d}}{x_d} ,
\label{FT_from_F2}
\\[1ex]
F_{Ld} \; &=& \; \frac{(-q^2) L^2}{(p_d q)^2} \; \frac{F_{2d}}{x_d} - 2 F_{1d}
\; = \; \frac{1}{x_d} \left[ 
\left( 1 + \frac{4 x_d^2 M_d^2}{Q^2} \right) F_{2d} - 2 x_d F_{1d} \right] ,
\\[2ex]
F_{Td} - F_{Ld} \; &=& \; 2 F_{1d} .
\ee
The third and fourth tensor structures in Eq.~(\ref{W_decomposition}) 
vanish when averaging over the orientation of the vector $T$ in the 
plane orthogonal to $L$ and $q$ and are present only for fixed momentum $p_p$. 

The contraction of the leptonic and deuteron tensors can be expressed 
in terms of the parameter
\beq
\epsilon \;\; \equiv \;\; \frac{\displaystyle w_{\mu\nu} \frac{L^\mu L^\nu}{L^2}}
{\displaystyle 
w_{\mu\nu} \left( \frac{L^\mu L^\nu}{L^2} + \frac{q^\mu q^\nu}{q^2} 
- g^{\mu\nu} \right)}
\;\; =\;\; \frac{\displaystyle 1 - y - \frac{x_d^2 y^2 M_d^2}{Q^2}}
{\displaystyle 1 - y + y^2/2 + \frac{x_d^2 y^2 M_d^2}{Q^2}} ,
\eeq
which can be interpreted as the ratio of the probabilities of longitudinal 
and transverse polarization of the virtual photon. To express the contractions of 
the $p_p$--dependent tensor structures in invariant form we expand the initial 
electron momentum as
\be
p_e &=& \frac{(p_e L) L}{L^2} + \frac{(p_e q) q}{q^2} + \Delta ,
\hspace{2em}
(\Delta L) \; = \; 0, \hspace{2em} (\Delta q) \; = \; 0,
\\[1ex]
(p_e T) &=& (\Delta T)  \;\; = \;\; \sqrt{-\Delta^2} \sqrt{-T^2} \; \cos\phi_p .
\label{phi_p_def}
\ee
The particular meaning of the angle $\phi_p$ in a collinear frame is described in
Sec.~\ref{subsec:collinear}. We obtain
\be
w_{\mu\nu} W^{\mu\nu} \; &=& \; 
\frac{Q^2}{1 - \epsilon}
\left[ 
\frac{1 + \epsilon}{1 - \epsilon} \; \frac{y^2}{(2 - y)^2} \; F_{Td}
\; - \; (1 - \epsilon) \; F_{Ld} 
+ \sqrt{2 \epsilon (1 + \epsilon)} \; \cos \phi_p \; F_{LT, d}
\; + \; \epsilon \; \cos (2 \phi_p) \; F_{TT, d}
\right]
\\[1ex]
&=& \; \frac{Q^2}{1 - \epsilon}
\left[ \frac{F_{2d}}{x_d} \; - \; (1 - \epsilon) \; F_{Ld} \; + \; \ldots 
\right] .
\ee
The scattered electron phase space element can easily be expressed in terms of the
invariants $x_d$ and $Q^2$ and the azimuthal angle around the incident electron 
momentum direction, $\phi_{e'}$. Altogether, the differential 
cross section for tagged inclusive scattering with unpolarized beams 
and recoil nucleon, Eq.~(\ref{dsigma_leptonic_hadronic}), becomes
\be
d\sigma [ed \rightarrow e'pX] \; &=& \; \frac{2\pi \alpha_{\rm em}^2 y^2}{Q^4 (1 - \epsilon)}
\; dx_d \, dQ^2 \, \frac{d\phi_{e'}}{2\pi} \nonumber \\
&\times& \; \left[ \frac{F_{2d}}{x_d} \; - \; (1 - \epsilon) F_{Ld}
\; + \; \sqrt{2\epsilon (1 + \epsilon)} \cos\phi_p F_{LT, d}
\; + \; \epsilon \cos (2 \phi_p) F_{TT, d} \right] \; 
d\Gamma_p ,
\label{dsigma_tagged}
\ee
where $\alpha_{\rm em} \equiv e^2/(4\pi) \approx 1/137$ is the fine structure constant.
The last two terms in the bracket drop out when the cross section is integrated over the recoil 
azimuthal angle $\phi_p$. Specific forms of the recoil momentum phase space element are described 
in Sec.~\ref{subsec:recoil}.
\subsection{Collinear frames}
\label{subsec:collinear}
\begin{figure}
\includegraphics[width=.5\textwidth]{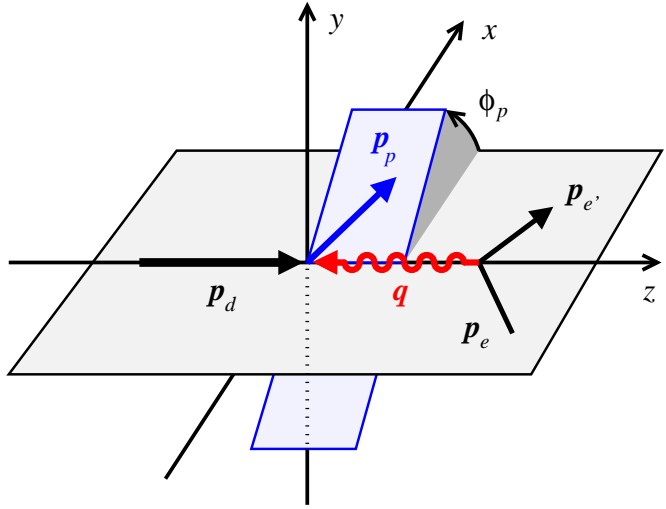}
\caption[]{Tagged DIS in the collinear frame, Eq.~(\ref{collinear_frame}). The
deuteron momentum $\bm{p}_d$ and the vector $\bm{q}$ are collinear and define the
$z$--axis, with $\bm{q}$ pointing in the negative $z$--direction. The initial and final electron 
momenta lie in the $xz$ plane and have positive $x$ component. $\phi_p$ is the
angle of the transverse ($xy$) component of the recoil momentum, measured relative to the
positive $x$ axis.}
\label{fig:collinear_frame}
\end{figure}

In the theoretical description of tagged DIS we consider the process Eq.~(\ref{dis_tagged}) 
in a frame where the deuteron momentum $\bm{p}_d$ and the momentum transfer $\bm{q}$
are collinear and define the $z$--axis of the coordinate system. 
This condition does not specify a unique frame, but rather a class of 
frames that are related by boosts along the $z$--axis (``collinear frames''). 
We specify the 4--momenta in this frame by their LF components
\beq
p^\pm \;\; \equiv \;\; p^0 \pm p^z, \hspace{2em}
\bm{p}_T \;\; \equiv \;\; (p^x, p^y) .
\label{lc_def}
\eeq
The LF components of $p_d$ and $q$ in the collinear frame are
\beq
\left.
\begin{array}{rclrclrcl}
p_d^+ &>& 0 \;\; \textrm{(arbitrary),} &
\hspace{2em} p_d^- &=& \displaystyle \frac{M_d^2}{p_d^+}, &
\hspace{2em} \bm{p}_{dT} &=& 0 , \hspace{2em}
\\[3ex]
q^+ &=& -\xi_d p_d^+, & 
q^- &=& \displaystyle \frac{q^2}{q^+} \; = \; \frac{Q^2}{\xi_d p_d^+}, &
\bm{q}_{T} &=& 0 .
\end{array}
\right\}
\label{collinear_frame}
\eeq
The parameter $\xi_d$ is fixed by the condition
\beq
2 p_d q \;\; = \;\; p_d^+ q^- + p_d^- q^+ \;\; = \;\; \frac{Q^2}{x_d} ,
\eeq
the solution of which is
\beq
\xi_d \;\; = \;\; \frac{2 x_d}{1 \pm \sqrt{1 + 4 M_d^2 x_d^2 / Q^2}}
\hspace{3em} \textrm{(choose +)}.
\eeq
We select the solution with the plus sign, which has the property that
in the scaling limit $Q^2 \gg M_d^2$
\beq
\xi_d \;\; = \;\; x_d \; + \; O(M_d^2/Q^2).
\eeq
With this choice the momentum transfer vector $\bm{q}$ 
points in the \textit{negative} $z$--direction (see Fig.~\ref{fig:collinear_frame}),
\beq
2 q^z \;\; = \;\; 
q^+ - q^- \;\; = \;\; -\xi_d p_d^+ - \frac{Q^2}{\xi_d p_d^+} \;\; < \;\; 0 .
\eeq
%
%

The LF components of the longitudinal auxiliary 4--vectors $L$ and $T$,
Eqs.~(\ref{L_def}) and (\ref{T_def}), are obtained as
\beq
\left.
\begin{array}{rclrclrcl}
L^+ &=& \displaystyle \left( 1 - \frac{\xi_d}{2 x_d} \right) p_d^+ , &
\hspace{2em}
L^- &=&  \displaystyle \left( M_d^2 + \frac{Q^2}{\xi_d} \right) \frac{1}{p_d^+} , &
\hspace{2em} \bm{L}_T &=& 0 ,
\\[3ex]
T^+ &=& 0, & 
T^- &=& 0, & 
\bm{T}_T &=&  \bm{p}_{pT} .
\end{array}
\hspace{2em} 
\right\}
\label{L_deuteron}
\eeq
The vector $L$ has only collinear components, while $T$ has only transverse components
and coincides with the recoil hadron transverse momentum in the collinear frame. Because the
momentum transfer $\bm{q}$ is along the (negative) $z$--direction, the initial and
final electron momenta have the same transverse components,
\beq
\bm{p}_{eT} \; = \; \bm{p}_{e'T} \hspace{2em} \textrm{(collinear frame)} ,
\eeq
such that they define a plane together with the $z$--axis (electron scattering plane). 
The recoil angle $\phi_p$, 
defined in terms of invariants in Eq.~(\ref{phi_p_def}), then is the azimuthal angle of 
$\bm{p}_p$, measured relative to the electron scattering plane. It is conventional
to choose the electron transverse momenta in the $x$--direction, such that the
electron scattering plane is defined by the $xz$ plane. In this case the angle 
$\phi_p$ becomes the conventional azimuthal angle of $\bm{p}_p$ in the $xy$ plane, 
$\cos\phi_p = p_p^x/|\bm{p}_{pT}|$ (see Fig.~\ref{fig:collinear_frame}).

The deuteron plus momentum $p_d^+ > 0$ in the above formulas remains arbitrary and 
defines a particular member of the class of collinear frames. Longitudinal boosts 
(along the $z$--axis) can be performed simply by changing the value of $p_d^+$ in 
the above formulas. Note that the class of collinear frames contains several special 
cases of interest: (a) the target rest frame, $p_d^+ = M_d$;
(b)~the Breit frame, $p_d^+ = \sqrt{Q^2}/\xi_d$, in which
$q^0 = (q^+ + q^-)/2 = 0$; (c) the center-of-mass frame of the virtual photon
and the deuteron, $p_d^+ = \sqrt{Q^2 + \xi_d M_d^2} / \sqrt{\xi_d (1 - \xi_d)}$,
in which $q^z = (q^+ - q^-)/2 = - p_d^z = -(p_d^+ - p_d^-)/2$ and thus $\bm{q} = -\bm{p}_d$.
For reference we note that the collinear frames used here are 
equivalent to the covariant formulation of the collinear expansion in terms of 
light--like vectors of Ref.~\cite{Ellis:1982cd}. 
\subsection{Recoil momentum variables}
\label{subsec:recoil}
The tagged structure functions of the deuteron in Eq.~(\ref{W_decomposition})
depend on the usual DIS variables (e.g. $W_d$ and $Q^2$) as well as the recoil nucleon 
momentum. The latter dependence involves two independent variables formed from $p_p$,
related to the two invariants $(p_p p_d)$ and $(p_p q)$; the dependence on the third
invariant $(p_p p_e)$ is encoded in the explicit $\phi_p$ dependence of the cross section. 
Here we describe several physically interesting choices of recoil momentum variables that are 
used in the subsequent calculations. We present their relation to the rest--frame recoil momentum, 
their kinematic limits, and the corresponding phase space elements. We assume isospin symmetry 
and define the nucleon mass as the average of the proton and neutron masses
\be
M_N &=& (M_p + M_n)/2 \; = \; 0.9389 \, \textrm{GeV} .
\ee
The deuteron binding energy and mass are taken at their exact values
\beq
\epsilon_d \; = \; 2.2 \, \textrm{MeV},
\hspace{2em} 
M_d \; = \; M_p + M_n - \epsilon_d \; = \; 2 M_N - \epsilon_d \; = \; 1.8756 \, \textrm{GeV} .
\eeq
Note that the relation between the deuteron binding energy and mass
is not affected when replacing the proton and neutron masses by their average.

In a collinear frame defined by Eqs.~(\ref{collinear_frame}) the tagged structure
functions can be regarded as functions of the LF plus momentum fraction of the recoil proton
and the transverse momentum modulus of the recoil momentum,
\be
\alpha_p \; &\equiv& \; \frac{2 p_p^+}{p_d^+} 
\;\; = \;\; \frac{2 (p_p^0 + p_p^z)}{p_d^0 + p_d^z},
\label{alpha_p_def}
\\[1ex]
|\bm{p}_{pT}| \; &\equiv& \; \sqrt{(p_p^x)^2 + (p_p^y)^2} .
\label{recoil_lf}
\ee
The definition of $\alpha_p$ in Eq.(\ref{alpha_p_def}), as the fraction relative to $p_d^+/2$, 
is natural and leads to simple expressions in the nuclear structure calculations below.
The kinematic limits of $\alpha_p$ are dictated by LF plus momentum conservation in the
scattering process and given by
\beq
\alpha_p/2 \; < \; 1 - \xi_d .
\eeq
The invariant phase space element in terms of $\alpha_p$ and $|\bm{p}_{pT}|$ takes the form
\beq
\frac{d^3 p_p}{E_p} \;\; = \;\; \frac{d\alpha_p}{\alpha_p} \; d^2p_{pT} .
\label{phase_space_alpha_pt}
\eeq

An important variable is the invariant momentum transfer between the initial--state deuteron 
and the final--state nucleon (see Fig.~\ref{fig:deut_tagged}),
\beq
t \;\; \equiv \;\; (p_p - p_d)^2 ,
\eeq
or the reduced variable
\beq
t' \;\; \equiv \;\; t - M_N^2 .
\label{t_prime_def}
\eeq
The theoretical analysis of tagged DIS Eq.~(\ref{dis_tagged}) relies essentially
on the analytic properties of the cross section in $t'$; see Secs.~\ref{sec:ia} and
\ref{sec:fsi} below. The invariant $t'$ is related in a simple manner to the energy
of the recoiling nucleon in the deuteron rest frame (we use ``RF'' to denote rest-frame
energy and momentum),
\beq
t' \;\; = \;\; M_d^2 - 2 M_d E_p(\textrm{RF}) ,
\hspace{2em}
[E_p(\textrm{RF}) \equiv {\textstyle\sqrt{|\bm{p}_p(\textrm{RF})|^2 + M_N^2}}] .
\label{tprime_from_E_p}
\eeq
The kinematic limit of $t'$ is attained at $\bm{p}_p(\textrm{RF}) = 0$,
\be
t' \;\; < \;\; t'_{\rm min} \; &\equiv& \; M_d^2 - 2 M_d M_N \;\; = 
\;\; - M_d \epsilon_d \; = \; -0.0041 \, \textrm{GeV}^2 .
\label{t_prime_min}
\ee
Inside the physical region the rest--frame momentum is obtained from $t'$ as
\be
|\bm{p}_p(\textrm{RF})|^2 &=& 
-\frac{t'}{2}\left(1 - \frac{t'}{2 M_d^2} \right) 
\; + \; \frac{M_d^2}{4} - M_N^2 .
\label{p_p_from_tprime}
\ee
A simpler relation is obtained if we neglect the $t'/(2 M_d^2)$ term in the parenthesis; this
approximation is well justified for typical values $|t'| \sim 0.1\, \textrm{GeV}^2$ 
and becomes exact in the limit $t' \rightarrow 0$. Namely,
\be
|\bm{p}_p(\textrm{RF})|^2 
\; &\approx& \; -\frac{t'}{2} \; + \; \frac{M_d^2}{4} - M_N^2
\;\; = \;\;  -\frac{t'}{2} \; + \; \frac{t'_0}{2} ,
\label{p_p_from_tprime_approx}
\\[2ex]
t'_0 \; &\equiv& \; \frac{M_d^2}{2} - 2 M_N^2 
\;\; = \;\; 
- M_d \epsilon_d \; - \frac{\epsilon_d^2}{2} 
\;\; = \;\; 
- 2 M_N \epsilon_d \; + \frac{\epsilon_d^2}{2} 
\;\;= \;\; t'_{\rm min} \; + \; O(\epsilon_d^2) .
\label{tprime_min_app}
\ee
The difference between $t'_0$ and the exact $t'_{\rm min}$, Eq.~(\ref{t_prime_min}),
is of the order $10^{-6} \, \textrm{GeV}^2$ and negligible for all practical purposes.
In this approximation the invariant $t'$ is the negative of twice the squared rest 
frame recoil momentum, minus a fixed small amount proportional to the deuteron binding energy,
\be
t' \; &\approx& \; - 2 \, |\bm{p}_p(\textrm{RF})|^2 \; + \; t'_0 .
\label{t_prime_from_p_p_approx}
\ee

The relation of the invariant $t'$ to the collinear variables $\alpha_p$ and $|\bm{p}_{pT}|$
can easily be established using the fact that the deuteron rest frame is a special 
collinear frame ($p_d^+ = M_d$). Thus the rest-frame energy and $z$--momentum can be calculated 
in terms of the plus and minus LF components as 
\beq
\left.
\begin{array}{r}
E_p(\textrm{RF})
\\[1ex]
p_p^z(\textrm{RF})
\end{array}
\right\}
 \;\; =\;\; \frac{p_p^+ \pm p_p^-}{2}
\;\; = \;\; \frac{\alpha_p M_d}{4} 
\pm \frac{|\bm{p}_{pT}|^2 + M_N^2}{\alpha_p M_d} ,
\label{pr3_alphar_prt}
\eeq
and $t'$ can be obtained from the above rest-frame formulas. Specifically, with
Eq.~(\ref{t_prime_from_p_p_approx}) we obtain
\beq
t' \;\; = \;\; - 2 \, \left[ |p_p^z(\textrm{RF})|^2 \; + \; |\bm{p}_{pT}|^2 \right] 
\; + \; t'_0 .
\label{tprime_alphar_prt_relation}
\eeq

%
%
\begin{figure}[t]
\includegraphics[width=.5\textwidth]{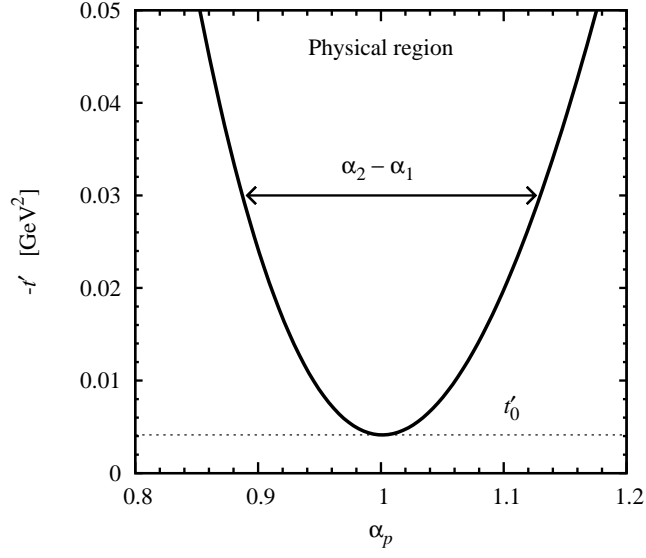}
\caption[]{Physical region of recoil proton momentum phase space in 
the variables $\alpha_p$ and $t'$.}
\label{fig:alphar_tprime}
\end{figure}
In the theoretical analysis below we use a representation in which $\alpha_p$
and $t'$ are independent variables. The physical region in these variables can
easily be established from Eq.~(\ref{tprime_alphar_prt_relation}). For a given 
$\alpha_p$ the kinematic limit in $t'$ is found by minimizing
Eq.~(\ref{tprime_alphar_prt_relation}) with respect to $|\bm{p}_{pT}|$
and given by (see Fig.~\ref{fig:alphar_tprime})
\be
t' \; &<& \; - \frac{2 M_d^2}{\alpha_p^2}
\left( \frac{\alpha_p^2}{4} - \frac{M_N^2}{M_d^2} \right)
\; + \; t'_0 .
\label{t_prime_min_alpha}
\ee
One sees that the minimum value of $-t'$ increases quadratically as $\alpha_p$
moves away from $2 M_N/M_d \approx 1$.
Conversely, for a given $t' < t'_0$ the allowed values of $\alpha_p$ are 
\be
\alpha_1 \; &<& \; \alpha_p \; < \; \alpha_2 ,
\\[1ex]
\alpha_{1, 2} \; &=& \; \frac{2}{M_d} \left[ 
E_p (\textrm{RF}) \mp |\bm{p}_p(\textrm{RF})| \right]
\; = \;  \frac{2}{M_d} \left[ 
\sqrt{\displaystyle \frac{t'_0 - t'}{2} + M_N^2}
\; \mp \; \sqrt{\displaystyle \frac{t'_0 - t'}{2}} \;
\right] .
\ee
The invariant phase space element in this representation is given by [cf.\ Eq.~(\ref{tprime_from_E_p})]
\be
\frac{d^3 p_p}{E_p} &=& \frac{M_d}{8 E_p (\textrm{RF})}
\; d(-t') \; d\alpha_p \; d\phi_p ,
\\[2ex]
E_p (\textrm{RF}) &=& {\textstyle\sqrt{|\bm{p}_p(\textrm{RF})|^2 + M_N^2}}
\;\; = \;\; \sqrt{\displaystyle \frac{t'_0 - t'}{2} + M_N^2} .
\label{phase_tprime_alpha}
\ee

Another physically interesting variable is the angle of the recoil momentum in the deuteron rest frame,
\beq
\cos \theta_p \;\; = \frac{p_p^z(\text{RF})}{|\bm{p}_p (\text{RF})|} .
\eeq
For a given modulus $|\bm{p}_p(\textrm{RF})|$ the angle is related to the LF fraction $\alpha_p$ as
\beq
\alpha_p \;\; = \;\; 
\frac{2}{M_d} \left( \sqrt{|\bm{p}_p (\text{RF})|^2 + M_N^2} \; + \; |\bm{p}_p (\text{RF})| 
\cos\theta_p \right) .
\label{alpha_from_cos_theta}
\eeq
The physical region for the angle $\theta_p$ is determined by the condition that
$0 < \alpha_p < 2$, which implies
\beq
-1 \;\; < \;\; \cos\theta_p \;\; < \;\; 
\frac{M_d - \sqrt{|\bm{p}_p (\text{RF})|^2 + M_N^2}}{|\bm{p}_p (\text{RF})|}
\eeq
The upper limit becomes less than unity only at $|\bm{p}_p (\text{RF})| > (M_d^2 - M_N^2)/(2 M_d) 
\approx 3 M_N/4$, which is much larger than the recoil momenta considered here, 
so that effectively all angles are allowed in our kinematics.
\section{Light-front quantum mechanics}
\label{sec:lightfront}
\subsection{Single--nucleon states}
In our theoretical calculations of the tagged DIS cross section we use methods of LF quantum mechanics. 
They permit a composite description of nuclear structure in high-energy processes in terms of 
nucleon degrees of freedom, which can be matched with deep-inelastic nucleon structure and preserves 
the partonic sum rules \cite{Frankfurt:1988nt,Frankfurt:1981mk}. 
In this section we summarize the description of nucleon single-particle states and the 
deuteron bound state in LF quantum mechanics and the correspondence with the non-relativistic
theory of the deuteron. The LF quantization axis chosen as the $z$--axis in the collinear frame 
of Sec.~\ref{subsec:collinear}. The specific dynamical considerations in the application 
to tagged DIS will be described in Sec.~\ref{sec:ia}.

In LF quantum mechanics plane-wave nucleon states are characterized by their LF plus and 
transverse momenta, $p_N^+ = p_N^0 + p_N^z$ and $\bm{p}_{NT} = (p_N^x, p_N^y)$
[cf.~Eq.~(\ref{lc_def})], while $p_N^- = p_N^0 - p_N^z$ plays the role of energy and is fixed by the 
mass-shell condition $p_N^2 = p_N^+ p_N^- - |\bm{p}_{NT}|^2 = M_N^2$,
\beq
| N, p_N \rangle \;\; \equiv \;\;  | N, p_N^+, \bm{p}_{NT}  \rangle,
\hspace{2em} p_N^- \; = \; \frac{|\bm{p}_{NT}|^2 + M_N^2}{p_N^+} .
\eeq
To simplify the notation we label the states by the 4--momentum $p_N$ and display the 
individual plus and transverse components only if necessary. The relativistic normalization 
condition for the states is [cf.\ Eq.~(\ref{relativistic_normalization})]
\beq
\langle N, p_N' | N, p_N \rangle
\;\; = \;\;  (2\pi)^3 \, 2 p_N^+ \, \delta(p_N^{+\prime} - p_N^+) \; 
\delta( \bm{p}_{NT}^{\prime} - \bm{p}_{NT}) .
\label{normalization_n}
\eeq
The invariant phase space integral over the nucleon LF momentum is
\be
\int d\Gamma_N \; (...) &=& \int \frac{d^4 p_N}{(2\pi)^4} \; 
2\pi \delta (p_N^2 - M_N^2) \; \theta (p_N^0) \; (...)
\nonumber \\
&=& \int \frac{dp_N^+ d^2 p_{NT}}{(2\pi)^3 2 p_N^+} \; \theta (p_N^+) \;
(...) \hspace{2em}  [p_N^- = (|\bm{p}_{NT}|^2 + M_N^2)/p_N^+]
\\
&\equiv& \int [dp_N] \; (...) .
\label{phase_space_lf}
\ee
The condition $p_N^+ > 0$ is satisfied for all physical nucleon momenta. The completeness of 
single--nucleon states can then be stated in the form
\beq
\int [dp_N] \; | N, p_N \rangle \langle N, p_N | \;\; = \;\; 1_{N} ,
\label{completeness_n}
\eeq
which is the unit operator in the single-nucleon space. For reference we note that for 
a general 4--momentum $p$ (not on mass--shell), the four--dimensional integral and the
four-dimensional delta function in LF components take the form
\beq
\int d^4p \; (...) \; = \; \frac{1}{2} \int_{-\infty}^\infty dp^+ \int_{-\infty}^\infty dp^- 
\int d^2 p_T \; (...) ,
\hspace{2em}
\delta^{(4)}(p) \; = \; 2 \delta(p^+) \delta(p^-) \delta^{(2)}(\bm{p}_T) .
\eeq
The formulas can easily be generalized to account for nucleon spin degrees of freedom.
Other hadronic states are described in a similar fashion.
\subsection{Deuteron wave function}
\begin{figure}
\includegraphics[width=.38\textwidth]{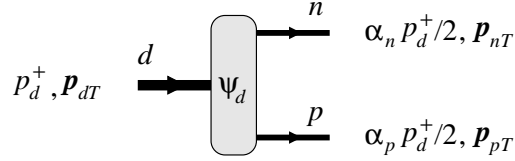}
\caption[]{Deuteron $NN$ LF wave function.}
\label{fig:deut_lfwf}
\end{figure}

In the LF description of high-energy processes nuclei are described as bound states of
nucleons and, possibly, non-nucleonic degrees of freedom ($\Delta$ isobars, 
pions) \cite{Frankfurt:1988nt,Frankfurt:1981mk}. Theoretical arguments show that for the deuteron 
the nucleonic ($pn$) component dominates over a wide range of excitation 
energies \cite{Frankfurt:1981mk} (see also Sec.~\ref{sec:introduction}), and we limit ourselves to 
this component in the present study. The deuteron is described as a bound state with relativistic 
normalization of the center-of-mass motion [cf.~Eq.(\ref{normalization_n})]
\be
|d, p_d \rangle, && p_d^- = \frac{|\bm{p}_{dT}|^2 + M_d^2}{p_d^+} ,
\\[1ex]
\langle d, p_d' | d, p_d \rangle
&=&  (2\pi)^3 \, 2 p_d^+ \, \delta (p_d^{+\prime} - p_d^+) \,
\delta^{(2)}(\bm{p}_{dT}' - \bm{p}_{dT}) .
\label{state_d_norm}
\ee
The expansion of the deuteron state in plane-wave nucleon states is described by
the LF wave function (see Fig.~\ref{fig:deut_lfwf})
\be
\langle p, p_p; n, p_n | d, p_d \rangle
&=& (2\pi)^3 \, 2 p_d^+ \, \delta (p_p^+ + p_n^+ - p_d^+) \,
\delta^{(2)}(\bm{p}_{pT} + \bm{p}_{nT} - \bm{p}_{dT}) \;
\nonumber \\[1ex]
&& \times \; (2\pi)^{3/2} \Psi_d (\alpha_p , \bm{p}_{pT} | \bm{p}_{dT}) ,
\label{lfwf_def}
\\[2ex]
\alpha_p &\equiv& \frac{2 p_p^+}{p_d^+}, 
\hspace{2em} \alpha_n \; \equiv \; \frac{2 p_n^+}{p_d^+} .
\ee
The factor $(2\pi)^{3/2}$ is conventional. The function $\Psi_d$ depends on the LF momentum
fraction and the transverse momentum of the proton, $\alpha_p$ and $\bm{p}_{pT}$, and the
deuteron transverse momentum $\bm{p}_{dT}$; it is independent of the total plus 
momentum $p_d^+$ because of longitudinal boost invariance \cite{Frankfurt:1981mk,Brodsky:1997de}. 
The delta functions in Eq.~(\ref{lfwf_def}) require that
\beq
\alpha_p + \alpha_n \; = \; 2,
\hspace{2em} 
\bm{p}_{pT} + \bm{p}_{nT} \; = \; \bm{p}_{dT} ,
\eeq
which in particular implies that
\beq
0 < \alpha_p, \alpha_n < 2 .
\eeq
%
%
In the present study we do not consider polarization phenomena at the deuteron or the nucleon
level and omit the spin quantum numbers in the states and in the wave function. At this level
the matrix element Eq.~(\ref{lfwf_def}) is symmetric with respect to interchange of the 
proton and neutron, and the wave function satisfies
\beq
\Psi_d (\alpha_p, \bm{p}_{pT} | \bm{p}_{dT}) 
\;\; = \;\;
\Psi_d (2 - \alpha_p , \bm{p}_{dT} - \bm{p}_{pT} | \bm{p}_{dT}) 
\;\; = \;\;
\Psi_d (\alpha_n , \bm{p}_{nT} | \bm{p}_{dT}) .
\label{lfwf_symmetry}
\eeq
The normalization of the deuteron wave function follows from the normalization condition
for the deuteron state Eq.~(\ref{state_d_norm}) and the completeness relation for 
the single-nucleon states Eq.~(\ref{completeness_n}). Inserting complete sets of single-nucleon 
intermediate states into Eq.~(\ref{state_d_norm}) and integrating out the delta functions
one obtains (we replace the arguments by $\alpha_N$ and $\bm{p}_{NT}$ for brevity)
\beq
\int \frac{d\alpha_p \; d^2 p_{pT}}{\alpha_p (2 - \alpha_p)} \;
|\Psi_d (\alpha_p, \bm{p}_{pT} | \bm{p}_{dT})|^2
\;\; = \;\; 1 .
\label{lfwf_normalization}
\eeq
In the calculations of deuteron structure in the collinear frame of Sec.~\ref{subsec:collinear}
we need the wave function at zero deuteron transverse momentum, which we denote by
\beq
\Psi_d (\alpha_p, \bm{p}_{pT})
\;\; \equiv \;\;
\Psi_d (\alpha_p, \bm{p}_{pT} | \bm{p}_{dT} = 0) .
\eeq
The symmetry relation Eq.(\ref{lfwf_symmetry}) for this function takes the simple form
\beq
\Psi_d (\alpha_p, \bm{p}_{pT})
\;\; = \;\;
\Psi_d (2 - \alpha_p, - \bm{p}_{pT})
\;\; = \;\;
\Psi_d (\alpha_n, - \bm{p}_{nT}) .
\label{lfwf_symmetry_rf}
\eeq
The above formulas can easily be generalized to account for deuteron and nucleon spin. The summation 
over spin quantum numbers will be performed implicitly in the calculation of matrix elements below.

For modeling the actual form of the deuteron LF wave function it is natural to consider the 
connection of the LF formulation with the non-relativistic description of deuteron structure.
In general this connection is rather complicated, because of the different symmetry groups of the 
dynamics in the two formulations. However, a simple connection can be established in the 
approximation where the deuteron's LF structure is restricted to the $pn$ component, which we
adopt here \cite{Frankfurt:1981mk,Frankfurt:1988nt}.
One starts with the LF version of the Lippmann-Schwinger equation for the two-body 
wave function (or Weinberg equation \cite{Weinberg:1966jm}) and imposes the condition that 
the scattering-state solutions give rotationally invariant on-shell $NN$ scattering 
amplitudes (angular conditions) \cite{Frankfurt:1981mk,Frankfurt:1988nt}. The resulting equation for 
bound states has a simple connection with the Schr\"odinger equation for the non-relativistic 
deuteron wave function, which one can use to construct an approximation of the
LF wave function in terms of the non-relativistic wave function (see Sec.~\ref{subsec:rotationally}). 
Methods for direct solution of the LF two-body bound-state equation have been described in 
Refs.~\cite{Cooke:2001kz,Cooke:2001dc}. For attempts to model deuteron LF structure beyond the 
$pn$ component, and for approximation methods for heavier nuclei, we refer to
Refs.~\cite{Miller:1999ap,Miller:2000kv} and references therein.
\subsection{Rotationally invariant representation}
\label{subsec:rotationally}
The LF wave function of a two-body bound state such as the deuteron can be expressed in a form
that exhibits 3-dimensional rotational 
invariance \cite{Terentev:1976jk,Frankfurt:1981mk,Frankfurt:1988nt}. This representation can be 
motivated by group-theoretical or dynamical considerations (see above) and is useful for 
several purposes: (a) it explains how rotational invariance is dynamically realized in LF 
quantum mechanics, where it is not manifest (angular conditions); (b) it enables an approximation 
of the LF wave function in terms of the 3-dimensional non-relativistic wave function; 
(c) it brings out the analytic properties of the LF wave function in the nucleon momentum.

The rotationally invariant momentum variable for the two-nucleon system can be introduced through 
an intuitive procedure, by identifying the $pn$ configurations in the deuteron LF wave function
with a free $pn$ system in its center-of-mass frame \cite{Frankfurt:1981mk}. One starts from 
a $pn$ configuration in its center-of-mass frame, with the proton having LF momentum $\alpha_p$ and 
$\bm{p}_{pT}$; calculates the invariant mass of the $pn$ configuration; and equates the invariant 
mass with the squared energy of a free $pn$ system with relative 3--momentum 
\beq
\bm{k} \;\; \equiv \;\; (\bm{k}_T, k^z), \hspace{2em} \bm{k}_T \; = \; \bm{p}_T .
\label{cm_from_lc}
\eeq
This leads to the equation
\beq
s_{pn} \;\; \equiv \;\; \frac{\bm{p}_{pT}^2 + M_N^2}{\displaystyle \frac{\alpha_p}{2} 
\left(1 - \frac{\alpha_p}{2}\right)} 
\;\; \stackrel{!}{=} \;\; 4 E_N(\bm{k})^2
\;\; = \;\; 4 \left( |\bm{k}|^2 + M_N^2 \right) .
\label{invariant_mass_from_k}
\eeq
One then determines the component $k^z$ as function of $\alpha_p$ and $|\bm{p}_{pT}|$ by solving
Eq.~(\ref{invariant_mass_from_k}),
\beq
k^z \;\; = \;\; (\alpha_p - 1) 
\left[\frac{\bm{p}_{pT}^2 + M_N^2}{\alpha_p (2 - \alpha_p)}\right]^{1/2} .
\label{k3_from_lc}
\eeq
Equations~(\ref{cm_from_lc}) and (\ref{k3_from_lc}) define the equivalent 3--momentum $\bm{k}$
in terms of the LF variables of the two-body system. The inverse relation is
\be
\alpha_p \; = \;  1 + \frac{k^z}{E_N}  \; = \;  \frac{2 (E_N + k^z)}{2 E_N},
\hspace{2em} E_N \; \equiv \; 
{\textstyle\sqrt{|\bm{k}|^2 + M_N^2}} .
\ee
Note that in this parametrization the nucleon plus momentum fraction is obtained by dividing 
$E_N + k^z$ by the {\em internal energy} of the $pn$ system, $2E_N$, not by the {\it external mass} 
of the bound state, as in the kinematic variable Eq.~(\ref{recoil_lf}). The invariant phase space 
elements in the two sets of variables are related as
\beq
\frac{d\alpha_p \; d^2 p_{pT}}{\alpha_p (2 - \alpha_p)}
\; = \; 
\frac{d^3k}{\sqrt{|\bm{k}|^2 + M_N^2}}
\;\; = \;\;
\frac{d^3k}{E_N(\bm{k})} .
\eeq
The rotationally invariant form of the deuteron LF wave function is then obtained by
demanding that
\beq
\Psi_d (\alpha_p, \bm{p}_{pT}) \;\; = \;\; \textrm{function}(|\bm{k}|) .
\label{wf_rotational}
\eeq
It was shown in Ref.~\cite{Frankfurt:1981mk} that this condition is sufficient to guarantee
rotational invariance in two-body bound state calculations. 

The rotationally symmetric form Eq.~(\ref{wf_rotational}) suggests a natural approximation of the 
rest-frame deuteron LF wave function in terms of the non-relativistic wave function:
\beq
\Psi_d (\alpha_p, \bm{p}_{pT}) \;\; \stackrel{\rm app}{=} \;\;
\sqrt{E_N(\bm{k})} \;\; \widetilde \Psi_d (\bm{k}) ,
\label{nonrel}
\eeq
where $\widetilde \Psi_d$ denotes the non-relativistic wave function and the arguments are related 
by Eqs.~(\ref{cm_from_lc}) and (\ref{k3_from_lc}). This approximation has the following properties:
(a) it becomes exact at small momenta $|\bm{k}| \ll M_N$, where $E_N(\bm{k}) \approx M_N$ is constant 
and the relation between $\alpha_N$ and $k^z$ becomes the standard non-relativistic approximation;
(b) it has correct overall normalization, because the normalization integrals are related as
[cf.~Eq.~(\ref{lfwf_normalization})]
\beq
\int \frac{d\alpha_p \; d^2 p_{pT}}{\alpha_p (2 - \alpha_p)} \;
|\Psi_d (\alpha_p, \bm{p}_{pT})|^2
\;\; = \;\; \int d^3 k \; |\widetilde \Psi_d (\bm{k})|^2 \;\; = \;\; 1 .
\eeq

The rotationally invariant representation Eq.~(\ref{wf_rotational}) is also sufficient for ensuring
the correct analytic properties of the LF wave function at small relative momenta (nucleon pole).
We can demonstrate this using the approximation Eq.~(\ref{nonrel}), which becomes exact at
small recoil momenta. On general grounds the non-relativistic deuteron wave function has a pole at 
small unphysical momenta of the form
\be
\widetilde \Psi_d (\bm{k}) \; &\sim& \; \frac{\Gamma}{|\bm{k}|^2 + a^2} \hspace{2em} 
(|\bm{k}| \rightarrow a),
\label{nucleon_pole_psitilde}
\\[1ex]
a^2 \; &=& \; M_N \epsilon_d - \frac{\epsilon_d^2}{4} .
\label{a2_def}
\ee
The pole results from the free propagation of the nucleons outside the range of $pn$ interaction
and controls the large-distance behavior of the coordinate--space wave function. 
[In the Bethe--Peierls approximation the entire deuteron wave function is 
given by Eq.~(\ref{nucleon_pole_psitilde}).] By expressing $|\bm{k}^2|$ in 
Eq.~(\ref{nucleon_pole_psitilde}) in terms of the LF momentum variables using 
Eq.~(\ref{invariant_mass_from_k}), one easily sees that the pole corresponds to a pole in the 
invariant mass $s_{pn}$ of the LF wave function,
\beq
\Psi_d (\alpha_p, \bm{p}_{pT}) \;\; \sim \;\; \frac{4 (M_N^2 - a^2)^{1/4} \Gamma}{s_{pn} - M_d^2} .
\label{nucleon_pole_lf}
\eeq
The singularities Eq.~(\ref{nucleon_pole_psitilde}) viz.\ Eq.~(\ref{nucleon_pole_lf}) give rise 
to the nucleon pole in the deuteron spectral function and play an essential role in the analysis 
of tagged DIS (see below).

For practical calculations one can use Eq.~(\ref{nonrel}) with a non-relativistic deuteron wave function 
obtained from realistic $NN$ potentials \cite{Wiringa:1994wb}. At the low momenta of interest here
($|\bm{k}| \lesssim 200 \, \textrm{MeV}$) an excellent approximation to the realistic wave functions 
is provided by a simple two-pole parametrization, which implements the nucleon pole and has correct 
analytic properties (see Appendix~\ref{app:deuteron}).
We use this parametrization in the numerical calculations below.
\section{Impulse approximation}
\label{sec:ia}
\subsection{LF current components}
\label{subsec:lf_current_components}
We now compute the cross section for tagged DIS on the deuteron using LF quantum mechanics.
The basic considerations in treating nuclear structure in high-energy scattering are described 
in Refs.~\cite{Frankfurt:1988nt,Frankfurt:1981mk} and summarized in Sec.~\ref{sec:introduction}. 
In LF quantization the effects of the off-shellness of the constituents in a bound state remain 
finite as the scattering energy becomes large, which makes possible a composite description of the 
nucleus in terms of nucleon degrees of freedom (see below). We use the collinear frame of 
Sec.~\ref{subsec:collinear} ($\bm{p}_{dT} = 0, \bm{q}_T = 0$), in which the initial nucleus and the 
DIS final state evolve along the same $z$--direction, as this permits a natural description of FSI 
with rotational invariance in the transverse plane. Non-collinear frames with $\bm{q}_T \neq 0$ 
can be used for LF calculations of the inclusive DIS cross section but are not suitable 
for FSI \cite{Frankfurt:1988nt}. In the collinear frame the momentum transfer to the nucleus 
has LF component $q^+ < 0$, Eq.~(\ref{collinear_frame}), so that the current cannot produce physical 
hadron states out of the vacuum, but can only couple to nucleons in the nuclear LF wave function.

In order to extract the tagged deuteron structure functions of Eq.~(\ref{W_decomposition}) in the 
collinear frame we must calculate both $+$ and $T$ components of the nuclear tensor. It is 
well-known that in LF quantization the different components of the current operator have different 
status as to how they involve the interactions of the system. This is seen explicitly in the LF 
quantization of quantum field theories, where only two components of the spin-1/2 Dirac field are 
independent canonical degrees of freedom, while the other two are dependent and must be eliminated 
through the equations of motion \cite{Kogut:1969xa,Cornwall:1971as}. 
The ``good'' current $J^+$ is formed out of canonical degrees of 
freedom and free of interactions; the ``bad'' current $\bm{J}_T$ is formed out of canonical and 
dependent degrees of freedom and involves explicit interactions; the ``worst'' current $J^-$ 
is formed entirely out of dependent degrees of freedom.\footnote{In the equivalent formulation
based on equal-time quantization in the infinite--momentum frame $|\bm{p}| \rightarrow \infty$, 
the ``good'' components are those that tend to a finite limit as $|\bm{p}| \rightarrow \infty$ 
(in the non-covariant normalization of states), while the other components vanish.} Following 
Refs.~\cite{Frankfurt:1988nt,Frankfurt:1981mk} we calculate the $J^+$ and $\bm{J}_T$ 
matrix elements in our approach (IA and FSI); the $J^-$ component can be eliminated through
the transversality condition in the collinear frame (current conservation) and does not need to be 
considered explicitly,
\beq
q^\mu \langle B | J_\mu | A \rangle
\;\; = \;\; \frac{q^+}{2} \langle B | J^- |A \rangle \; + \; 
\frac{q^-}{2} \langle B | J^+ |A \rangle \;\; = \;\; 0 .
\eeq
The use of the component $\bm{J}_T$ for structure function calculations represents an approximation,
whose accuracy cannot be established from first principles in our phenomenological approach. 
In inclusive DIS, comparison between a good-current calculation in a non-collinear frame and the 
good-and-bad-current calculation in the collinear frame shows that the two schemes give the same 
results in the DIS limit $(\textrm{mass}^2)/W^2 \rightarrow 0$ \cite{Frankfurt:1988nt}. This indicates
that the collinear method should be reliable for the leading-twist tagged structure functions 
$F_{2d}$ and $F_{Ld}$ calculated in this work. A further test of the method will be performed 
in Sec.~\ref{subsec:structure}. The applicability to higher-twist structure functions, 
which represent power-suppressed structure in the tagged cross section, remains to be investigated. 
\subsection{IA current}
\label{subsec:ia}
%
%
\begin{figure}[t]
\includegraphics[width=.24\textwidth]{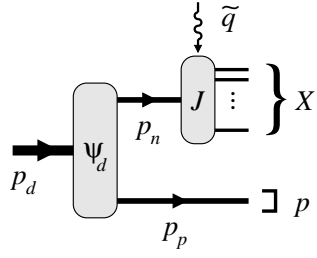}
\caption[]{Current matrix element of tagged DIS on the deuteron in the IA.}
\label{fig:deut_ia}
\end{figure}
The starting point of the nuclear structure calculation is the IA.
Its physical assumptions are: (a) the current operator is the sum of one--body nucleon currents;
(b) the final state produced by the one--body nucleon current evolves independently of 
nuclear remnant (see Fig.~\ref{fig:deut_ia}) \cite{Benhar:2005dj}.
In the IA we consider the nuclear current matrix element in Eq.~(\ref{w_mu_nu_current}) in the
collinear frame ($\bm{p}_{dT} = 0$) and insert plane-wave proton and neutron states between the
deuteron state and the current operator. Taking the proton as the spectator, and the neutron
as coupling to the current, we obtain (see Fig.~\ref{fig:deut_ia})
\be
\langle p X | \hat J^\mu (0) | d \rangle [\textrm{IA}]
\; &=& \;
\int [dp_p] \; \int [dp_n]
\;
\langle X | \hat J^\mu (0) | n, p_n \rangle \;
\langle p, p_p | p, p_{p1} \rangle \;
\langle p, p_p ; n, p_n | d, p_d \rangle
\label{current_ia_1}
\\
&=& \;
(2\pi)^{3/2} \; \frac{p_d^+}{p_n^+} \;
\langle X | \hat J^\mu (0) | n, p_n \rangle \; 
\Psi_d (\alpha_p , \bm{p}_{pT})
\hspace{2em} (p_n^+ = p_d^+ - p_p^+, \; \bm{p}_{nT} = - \bm{p}_{pT}) .
\label{current_ia_2}
\ee
The deuteron tensor Eq.~(\ref{w_mu_nu_current}) then becomes 
\be
W^{\mu\nu}_d (p_d, q; p_p) \; &=& \; \left(\frac{p_d^+}{p_n^+}\right)^2 \; 
(2\pi)^3 \; |\Psi_d (\alpha_p , \bm{p}_{pT} )|^2
\nonumber \\[1ex]
&\times& \; (4\pi)^{-1} \sum_{X} \; (2\pi)^4 \; \delta^{(4)} (q + p_d - p_p - p_X) \; 
\langle n, p_n | \hat J^\mu (0) | X \rangle \; \langle X | 
\hat J^\nu (0) | n, p_n \rangle
\nonumber
\\[1ex]
&& \; (p_n^+ = p_d^+ - p_p^+, \; \bm{p}_{nT} = - \bm{p}_{pT}) .
\label{tensor_ia}
\ee
The expression on the second line has a form suggestive of the scattering tensor for inclusive 
scattering on the neutron. However, we must take into account that in LF quantum mechanics 4--momenta 
are not conserved, and that
\beq
p_d - p_p \;\; \neq \;\; p_n \hspace{2em} \textrm{(4-vectors)}
\eeq
in the argument of the 4-dimensional delta function. This is because the LF energy of the
neutron is determined by the mass shell condition
\beq
p_n^- \; = \; \frac{|\bm{p}_{nT}|^2 + M_N^2}{p_n^+} \;\neq \; 
p_d^- - p_p^- .
\label{offshellness_minus}
\eeq
The expression in Eq.~(\ref{tensor_ia}) can therefore not be regarded as the neutron
scattering tensor with the original 4--momentum transfer $q$, which is fixed kinematically 
by the electron 4--momenta. To write it as a proper scattering tensor we define an 
effective 4--momentum transfer as
\be
\widetilde{q} \; &\equiv& \; q + p_d - p_n - p_p, 
\hspace{2em} \textrm{or} \hspace{2em}
p_n + \widetilde{q} \;\; = \;\; q + p_d - p_p 
\nonumber
\\[2ex]
&&
\left[ p_n^+ = p_d^+ - p_p^+, \; \bm{p}_{nT} = - \bm{p}_{pT}, \;
p_n^- = (|\bm{p}_{nT}|^2 + M_N^2)/p_n^+ \right] .
\label{qtilde_def}
\ee
The vector $\widetilde{q}$ has the same plus and transverse components as the original
$q$ and differs only in its minus component, which is not conserved in LF quantum mechanics.
The difference accounts for the fact that the $pn$ configurations in the deuteron are off the LF 
energy shell and participate in the scattering process with shifted kinematics. 
With the effective momentum transfer Eq.~(\ref{qtilde_def}) the deuteron tensor 
Eq.~(\ref{tensor_ia}) can then be expressed in terms of the effective 
neutron tensor [we use $p_d^+/p_n^+ = 2/(2 - \alpha_p)$] 
\be
W^{\mu\nu}_d (p_d, q, p_p) \; &=& \;
\left(\frac{p_d^+}{p_n^+}\right)^2 \; (2\pi)^3 \;
|\Psi_d (\alpha_p , \bm{p}_{pT})|^2 \; 
W_n^{\mu\nu} (p_n, \widetilde{q})
\\[1ex]
&=& \;
\frac{4 (2\pi)^3}{(2 - \alpha_p)^2} \; |\Psi_d (\alpha_p , \bm{p}_{pT})|^2 \;
W_n^{\mu\nu} (p_n, \widetilde{q})
\label{w_impulse}
\\[1ex]
W_n^{\mu\nu} (p_n, \widetilde{q}) \; &\equiv& \; (4\pi)^{-1} \sum_{X} 
\; (2\pi)^4 \; \delta^{(4)} (\widetilde{q} + p_n - p_{X})
\; \langle n, p_n | J^\mu (0) | X \rangle \;
\langle X | J^\nu (0) | n, p_n \rangle .
\ee
Equation~(\ref{w_impulse}) represents the ``master formula'' for tagged DIS in the
LF IA and expresses the factorization of deuteron and nucleon structure.

The assignment of the active nucleon 4-momentum $p_n$ as in Eq.~(\ref{offshellness_minus}),
and of the 4-momentum transfer $\widetilde q$ as in Eq.~(\ref{qtilde_def}), are dictated by
LF quantum mechanics, in which the LF $+$ and $T$ momenta are conserved and the particles are 
on mass-shell, but the total LF energy of the intermediate states is different from that of the 
initial and final states. A crucial point is that in this scheme the non-conservation of 4-momentum 
does not give rise to any large invariants in the DIS limit 
$W^2 \rightarrow \infty, \, Q^2 \rightarrow \infty, \, Q^2/W^2 \; \textrm{fixed}$. 
The off-shellness of the minus component of the nucleon 
4-momentum implied by Eq.~(\ref{offshellness_minus}) is
\be
p_n^- - p_d^- + p_p^-
\; &=& \; \frac{|\bm{p}_{pT}|^2 + M_N^2}{p_n^+} - \frac{M_d^2}{p_d^+}
+ \frac{|\bm{p}_{pT}|^2 + M_N^2}{p_p^+}
\\[1ex]
&=& \; \frac{1}{p_d^+} \left[ \frac{4(|\bm{p}_{pT}|^2 + M_N^2)}{\alpha_p (2 - \alpha_p)} - M_d^2 \right] ,
\ee
where we have used the explicit expressions for the LF momentum components in the 
collinear frame of Sec.~\ref{subsec:collinear}. The plus component of the momentum
transfer is $q^+ = -\xi_d p_d^+$, cf.~Eq.(\ref{collinear_frame}). The variables $\alpha_p, |\bm{p}_{pT}|$ 
and $\xi_d$ remain finite in the DIS limit, and $p_d^+$ is a finite boost parameter.
One therefore has
\beq
2 q (p_n - p_d + p_p) \;\; = \;\; q^+ (p_n^- - p_d^- + p_p^-) \;\; = \;\; 
O\left\{ |\bm{p}_{pT}|^2, (\textrm{mass})^2 \right\} ,
\label{offshell_power_suppressed}
\eeq
i.e., the invariant remains finite and does not grow as $W^2$ or $Q^2$. (Note that individually 
$qp_d \sim W^2$ and $qp_n \sim W^2$, because $p_d$ and $p_n$ have non-zero plus components.) 
It implies that the effects caused by the LF energy off-shellness are power-suppressed as 
$\sim |\bm{p}_{pT}|^2/W^2$ or $\sim (\textrm{mass})^2/W^2$ in the DIS limit.
This circumstance is unique about LF quantization and is the reason for the use this
approach in high-energy scattering.
\subsection{Structure functions}
\label{subsec:structure}
Expressions for the tagged deuteron structure functions are obtained from Eq.~(\ref{w_impulse})
by substituting the specific form of the neutron tensor and projecting the 
tensor equation on the structures of Eq.~(\ref{W_decomposition}). The decomposition of the 
neutron tensor is analogous to that of the deuteron tensor Eq.~(\ref{W_decomposition}), but with 
the target 4--momentum given by $p_n$, and the 4--momentum transfer given by $\widetilde{q}$,
\be
W_n^{\mu\nu} (p_n, \widetilde{q})
&=& \left( g^{\mu\nu} - \frac{\widetilde q^\mu \widetilde q^\nu}{\widetilde q^2} \right) 
\frac{F_{Ln}(\widetilde x, \widetilde Q^2)}{2}
\; + \; \left( \frac{\widetilde L_n^\mu \widetilde L_n^\nu}{\widetilde L_n^2} 
+ \frac{\widetilde q^\mu \widetilde q^\nu}{\widetilde q^2} 
- g^{\mu\nu} \right) \frac{F_{Tn} (\widetilde x, \widetilde Q^2)}{2} ,
\label{nucleon_tensor_1}
\\[2ex]
F_{Tn} &=& \frac{(-\widetilde q^2) L_n^2}{(p_n \widetilde q)^2} \; \frac{F_{2n}}{\widetilde x}
\; = \; \left( 1 + \frac{4 \widetilde x^2 M_N^2}{\widetilde Q^2} \right) \; \frac{F_{2n}}{\widetilde x} ,
\label{nucleon_FT_from_F2}
\\[2ex]
\widetilde L_n^\mu &\equiv& p_n^\mu - \frac{(p_n \widetilde q) \widetilde q^\mu}{\widetilde q^2}, 
\hspace{2em}
\widetilde{x} \; \equiv \; 
\frac{-\widetilde{q}^2}{2 (p_n \widetilde q)},
\hspace{2em}
\widetilde{Q}^2 \; \equiv \; -\widetilde{q}^2 .
\label{nucleon_tensor_2}
\ee
Equations for the structure functions can then be derived by considering the $+$ and $T$
tensor components in the collinear frame (see Appendix~\ref{app:projection}).
They take on a simple form in the DIS limit, where one can neglect terms of the order 
$|\bm{p}_{pT}|^2/W^2$ and $(\textrm{mass})^2/W^2$, so that off-shell
effects are suppressed [cf.~Eq.~(\ref{offshell_power_suppressed})]. In particular, 
in this limit 
\beq
\widetilde{Q}^2 \;\; = \;\; Q^2, 
\hspace{2em}
\widetilde{x} \;\; = \;\; \frac{x}{2 - \alpha_p},  
\eeq
up to power corrections, i.e., the nucleon structure functions are evaluated at the kinematically 
given $Q^2$, and at an effective value of $x$ that accounts for the longitudinal motion of the 
nucleons in the bound state. Altogether we obtain
\be
F_{2d} (x, Q^2; \alpha_p, p_{pT}) \; &=& \; \frac{|\Psi_d(\alpha_p, \bm{p}_{pT})|^2}{2 - \alpha_p} 
\; F_{2n} \left(\widetilde{x}, Q^2 \right) ,
\label{IA_F2_wf}
\\[2ex]
F_{Ld} (x, Q^2; \alpha_p, p_{pT}) \; &=& \; 
\frac{2 \, |\Psi_d(\alpha_p, \bm{p}_{pT})|^2}{(2 - \alpha_p)^2} 
\; F_{Ln} \left(\widetilde{x}, Q^2 \right) .
\label{IA_FL_wf}
\ee
These formulas express the deuteron DIS structure functions with tagged proton in terms of the 
deuteron LF momentum density and the active neutron inclusive structure functions. The case of
tagged neutron and active proton is described by the same formulas with the proton and neutron 
labels interchanged.

Our calculation in the collinear frame uses both good and bad LF current components to identify 
the structure functions (cf.~\ Sec.~\ref{subsec:lf_current_components}). The results for the
bad current component in the LF IA are generally sensitive to the energy off-shellness (4-momentum 
nonconservation) in the intermediate state. These effects are related to those of explicit interactions 
in the bad current component operators. In a complete dynamical theory both could be treated 
consistently starting from the microscopic interaction. To assess their influence within our 
phenomenological approach we perform a simple test, following Ref.~\cite{Frankfurt:1988nt}. 
We evaluate Eq.~(\ref{w_impulse}) with the 
neutron tensor $W^{\mu\nu}_n$ taken at the off-mass-shell 4-momentum $\widetilde p_n \equiv
p_d - p_p$ with $\widetilde p_n^2 \neq M_N^2$, as would by obtained from the external 4-momenta
using 4-momentum conservation, and at the original momentum transfer $q$ (``virtual nucleon''). 
We compare the results with those of the LF prescription, where 
$W^{\mu\nu}_n$ is evaluated at $p_n$ and $\widetilde q$, Eqs.~(\ref{offshellness_minus}) and 
(\ref{qtilde_def}). The differences in the leading-twist tagged structure functions $F_{2d}$ 
and $F_{Ld}$ turn out to be of the order $|\bm{p}_{pT}|^2/W^2$ and $(\textrm{mass})^2/W^2$ and are
thus power-suppressed in the DIS limit [cf.~Eq.~(\ref{offshell_power_suppressed})]. 
This suggests that our collinear LF calculation is safe in the DIS limit.

In addition to the kinematic off-shell effects discussed so far, nuclear binding causes dynamical 
modifications of the structure of the nucleon, which manifest themselves e.g.\ in the suppression 
of the nuclear structure functions at $x > 0.3$ compared to the sum of the corresponding nucleon 
structure functions (EMC effect). Theoretical analysis shows that to first order in the nuclear
binding these modifications are proportional to the LF energy off-shellness of the nuclear 
configurations (or the nucleon virtuality in the virtual nucleon formulation), which in turn is 
proportional to the non-relativistic kinetic energy of the 
nucleons \cite{Frankfurt:1988nt,CiofidegliAtti:2007ork}. The modifications are therefore much smaller 
in the deuteron than in heavy nuclei. Simple scaling arguments suggest that in average configurations 
in the deuteron the EMC-like modifications should be at the level of $\sim 2-3\%$. They are reduced 
further when selecting configurations with proton recoil momenta less than the typical nucleon 
momentum in the deuteron (the median momentum is $\sim 70 \, \textrm{MeV}$; see Fig.~\ref{fig:nonrel} 
and Appendix~\ref{app:deuteron}). The modifications can be eliminated entirely by performing on-shell 
extrapolation in the recoil momentum, which effectively turns the deuteron into a free $pn$ system
(see Sec.~\ref{subsec:analytic_ia}).
\subsection{Spectral function}
\label{subsec:spectral}
The IA for the deuteron tensor in tagged DIS, Eq.~(\ref{w_impulse}), is conveniently expressed in 
terms of the deuteron spectral function, defined as
\be
S_d (\alpha_p, \bm{p}_{pT})
&\equiv& \frac{|\Psi_d(\alpha_p, \bm{p}_{pT})|^2}{2 - \alpha_p} .
\label{spectral_impulse}
\ee
It is a function of the LF momentum variables of the recoil proton and satisfies the sum rules
\be
\int_0^2\frac{d\alpha_p}{\alpha_p} \int d^2 p_{pT} \; 
S_d (\alpha_p, \bm{p}_{pT}) &=& 
\int\frac{d\alpha_p \; d^2 p_{TR}}{\alpha_p (2 - \alpha_p)} 
\; |\Psi_d (\alpha_p, \bm{p}_{pT})|^2 
\;\; = \;\; 1 ,
\label{spectral_number_sumrule}
\\[2ex]
\int_0^2 \frac{d\alpha_p}{\alpha_p} \int d^2 p_{pT} \; (2 - \alpha_p) \; S_d (\alpha_p, \bm{p}_{pT})
&=& \int \frac{d\alpha_p \; d^2 p_{pT}}{\alpha_p (2 - \alpha_p)} \;\;
(2 - \alpha_p) \; |\Psi_d (\alpha_p, \bm{p}_{pT})|^2 
\nonumber
\\[2ex]
&=& \int \frac{d\alpha_p \; d^2 p_{pT}}{\alpha_p (2 - \alpha_p)} \;\;
\alpha_p \; |\Psi_d (\alpha_p, \bm{p}_{pT})|^2 
\nonumber
\\[2ex]
&=& \int \frac{d\alpha_p \; d^2 p_{pT}}{\alpha_p (2 - \alpha_p)} \;
|\Psi_d (\alpha_p, \bm{p}_{pT})|^2 \;\; = \;\; 1.
\label{spectral_momentum_sumrule}
\ee
The first sum rule, Eq.~(\ref{spectral_number_sumrule}), follows from the normalization
condition of the deuteron LF wave function, Eq.~(\ref{lfwf_normalization}), and reflects
the total number of nucleons in the bound state (nucleon number sum rule). 
The second sum rule, Eq.~(\ref{spectral_momentum_sumrule}), follows from the symmetry of 
the two-body LF wave function in the transverse rest frame, Eq.~(\ref{lfwf_symmetry_rf}), and
expresses the conservation of the LF plus momentum (momentum sum rule).
The physical implications of these sum rules will be explained in the following.
In terms of the spectral function the IA result for the tagged structure functions,
Eqs.~(\ref{IA_F2_wf}) and (\ref{IA_FL_wf}), are now expressed as
\be
F_{2d} (x, Q^2; \alpha_p, p_{pT}) &=& 
S_d(\alpha_p, \bm{p}_{pT}) \; F_{2n} \left(\widetilde{x}, Q^2 \right) ,
\label{IA_F2}
\\[2ex]
F_{Ld} (x, Q^2; \alpha_p, p_{pT}) &=& 
\frac{2 \, S_d(\alpha_p, \bm{p}_{pT})}{2 - \alpha_p} 
\; F_{Ln} \left(\widetilde{x}, Q^2 \right) .
\label{IA_FL}
\ee

It is instructive to consider the integral of the tagged deuteron structure function over the 
recoil momentum
\beq
F_{2d}^{\rm int} (x, Q^2) \;\; \equiv \;\; \int_0^{2 - x} \frac{d\alpha_p}{\alpha_p}
\int d^2 p_{pT} \; F_{2d} (x, Q^2; \alpha_p, p_{pT}) .
\eeq
The restriction $\alpha_p < 2 - x$ results because the recoil proton plus momentum cannot
exceed the total plus momentum of the DIS final state.
Notice that this integral over the LF variables corresponds to the integral over the invariant
recoil momentum phase space, Eq.~(\ref{phase_space_alpha_pt}). With the IA expression
Eq.~(\ref{IA_F2}) the integrated structure function becomes
\be
F_{2d}^{\rm int} (x, Q^2) 
& \equiv & \int_0^{2 - x} \frac{d\alpha_p}{\alpha_p}
\int d^2 p_{pT} \; 
S_d(\alpha_p, \bm{p}_{pT}) \; F_{2n} ( \widetilde x, Q^2 ) 
\\[1ex]
& = & \int_0^{2 - x} \frac{d\alpha_p}{\alpha_p (2 - \alpha_p)}
\int d^2 p_{pT} \; 
|\Psi_d (\alpha_p, \bm{p}_{pT})|^2 \; F_{2n} ( \widetilde x, Q^2 ) 
\\[2ex]
\nonumber 
&& [\widetilde x = x/(2 - \alpha_p)].
\label{integrated_ia}
\ee
Equation~(\ref{integrated_ia}) has several interesting properties. First, using the
symmetry of the LF wave function, Eq.~(\ref{lfwf_symmetry_rf}), the integral can equivalently 
be expressed as an integral over the active neutron fraction $\alpha_n = 2 - \alpha_p$, whereupon it 
takes the form of a standard partonic convolution formula,
\be
F_{2d}^{\rm int} (x, Q^2) 
&=& \int_x^2 \frac{d\alpha_n}{\alpha_n}
\int d^2 p_{nT} \; 
|\Psi_d (\alpha_n, \bm{p}_{nT})|^2 \; F_{2n} ( \widetilde x, Q^2 ) 
\hspace{2em}  (\widetilde x \equiv x/\alpha_n).
\ee
Second, using the momentum sum rule for the spectral function, Eq.~(\ref{spectral_momentum_sumrule}), 
and changing the order of the integrations, one easily shows that
\beq
\int_0^2 dx \; F_{2d}^{\rm int} (x, Q^2) \;\; = \;\; 
\int_0^1 d\widetilde{x} \; F_{2n} (\widetilde{x}, Q^2) .
\label{sumrule_f2}
\eeq
A similar formula applies to the case of tagged neutron and active proton. Together, they 
imply that the LF momentum sum rule for the deuteron is satisfied exactly in the IA if one
adds the contributions from proton and neutron tagging, i.e., from scattering on the 
active neutron and proton, 
\beq
\int_0^2 dx \; \left[ F_{2d}^{\rm int} (p \; \textrm{tagged})  + 
F_{2d}^{\rm int} (n \; \textrm{tagged}) \right] (x, Q^2)
\;\; = \;\; \int_0^1 d\widetilde{x} \; \left[ F_{2n} + F_{2p} \right] (\widetilde{x}, Q^2) .
\label{sumrule_f2_sum}
\eeq
Third, for non-exceptional values of $x$
the integral over $\alpha_p$ in Eq.~(\ref{integrated_ia}) is dominated by the region $\alpha_p \sim 1$, 
so that one can neglect the variation of $\widetilde x = x/(2 - \alpha_p)$ under the integral and 
evaluate the structure function at $\alpha_p = 1$ (peaking approximation),
\beq
F_{2d}^{\rm int} (x, Q^2) \;\; \approx \;\; F_{2n} (x, Q^2) \; 
\int_0^2 \frac{d\alpha_p}{\alpha_p} \int d^2 p_{pT} \; S_d(\alpha_p, \bm{p}_{pT}) 
\;\; = \;\; F_{2n} (x, Q^2) .
\label{peaking_ia}
\eeq
In the second step we have used the number sum rule for the spectral function, 
Eq.~(\ref{spectral_number_sumrule}). Again a similar formula applies to the case of tagged 
neutron and active proton. Thus the sum of proton-tagged and neutron-tagged deuteron structure 
functions in the peaking approximation reduces to the sum of the free neutron and proton 
structure functions, as it should be.

Some comments are in order regarding our definition of the spectral function 
Eq.~(\ref{spectral_impulse}). In the IA for a complex nucleus $(A > 2)$ the spectral function 
describes the probability for removing a nucleon, leaving the $A - 1$ remnant system $R$ in 
a state with given momentum $\bm{p}_R$ and total energy $E_R$, which includes the energy of 
the excitation and/or internal motion of the system. In the IA for the deuteron 
($A = 2$), assuming that it can be described as an $pn$ system (neglecting $NN\pi$ and 
$\Delta\Delta$ components), the recoiling system is a single nucleon, and its energy is fixed
by the energy-momentum relation (there is no excitation or internal motion), so that the spectral
function depends on the momentum variables only. In fact, the proton-tagged spectral function 
defined by Eq.~(\ref{spectral_impulse}) is related in a simple way to the neutron LF momentum 
density in the deuteron \cite{Frankfurt:1981mk}, cf.\ Eq.~(\ref{lfwf_symmetry_rf}),
\be
S_d (\alpha_p, \bm{p}_{pT}) &=& 
\frac{\alpha_p}{\alpha_n} \; \rho_d (\alpha_n, \bm{p}_{nT}) ,
\label{spectral_density_deuteron}
\\[1ex]
\rho_d (\alpha_n, \bm{p}_{nT}) &\equiv& \frac{|\Psi_d(\alpha_n, \bm{p}_{nT})|^2}{2 - \alpha_n} 
\;\; = \;\; \frac{|\Psi_d(\alpha_p, \bm{p}_{pT})|^2}{\alpha_p} 
\\[1ex]
\nonumber
&& (\alpha_n = 2 - \alpha_p, \; \bm{p}_{nT} = - \bm{p}_{pT}) .
\ee
The density is regarded as a function of the neutron LF momentum variables and satisfies
the normalization condition
\beq
\int \frac{d\alpha_n}{\alpha_n} \int d^2 p_{nT} \; \rho_d (\alpha_n, \bm{p}_{nT}) 
\;\; = \;\; 1 .
\eeq
In this sense we could express the IA result (and the distortion effects due to FSI considered below)
as well in terms of the active neutron density. We choose to express them in terms of the spectral 
function Eq.~(\ref{spectral_impulse}), as this function depends on the observable recoil proton momentum.
\subsection{Nonrelativistic limit}
\label{subsec:nonrelativistic}
A remarkable property of the IA in LF quantum mechanics is that it coincides with the
non-relativistic approximation in the limit of small proton recoil momentum in the deuteron 
rest frame. This coincidence is not trivial, as the LF expression Eq.~(\ref{w_impulse}) 
involves a wave function and a flux factor that refer explicitly to the direction of the quantization
axis set by the high-energy process. To demonstrate it, we consider the function
\beq
\frac{2\, S_d (\alpha_p, \bm{p}_{pT})}{2 - \alpha_p} 
\;\; = \;\; 
\frac{2\, |\Psi_d (\alpha_p, \bm{p}_{pT})|^2}{(2 - \alpha_p)^2} 
\;\; = \;\; 
\frac{2 E_N(\bm{k}) \, |\widetilde\Psi_d (\bm{k})|^2}{(2 - \alpha_p)^2} 
\label{function_lf}
\eeq
in the deuteron rest frame, where the LF variables $\alpha_p$ and $\bm{p}_{pT}$ and the equivalent
3--momentum variable $\bm{k}$ (see Sec.~\ref{subsec:rotationally}) are given in terms of the 
proton recoil momentum $\bm{p}_p$, and expand the function in powers of the recoil momentum, 
\beq
|\bm{p}_p|, \; p_p^z \;\; \ll \;\; M_N .
\eeq
To simplify the expressions we also expand in the deuteron binding energy $\epsilon_d = 2 M_N - M_d$, 
counting $\epsilon_d M_N = O(p_p^2)$, which allows us to study the deuteron wave function at momenta 
near the nucleon pole Eq.~(\ref{nucleon_pole_lf}). The proton LF momentum fraction in the rest frame 
is given by
\beq
\frac{\alpha_p}{2} \; = \; \frac{\sqrt{|\bm{p}_p|^2 + M_N^2} + p_p^z}{M_d}
\; = \; \frac{\sqrt{|\bm{p}_p|^2 + M_N^2} + p_p^z}{2 M_N - \epsilon_d}
\eeq
and can be expanded to the necessary order. The flux factor in Eq.~(\ref{function_lf}) becomes
\beq
2 - \alpha_p \;\; = \;\; 1 - \frac{p_p^z}{M_N} \; + \; O(p_p^2) .
\label{nonrel_alphap}
\eeq
The modulus $|\bm{k}|$ is defined by Eq.~(\ref{invariant_mass_from_k}), and the expansion gives
\be
|\bm{k}|^2 + a^2 \;\; = \;\; (|\bm{p}_p|^2 + a^2) \left( 1 + \frac{p^z}{M_N}\right) \; + \; O(p_p^3)
\hspace{2em} (a^2 \equiv \epsilon M_N) .
\label{nonrel_k2}
\ee
Combining them to form the function of Eq.~(\ref{function_lf}) we obtain
\be
\frac{2\, |\Psi_d(\alpha_p, \bm{p}_{pT})|^2}{(2 - \alpha_p)^2} 
\; &=& \; 
\frac{2 E_N(\bm{k}) \, |\widetilde\Psi_d (\bm{k})|^2}{(2 - \alpha_p)^2} 
\nonumber
\\[1ex]
\; &\sim& \; \frac{2 M_N \, \Gamma^2}{(2 - \alpha_p)^2 (|\bm{k}|^2 + a^2)^2} 
\nonumber
\\[1ex]
\; &=& \;
\frac{2 M_N \Gamma^2}{(1 - p_p^z/M_N)^2 \;
(1 + p_p^z/M_N)^2 \; (|\bm{p}_p|^2 + a^2)^2} \; + \; O(p_p^2)
\nonumber
\\[1ex]
\; &=& \;
\frac{2 M_N \Gamma^2}{(|\bm{p}_p|^2 + a^2)^2} \; + \; O(p_p^2)
\nonumber
\\[2ex]
\; &=& \;
2 M_N \, |\widetilde\Psi (\bm{p}_p)|^2 \; + \; O(p_p^2) .
\label{function_lf_expanded}
\ee
Both the flux factor Eq.~(\ref{nonrel_alphap}) and the wave function argument Eq.~(\ref{nonrel_k2}) 
involve corrections linear in $p_p^z$, which refer explicitly to the LF direction and break 
rotational symmetry. In the function Eq.~(\ref{function_lf_expanded}), however, the linear corrections 
cancel, and the first corrections are quadratic in the recoil momentum components. This means that 
rotational invariance is effectively restored in the LF formulation at small recoil momenta. 
It implies that the results of the LF IA are numerically close to those of the conventional 
non-relativistic IA at recoil momenta $|\bm{p}_p|, p_p^z \ll M_N$. It also ensures proper analyticity 
of the LF expressions in $t'$ (see Sec.~\ref{subsec:analytic_ia}).
\subsection{Analytic properties}
\label{subsec:analytic_ia}
We now want to study the analytic properties of the IA spectral function in the invariant
momentum transfer $t'$. For this purpose it is natural to use as independent variables $t'$
and the proton LF fraction $\alpha_p$; the relation of these variables to $\alpha_p$ and $\bm{p}_{pT}$ 
is given by Eq.~(\ref{tprime_alphar_prt_relation}). The analytic properties of the spectral function
are governed by the nucleon pole of the deuteron LF wave function, Eq.~(\ref{nucleon_pole_lf}).
The invariant mass difference in Eq.~(\ref{nucleon_pole_lf})
is expressed in terms of $\alpha_p$ and $t'$ as
\beq
s_{pn} - M_d^2 \;\; = \;\; \frac{-2 t'}{2 - \alpha_p} .
\label{invariant_mass_tprime}
\eeq
One sees that the LF spectral function Eq.~(\ref{spectral_impulse}) in the limit 
$t' \rightarrow 0$ at fixed $\alpha_p$ behaves as
\be
S_d (\alpha_p, \bm{p}_{pT})
\; &\sim& \; \frac{R}{(t')^2} \; + \; \textrm{terms} \; O(t'^{-1})
\hspace{2em} 
(t' \rightarrow 0, \; \alpha_p \; \textrm{fixed}),
\label{spectral_pole}
\\[1ex]
R \;\; \equiv \;\; R(\alpha_p) \; &=& \; 4 {\textstyle\sqrt{M_N^2 - a^2}} \, \Gamma^2 (2 - \alpha_p) .
\ee
The spectral function has a pole at $t' = 0$, whose residue depends on $\alpha_p$ and is
calculable in terms of the residue of the pole of the 3--dimensional deuteron wave function, $\Gamma$. 
We note that (a) the nucleon pole is a general feature and relies only on rotational invariance and the 
analytic properties of the rest--frame wave function; (b) the pole in the spectral function
is reproduced by the relativistically invariant formulation of high-energy scattering on the deuteron
(Feynman diagrams, virtual nucleon approximation), where it corresponds to ``nucleon exchange''
between the deuteron and the electromagnetic current;
(c) the pole Eq.~(\ref{spectral_pole}) represents the leading singularity in the limit 
$t' \rightarrow$ and is contained in the IA cross section; FSI modify only subleading singularities 
in $t'$, as was proven in general in Ref.~\cite{Sargsian:2005rm} and will be demonstrated 
explicitly using the specific model of FSI derived in Sec.~\ref{sec:fsi}.

In the limit $t' \rightarrow 0$ the invariant mass difference in deuteron LF wave function 
tends to zero, Eq.~(\ref{invariant_mass_tprime}). This implies that the LF energy off--shellness 
of the $pn$ system in the IA vanishes [cf.\ Eq.~(\ref{offshellness_minus}) and (\ref{qtilde_def})]
\beq
p_d^+ (\widetilde{q}^- - q^-) \;\; = \;\;
p_d^+ (p_d^- - p_n^- - p_p^-) \;\; = \;\;
- (s_{pn} - 4 M_N^2) \;\; \rightarrow 0.
\eeq
The kinematic shift in the 4--momentum transfer, $\widetilde q - q$, Eq.~(\ref{qtilde_def}), 
therefore disappears at the pole, and the IA effectively describes the scattering from a
free on-shell neutron.

%
%
\begin{figure}[t]
\includegraphics[width=.55\textwidth]{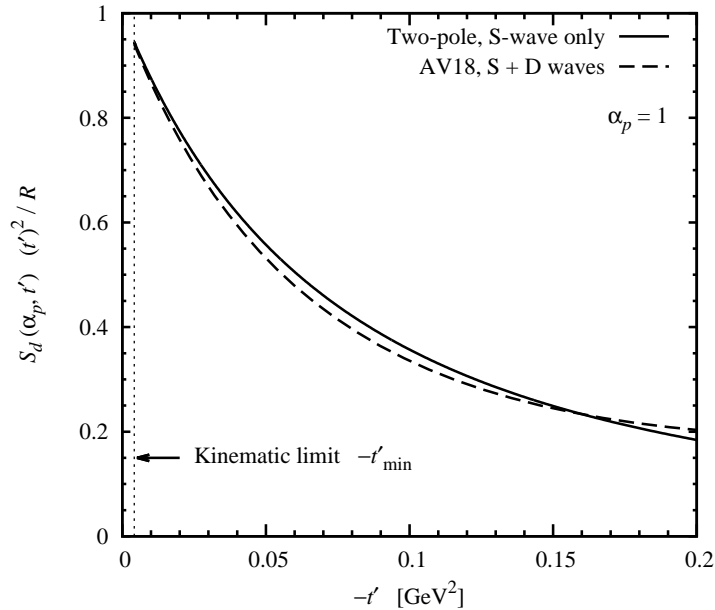}
\caption[]{The deuteron spectral function in the IA, $S_d[\textrm{IA}]$, with the pole factor 
$R/(t')^2$ extracted, as function of $t'$, for $\alpha_p = 1$. Solid line: Two--pole parametrization
of deuteron wave function, Eq.~(\ref{twopole}) (S--wave only).
Dashed line: AV18 wave function (S + D waves) \cite{Wiringa:1994wb}.
The momentum densities are shown in Fig.~\ref{fig:nonrel}a.}
\label{fig:spectral_ia}
\end{figure}
The analytic properties of the LF spectral function suggest a natural method for extracting
the free neutron structure functions from proton-tagged DIS measurements on the deuteron.
One measures the proton-tagged structure function at fixed $Q^2$ as a function of $x$ and the recoil 
proton momentum $|\bm{p}_p|$. One then tabulates the tagged structure function data in $\alpha_p$ and $t'$, 
which extends over the physical region $t' < t'_{\rm min}$. The free neutron structure function
is then obtained by multiplying the tagged structure function data by $(t')^2/R$ (i.e., extracting 
the pole factor of the spectral function) and extrapolating the resulting data to $t' \rightarrow 0$
(on-shell extrapolation). The procedure gives the residue of the tagged structure function at the
pole (with the residue of the spectral function removed), which by definition is the free neutron 
structure function. Nuclear binding and FSI only modify the tagged structure function at $t' \neq 0$
but drop out at the pole, so that the procedure is exact in principle. In practice its
accuracy is determined by the variation of the tagged structure function in $t'$ away from the pole.
This question will be addressed with the specific model of FSI developed in Sec.~\ref{sec:fsi}.

Figure~\ref{fig:spectral_ia} shows the $t'$ dependence of the IA spectral function
after extraction of the pole factor $R/(t')^2$. One sees that the dependence is
smooth over a broad region $|t'| \lesssim 0.1 \, \textrm{GeV}^2$, suggesting that
a polynomial fit would permit accurate extrapolation to $t' = 0$. (The minimum value 
$|t'_{\rm min}|$ is indicated on the graph.) The plot shows the IA spectral functions
obtained with two different deuteron wave functions --- the two-pole parametrization
of Appendix~\ref{app:deuteron} (S wave only), and the AV18 wave function 
(S + D waves) \cite{Wiringa:1994wb}; the differences are very small over the 
$t'$ range shown here.
\section{Final--state hadron distributions}
\label{sec:hadron}
\subsection{Kinematic variables}
\label{subsec:momentum_distribution}
FSI in tagged DIS arise from interactions of the spectator nucleon with ``slow'' hadrons 
produced by the DIS process on the active nucleon (rest frame momenta 
$|\bm{p}_h| \lesssim 1\, \textrm{GeV}$, or target fragmentation region; see Sec.~\ref{sec:introduction}).
In order to calculate these effects we need to study the properties of the slow hadron 
distributions in DIS on the nucleon and parametrize them for our purposes. In this section we discuss 
the kinematic variables characterizing the final--state hadron distributions, the conditional 
structure functions, and the basic features of experimental distributions. 

For the theoretical description of DIS on the nucleon ($N = p, n$) we use a frame where the 
nucleon momentum $\bm{p}_N$ and the momentum transfer $\bm{q}$ are collinear and 
define the $z$--axis of the coordinate system (cf.\ Sec.~\ref{subsec:collinear}).\footnote{In the
calculation of FSI in tagged DIS on the deuteron below we shall neglect the effect of the
active nucleon's transverse momentum on the final-state hadron spectrum, so that the
$z$--axis axis of the virtual photon-nucleon frame coincides with that of the virtual photon-deuteron
frame.} In such a frame the LF components of the nucleon 4--momentum $p_N$ and the 4--momentum 
transfer $q$ are
\beq
\left.
\begin{array}{rclrclrcl}
p_N^+ &>& 0 \;\; \textrm{(arbitrary),} &
\hspace{2em} p_N^- &=& \displaystyle \frac{M_N^2}{p_N^+}, &
\hspace{2em} \bm{p}_{NT} &=& 0 ,
\\[2ex]
q^+ &=& -\xi p_N^+, & 
q^- &=& \displaystyle \frac{Q^2}{\xi p_N^+}, &
\bm{q}_{T} &=& 0 ,
\end{array}
\hspace{2em}
\right\}
\eeq
where $p_N^+$ is arbitrary and defines the particular frame, and the
variable $\xi$ is determined by
\beq
\xi \;\; = \;\; \frac{2 x}{1 + \sqrt{1 + 4 M_N^2 x^2 / Q^2}}
\;\; = \;\; x \; + \; O(M_N^2/Q^2) \hspace{2em} (Q^2 \gg M_N^2) .
\label{xi_nucleon_def}
\eeq
With this choice of components the momentum transfer vector 
$\bm{q}$ points in the \textit{negative} $z$--direction, $2 q^z = q^+ - q^- < 0$.

\begin{figure}
\includegraphics[width=.35\textwidth]{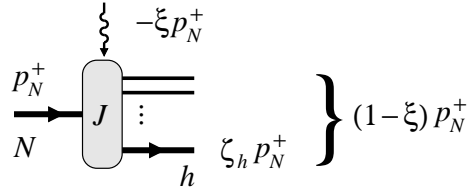}
\caption[]{Current matrix element in DIS on the nucleon with an
identified hadron $h$ in the nucleon fragmentation region,
$e N \rightarrow e' + h + X'$. The LF plus momenta of the final
state are expressed as fractions of $p_N^+$.}
\label{fig:nucleon_frag}
\end{figure}

An identified hadron $h$ in the DIS final state is characterized by its 
LF momentum $p_h^+ \equiv \zeta_h p_N^+$ and transverse momentum 
$\bm{p}_{hT}$ (see Fig.~\ref{fig:nucleon_frag}). Because the hadron LF momentum 
cannot exceed the total LF momentum of the DIS final state, $p_N^+ + q^+ = 
(1 - \xi) p_N^+$, the hadron fraction $\zeta_h$ is restricted to
\be
0 \; < \; \zeta_h \; < \; 1 - \xi .
\label{zeta_limit_nucleon}
\ee
In particular, values $\zeta_h \sim 1$ become kinematically accessible only for $x \sim \xi \ll 1$. 
%
%

It is instructive to consider the hadron momentum distribution in the nucleon rest 
frame in terms of the ordinary momentum variables. The connection with the
LF momentum distribution can established easily, because the nucleon rest frame is
a special collinear frame with $p_N^+ = M_N$. The fraction $\zeta_h$ is
related to the 3--component of the hadron rest frame momentum, $p_h^z$, by
\be
\zeta_h &=& \frac{\sqrt{(p_h^z)^2 + |\bm{p}_{hT}|^2 + M_h^2} + p_h^z}{M_N} ,
\label{zeta_from_pz}
\\[1ex]
p_h^z &=& - \frac{|\bm{p}_{hT}|^2 + M_h^2 - \zeta_h^2 M_N^2}{2 \zeta_h M_N} .
\label{pz_from_zeta}
\ee
One observes: (a) If $M_h \geq M_N$ (e.g., if the identified hadron is a nucleon)
the hadron $z$--momentum is negative, i.e., along the direction of the 
$\bm{q}$--vector. Such hadrons 
always go ``forward'' in the rest frame, meaning in the direction of 
the $\bm{q}$--vector. The momentum distribution is a cone opening 
in the negative $z$--direction. (b)
If $M_h < M_N$ (e.g., if the identified hadron is a pion) the hadron
$z$--momentum can be positive for sufficiently small $|\bm{p}_{hT}|$, i.e.,
opposite to the direction of the $\bm{q}$--vector. Such hadrons can go
``backwards'' in the nucleon rest frame.

\begin{figure}
\includegraphics[width=.66\textwidth]{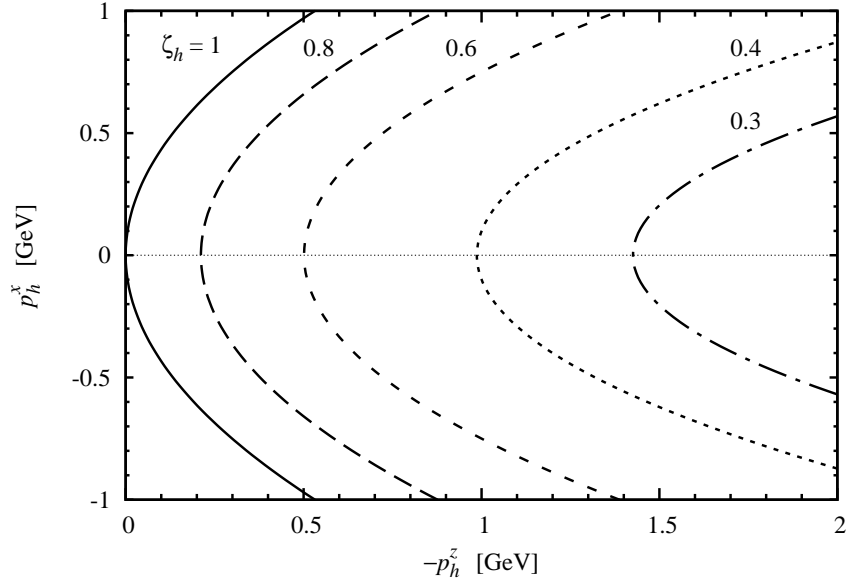}
\caption[]{Momentum distributions of nucleons ($M_h = M_N$) in the nucleon fragmentation region 
in DIS, $e N \rightarrow e' + h + X$. The contours show the allowed values of $p_h^z$ and $p_h^x$ 
(with $p_h^y = 0$, so that $|\bm{p}_{hT}| = |p_h^x|$) for given constant values of $\zeta_h$. 
The contours thus describe the intersection of the allowed 3-momentum cones with the
transverse $x$--plane.}
\label{fig:slow}
\end{figure}

\begin{figure}
\includegraphics[width=.5\textwidth]{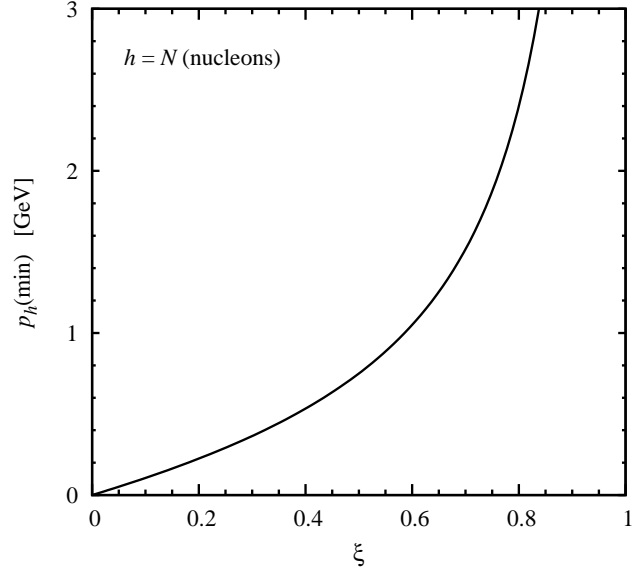}
\caption[]{Minimal 3--momentum $|\bm{p}_h|(\textrm{min})$ of nucleons ($M_h = M_N$) in the 
nucleon fragmentation region in DIS, Eq.~(\ref{p_nucleon_min}), as a function of the
variable $\xi = x + O(M_N^2/Q^2)$, Eq.~(\ref{xi_nucleon_def}).}
\label{fig:slow_pmin}
\end{figure}

Figure \ref{fig:slow} shows the momentum distribution of nucleons ($M_h = M_N$) 
in the nucleon fragmentation region for fixed values of $\zeta_h$. One sees that
small longitudinal momenta $p_h^z \rightarrow 0$ correspond to LF fractions 
$\zeta_h \rightarrow 1$, and that the cones are shifted to larger longitudinal
momenta as $\zeta_h$ deviates from unity. Note that $\zeta_h$ is kinematically
restricted by Eq.~(\ref{zeta_limit_nucleon}). Figure \ref{fig:slow_pmin}
shows the minimal 3--momentum of nucleons in the nucleon fragmentation region
as a function of $\xi$. The minimal value of the 3--momentum is attained for 
$\bm{p}_{hT} = 0$ [see Eq.~(\ref{pz_from_zeta}) and Fig.~\ref{fig:slow}] and 
given by
\beq
|\bm{p}_h|(\textrm{min}) \;\; = \;\; \frac{\xi (1 - \xi/2) M_N}{1 - \xi} 
\hspace{2em} (h = N).
\label{p_nucleon_min}
\eeq
One sees that nucleons with $|\bm{p}_h| \lesssim 1\,\textrm{GeV}^2$ appear only if $x \sim \xi \ll 1$.
Note that Eq.~(\ref{p_nucleon_min}) gives only the {\em kinematic limit}, and that the {\em average} 
values of the nucleon momenta in the fragmentation region are substantially larger, because the 
phase space opens with transverse momentum.
%
%
%
%

The LF variables $\zeta_h$ and $\bm{p}_{hT}$ can be related to other 
variables used to characterize experimental hadron distributions in DIS.
One commonly used variable is the fraction of the rest-frame energy transfer
carried by the hadron, 
\beq
z_h \; \equiv \; \frac{E_h}{\nu},
\hspace{3em}
\nu \; = \; \frac{Q^2}{2 M_N x} \;\; = \;\; \frac{Q^2}{2 M_N \xi} 
\left( 1 - \frac{\xi^2 M_N^2}{Q^2} \right) .
\eeq
Using Eq.~(\ref{zeta_from_pz}) and setting 
\beq
p_h^z \; = \; - {\textstyle \sqrt{E_h^2 - M_{hT}^2}},
\hspace{3em} M_{hT}^2 \; \equiv \; |\bm{p}_{hT}|^2 + M_h^2 ,
\eeq
one obtains
\be
\zeta_h &=& \frac{z\nu - \sqrt{(z\nu)^2 - M_{hT}^2}}{M_N} 
\;\; \approx \;\; \frac{M_{hT}^2}{2 z \nu M_N}
\;\; = \;\; \frac{x M_{hT}^2}{z Q^2} \hspace{2em} (z\nu \gg M_{hT}),
\\[1ex]
z_h &=& \frac{\zeta_h^2 M_N^2 + M_{hT}^2}{2\zeta_h M_N \nu} .
\ee
One sees that LF fractions $\zeta_h = O(1)$ correspond to parametrically
small energy fractions $z = O(M_{hT}/\nu) \ll 1$ in the 
nucleon rest frame (slow hadrons).
Another commonly used variable is the hadron's longitudinal momentum in the
center--of--mass (CM) frame of the virtual photon--nucleon collision, in which
$\bm{q} + \bm{p}_N = 0$. It is usually expressed in terms of the
Feynman scaling variable \footnote{The variable $x_{\rm F}$ in electroproduction is
conventionally defined such that hadrons moving in virtual photon direction 
have $x_{\rm F} > 0$, and hadrons moving in the target direction 
have $x_{\rm F} < 0$. In our convention $q^z < 0$ and $p^z_N > 0$; hence the
minus sign in Eq.~(\ref{x_feynman_def}).}
\beq
x_{\rm F} \;\; \equiv \;\; - \frac{p_h^z}{p^z_{h, {\rm max}}} .
\label{x_feynman_def}
\eeq
The connection with the LF variables is established by noting that the CM
frame is the special collinear frame with
\beq
p_N^+ \;\; = \;\; \sqrt{\frac{Q^2 + \xi M_N^2}{\xi (1 - \xi)}} .
\eeq
The hadron longitudinal CM momentum is
\beq
p_h^z \;\; = \;\; \frac{p_h^+ + p_h^-}{2}
\;\; = \;\; \frac{\zeta_h^2 (p_N^+)^2  - M_{hT}^2}{2 \zeta_h p_N^+} .
\eeq
The maximum (positive) value is attained for $\zeta_h = 1 - \xi$ and $\bm{p}_{hT} = 0$,
\beq
p_{h, {\rm max}}^z \;\; = \;\; \frac{(1 - \xi)^2 (p_N^+)^2  - M_h^2}{2 (1 - \xi) p_N^+} .
\eeq
The scaling variable is thus obtained as
\be
x_{\rm F} &=& - \frac{(1 - \xi) \; [\zeta_h^2 (p_N^+)^2  - M_{hT}^2]}
{\zeta_h \; [(1 - \xi)^2 (p_N^+)^2  - M_h^2]}
\;\; \approx \;\; -\frac{\zeta_h}{1 - \xi}
\hspace{2em} (\zeta_h p_N^+ \gg M_{hT}) .
\label{xf_from_zeta}
\ee
The latter condition is fulfilled if $Q^2 \gg M_{hT}^2$ (DIS limit) and
$\zeta_h = O(1)$. One concludes that $-x_{\rm F}$ in the nucleon fragmentation region
(from $-1$ to approximately $-0.5$) can be identified directly with the normalized
hadron LF fraction $\zeta_h/(1 - \xi)$. We use this relation in our analysis
of experimental slow hadron spectra below.
\subsection{Multiplicity distributions}
The hadronic tensor and differential cross section for DIS on the nucleon with
an identified final-state hadron $h$ are described by expressions analogous to those
for DIS on the deuteron with an identified nucleon in Sec.~\ref{subsec:cross_section};
see Eqs.~(\ref{w_mu_nu_current}) and (\ref{dsigma_tagged}). The hadronic tensor is
parametrized by conditional nucleon structure functions
\beq
F_{2N, h}(x, Q^2; \zeta_h, \bm{p}_{hT}), \hspace{2em} \textit{etc.},
\eeq
which depend on the identified hadron's LF momentum fraction $\zeta_h$ and transverse 
momentum $\bm{p}_{hT}$. It is convenient to extract the inclusive structure functions
and write the conditional structure functions in the form
\be
F_{2N, h}(x, Q^2; \zeta_h, \bm{p}_{hT}) &=& F_{2N}(x, Q^2) \;
D_h(x, Q^2; \zeta_h, \bm{p}_{hT}) ,
\hspace{2em} \textit{etc.}
\label{F2Nh_factorized}
\ee
The the function $D_h$ describes the normalized differential multiplicity distribution 
of the hadron $h$, i.e., the differential number of hadrons $dN_h$ per DIS event observed 
in a phase space element $d\Gamma_h$:
\beq
\frac{dN_h}{N_{\rm incl}}
\;\; = \;\; D_h \; d\Gamma_h ,
\hspace{3em}
d\Gamma_h \;\; = \;\; \frac{d\zeta_h \; d^2 p_{hT}}{2 (2\pi)^3 \zeta_h} .
\label{differential_multiplicity}
\eeq
As such it can be directly extracted from the experimental multiplicity distributions.
In particular, the $p_{hT}$--integrated LF momentum distribution of the hadron is
\be
\frac{1}{N_{\rm incl}} \frac{dN_h}{d\zeta_h}
&=& \frac{1}{2 (2\pi)^3 \zeta_h} \;
\int d^2 p_{hT} \; D_h .
\label{multiplicity_ptint}
\ee
It can be identified with the $x_{\rm F}$ distribution in the nucleon fragmentation region,
cf.~Eq.~(\ref{xf_from_zeta}). Note the factor $1/\zeta_h$ on the right-hand side,
which results from the definition of the invariant phase space element 
Eq.~(\ref{differential_multiplicity}).
\subsection{Experimental distributions}
\label{subsec:final_state_experimental}
Measurements of hadron multiplicity distributions in the target fragmentation 
region in DIS on the nucleon have been reported by several fixed-target experiments 
using electron beams (Cornell Synchrotron \cite{Hanson:1975zi}) and muon beams 
(CERN EMC \cite{Arneodo:1984rm,Arneodo:1984nf,Arneodo:1986yb}, FNAL E665 \cite{Adams:1993wv}),
as well as at the HERA electron-proton 
collider \cite{Alexa:2013vkv,Andreev:2014zka,Aaron:2010ab,Chekanov:2008tn}. 
Slow hadron distributions were also measured in neutrino-proton DIS 
experiments \cite{Derrick:1977zi,Allen:1980ip,Allen:1982jp}. While the kinematic coverage is 
far from complete, these data roughly cover the $x$--region of interest for
our study and allow us to infer the basic features of the multiplicity distributions. 
Unfortunately many data are not separated according to hadron species, as few 
dedicated studies of the target fragmentation region have been performed so far.
We now briefly review the main features of the data and their theoretical
interpretation.

The multiplicity distributions of hadrons with 
$x_{\rm F} \lesssim -0.2$ are approximately independent of $Q^2$ for fixed $x$. 
Scaling of the distributions is observed in all quoted electron and muon 
experiments, covering the valence region $x > 0.2$ \cite{Hanson:1975zi}, 
the region $x \lesssim 0.1$
\cite{Arneodo:1984rm,Arneodo:1984nf,Arneodo:1986yb,Adams:1993wv}, 
and the small--$x$ region $x < 10^{-2}$
\cite{Alexa:2013vkv,Andreev:2014zka,Aaron:2010ab,Chekanov:2008tn}.
This behavior is consistent with theoretical expectations based on QCD
factorization of the conditional DIS cross sections in the target
fragmentation region \cite{Trentadue:1993ka,Collins:1997sr}.
The multiplicity distributions in the target fragmentation region show only 
weak variation with $x$ in the region $x \lesssim 0.1$. This indicates that
the hadronization of the target remnant is largely independent of the dynamics
producing the parton distributions in the nucleon in this region of $x$ (sea quarks, gluons).

The $x_{\rm F}$ distributions of protons (integrated over transverse momentum) 
in DIS on the proton are approximately flat 
for $x_{\rm F} < -0.3$. A value of $(1/N_{\rm incl}) \, dN_p/dx_{\rm F} \sim 0.5-0.6$ at
$x_{\rm F} = (-0.7, -0.3)$ was measured by EMC at $\langle x \rangle = 0.1$ \cite{Arneodo:1984nf}.
A value $(1/N_{\rm incl}) \, dN_p/dx_{\rm F} \sim 0.4$ at $x_{\rm F} = (-0.8, -0.4)$ was obtained 
by the HERA experiments at $x \lesssim 0.01$ \cite{Alexa:2013vkv,Chekanov:2008tn}.
(At larger negative $x_{\rm F}$ diffraction gives rise to a distinct contribution to the 
proton spectrum at HERA; this mechanism is marginal in the kinematic region 
considered here.) The $x_{\rm F}$ distribution of neutrons measured in DIS on the proton 
at HERA \cite{Andreev:2014zka,Aaron:2010ab} is also flat and has a value
of $(1/N_{\rm incl}) \, dN_p/dx_{\rm F} \sim 0.2$ at $x_{\rm F} = (-0.8, -0.4)$. The sum of
proton and neutron multiplicity distributions is thus 
$(1/N_{\rm incl}) \, dN_{p + n}/dx_{\rm F} \sim 0.6$ at $x_{\rm F} = (-0.8, -0.4)$.
That this value is significantly less than 1 shows that part of the baryon number 
is transported to smaller $x_{\rm F}$ and materializes outside the target region.
We note that both the flatness of the distributions and the baryon number transport
are reproduced by string models of the fragmentation mechanism.

The transverse momentum distributions of protons and neutrons in the target 
fragmentation region drop steeply with $p_{T, h}$ and can be approximated by 
Gaussian distributions $\sim \exp(-B_h p_{hT}^2) \; (h = p, n)$, 
where the slope $B_h$ determines the average squared transverse momentum 
as $\langle p_{hT}^2 \rangle = B_h^{-1}$.
The empirical slope for protons is $B_p \sim 4\, \textrm{GeV}^{-2}$ at $x > 0.2$ 
(Cornell) \cite{Hanson:1975zi} and $B_p =$ 6--8 $\textrm{GeV}^{-2}$ at $x < 10^{-2}$ (HERA)
\cite{Alexa:2013vkv,Chekanov:2008tn}. A values $B_p \sim$ 6 $\textrm{GeV}^{-2}$ was
also observed in neutrino DIS at $W^2 < 50\, \textrm{GeV}^2$ \cite{Derrick:1977zi}.

The multiplicity distribution of charged pions shows very different behavior from that 
of protons and neutrons. The $x_{\rm F}$ distribution of pions are significantly smaller than 
those of protons at $x_{\rm F} < -0.3$ but rise strongly at $x_{\rm F} > -0.3$ \cite{Arneodo:1984nf}.
Pion production thus happens mainly in the central region of the DIS process and is
governed by other dynamical mechanisms than target fragmentation.
\subsection{Implications for FSI}
\label{subsec:implications}
The experimental results described in Sec.~\ref{subsec:final_state_experimental}
characterize the slow hadron distributions causing FSI in tagged DIS on the deuteron. 
We now want to summarize the implications and formulate a simple parametrization
of the slow hadron distribution for our subsequent calculations.

The dominant hadrons produced in electron-nucleon DIS at $\zeta_h > 0.2$ are protons and 
neutrons emerging from the hadronization of the remnant of the active nucleon. These 
protons and neutrons can interact with the spectator nucleon with the large $NN$ cross 
section of $\sim 40$ mb at momenta $|\bm{p}_h| \sim$ 1--2 GeV (see Fig.~\ref{fig:slow} 
and Appendix~\ref{app:rescattering_amplitude}). We therefore suppose that the dominant 
FSI in tagged DIS at $x \sim 0.1-0.5$ arises from 
such protons and neutrons in the target fragmentation region of the active nucleon.
FSI induced by pions could in principle be treated within the same picture but are 
expected to be small.

If the active nucleon in the deuteron is the proton (i.e., if the neutron is tagged), 
the multiplicity distributions of slow nucleons (protons plus neutrons) can be inferred 
directly from the proton DIS data. If the active nucleon is a neutron (proton tagged), 
we suppose that at $x \sim 0.1$ the distribution of slow nucleons (protons plus neutrons) 
is approximately the same as in DIS on the proton, because the deep-inelastic process occurs 
mainly on singlet sea quarks produced by gluon radiation and does not change the flavor 
structure of the baryon remnant system (we neglect the effect of the flavor asymmetry of
sea quarks in this context). Since furthermore the nucleon-nucleon cross section at 
momenta $\sim$ few GeV is approximately the same for $pp, pn$ and $nn$ scattering 
(see Appendix~\ref{app:rescattering_amplitude}), the FSI effect is the same for deuteron 
DIS with active proton or active neutron. These approximations permit a model-independent
estimate of FSI effects at $x \sim 0.1$ and will be used in our subsequent calculations. 
The physical picture of FSI and the formulas derived in the following are valid 
also at larger $x$ ($\lesssim$ 0.5), where the scattering primarily occurs mainly
on valence quarks; in this region they should be evaluated with a detailed model of 
the quark flavor dependence of the slow hadron multiplicity distributions.

In our numerical studies of FSI in tagged deuteron DIS we use a simple parametrization 
of the multiplicity distribution of slow protons and neutrons, $D_h \; (h = p, n)$, 
which reflects the basic features of the experimental distributions and offers sufficient 
flexibility to study the dependence of FSI on the slow hadron distribution.
We parametrize the distribution in the form
\be
D_h (x, Q^2; \zeta_h, p_{hT}) &=& [2 (2\pi)^3] \; \zeta_h \; f_h (\zeta_h) \; g_h (p_{hT})
\hspace{2em} (h = p, n).
\label{D_model}
\ee
The function $f_h (\zeta_h)$ describes the $\zeta_h$ distribution and can be
identified with the $p_{hT}$--integrated multiplicity distribution 
Eq.~(\ref{multiplicity_ptint}). We choose it such that
\be
f_h (\zeta_h) &\sim& c_h \; = \; \textrm{const} \hspace{2em} (1 > \, \zeta_h \, \gtrsim \zeta_0 ),
\\[2ex]
f_h (\zeta_h) &\rightarrow& 0 \hspace{4em} (\zeta_h \rightarrow 0 ).
\ee
The constant $c_h$ can be inferred from the experimental proton/neutron $\zeta_h$ 
(or $x_{\rm F}$) distributions in the ``flat'' region. For the sum of proton and neutron 
distributions it is
\beq
\sum_{h = p, n} c_h \; = 
c_p + c_n \; = \; 0.6-0.8 .
\eeq
The cutoff at $\zeta_h \rightarrow 0$ 
limits the distribution to slow hadrons in the nucleon rest frame, which are fully formed 
inside the deuteron and interact with the spectator with the $NN$ cross section. A value
$\zeta_0 \sim 0.2$ corresponds to rest-frame momenta 
$|\bm{p}_h| \lesssim 2$ GeV (see Fig.~\ref{fig:slow}).
A simple choice for $f_h (\zeta_h)$ is
\beq
f_h(\zeta_h ) \;\; = \;\; \frac{c_h (\zeta_h/\zeta_0)^n}
{1 + (\zeta_h/\zeta_0)^n} ,
\eeq
where we choose $n$ = 3--5; the results are not sensitive to the details of the cutoff.
The function $g_h(p_{hT})$ describes the normalized $p_{hT}$ distribution of the
protons/neutrons and is modeled by a Gaussian,
\beq
g_h(p_{hT}) \; = \; \frac{B_h}{\pi} \exp(-B_h p_{hT}^2), 
\hspace{2em}
\int d^2 p_{hT} \; g_h (p_{hT}) \; = \; 1, 
\eeq
with an empirical slope 
\beq
B_h \; = \; 6-8 \, \textrm{GeV}^{-2} \hspace{2em} (h = p, n).
\eeq
As explained above, this distribution is used for the slow protons/neutrons in DIS
on either the proton or neutron in the deuteron.

In our treatment of FSI we describe the interactions of the slow protons/neutrons with 
the spectator nucleon as on-shell rescattering with an effective interaction. Off-shell 
effects are physically related to effects of the finite hadron formation time and can 
be modeled as a modification of the on-shell effective interaction and the slow 
proton/neutron distribution. The on-shell effective interaction (scattering amplitude)
can be determined from the $NN$ total and elastic cross section data. The main features
of the data and and a simple parametrization of the amplitude at incident momenta 
$|\bm{p}_h| \lesssim 2\, \textrm{GeV}$ are described in 
Appendix~\ref{app:rescattering_amplitude}.
\section{Final--state interactions}
\label{sec:fsi}
\subsection{FSI and IA currents}
We now proceed to calculate the tagged DIS cross section including FSI between the spectator
nucleon and the slow hadrons (protons/neutrons) emerging from the fragmentation of the
active nucleon, in the physical picture described in Secs.~\ref{sec:introduction} 
and \ref{subsec:implications}. The calculation is performed in LF quantum mechanics 
in the collinear frame $\bm{p}_{dT} = 0$ as in Sec.~\ref{sec:ia} and identifies corrections 
to the IA current and the deuteron tensor resulting from FSI.

To properly account for the configurations in which FSI can and cannot occur, we separate
the multi-hadron states $X$ produced in DIS on the nucleon into two classes:
\begin{itemize}
\item[(a)] Multi-hadron states not containing a slow hadron capable of inducing FSI,
which we denote by $X_0$.
\item[(b)] Multi-hadron states containing a slow hadron $h$ capable of inducing FSI,
which we denote by $X_1$. Their state vectors are of the form
\beq
|X_1 \rangle \;\; = \;\; |h, X' \rangle \;\; \equiv \;\; 
|h \rangle |X' \rangle ,
\eeq
where $X'$ is the product state of the remaining hadrons. The summation over these 
states is performed as
\beq
\sum_{X_1} \;\; \equiv \;\; \int d\Gamma_h \; \sum_{X'} .
\eeq
\end{itemize}
By construction the classes $X_0$ and $X_1$ then exhaust all possible multi-hadron
states, and the summation over all states becomes
\beq
\sum_X \; = \; \sum_{X_0} \; + \; \sum_{X_1} .
\label{x_x0_x1}
\eeq
The separation is possible because the average slow hadron multiplicity is $< 1$
(cf.~Sec.~\ref{subsec:implications}); i.e., 
we can assume that the final state contains zero or one slow hadrons, but not more. 

%
%
\begin{figure}
\includegraphics[width=.7\textwidth]{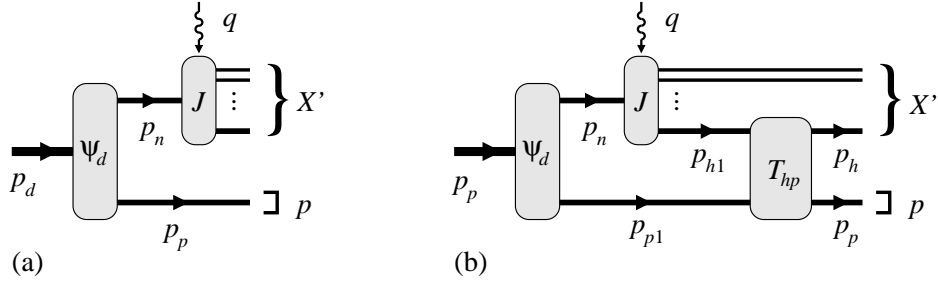}
\caption[]{Current matrix elements in tagged DIS on the deuteron. (a) IA current.
(b) FSI between a slow DIS hadron $h$ and the spectator.}
\label{fig:deut_ia_fsi}
\end{figure}
We now consider tagged DIS on the deuteron separately for final states $X_0$ and $X_1$,
\beq
e + d \;\; \rightarrow \;\; e' + p + (X_0 \; \textrm{or} \; X_1).
\eeq
For final states of type $X_0$ FSI cannot occur, and the transition current is identical 
to that obtained in the IA, Eqs.~(\ref{current_ia_1}) and (\ref{current_ia_2}),
\beq
\langle p X_0 | \hat J^\mu (0) | d \rangle \;\; = \;\; 
\langle p X_0 | \hat J^\mu (0) | d \rangle[\textrm{IA}] .
\label{current_x0}
\eeq
For final states of type $X_1$ the transition current is computed by inserting plane-wave 
nucleon and slow hadron intermediate states into the current matrix 
element (cf.\ Fig.~\ref{fig:deut_ia_fsi}), 
\be
\langle p X_1 | \hat J^\mu (0) | d \rangle 
\; &\equiv& \; \langle p h X' | \hat J^\mu (0) | d \rangle 
\nonumber
\\[1ex]
&=& \; \int [dp_n] \;
\int [dp_{p1}] \;
\int [dp_{h1}] \;
\langle p, p_p; h, p_h | p, p_{p1}; h, p_{h1} \rangle 
\nonumber
\\[1ex]
&\times& \;
\langle X'; h, p_{h1} | \hat J^\mu (0) | n, p_n \rangle \;
\langle p, p_{p1} ; n, p_n | d, p_d \rangle .
\label{current_fsi_general}
\ee
FSI between the slow hadron and the spectator are now incorporated by taking the 
hadron--spectator final state not as a product state (as is done in the IA) but as the scattering 
state generated by interactions between them. The boundary conditions for the scattering
state correspond to the incoming--wave solution \cite{Arriaga:2007tc},
\beq
\langle p, p_p; h, p_h | \;\; \rightarrow \;\; {}_{\rm in}\langle p, p_p; h, p_h | .
\eeq
Following standard practice in nuclear physics we can express the wave function
of this scattering state in terms of the effective interaction operator $\hat T$ 
(or $T$--matrix) corresponding to the interaction,
\beq
{}_{\rm in}\langle p, p_p; h, p_h | p, p_{p1}; h, p_{h1} \rangle \;\; = \;\;
\langle p, p_p; h, p_h | p, p_{p1}; h, p_{h1} \rangle 
\; - \; \frac{\langle p, p_p; h, p_h | \; \hat{T} \; |p, p_{p1}; h, p_{h1} \rangle}
{\frac{1}{2}(p_p^- + p_h^- - p_{p1}^- - p_{h1}^- + i0)} .
\label{scattering_wf}
\eeq
Equation~(\ref{scattering_wf}) represents the LF analogue of the Lippmann--Schwinger 
equation in non-relativistic quantum mechanics. The denominator is the difference
of LF energies between the initial and final states (the LF Hamiltonian in our 
convention is $\hat H_{\rm LF} = \frac{1}{2} \hat P_{\rm tot}^{-}$; see Appendix 1 of
Ref.~\cite{Granados:2015rra}).
In the context of a time-dependent formulation Eq.~(\ref{scattering_wf}) can be regarded 
as the matrix element of the LF time evolution operator corresponding to the 
hadron--spectator interaction between LF time $x^+ = 0$ (when the current 
creates the state $h$) and $x^+ \rightarrow +\infty$ (when the interactions are switched off),
\beq
{}_{\rm in} \langle p, p_p; h, p_h | p, p_{p1}; h, p_{h1} \rangle 
\;\; = \;\; 
\langle p, p_p; h, p_h | \, \hat U(\infty, 0) \, | p, p_{p1}; h, p_{h1} \rangle .
\label{time_evolution_operator}
\eeq
This representation makes it obvious that the scattering state obeys incoming-wave 
boundary conditions.\footnote{The formal operator generating a two-body scattering state 
from the product states [cf.~Eq.~(\ref{scattering_wf})] is known as the M\o ller operator
and can be defined in a general context. Its representation as the limit of a time evolution 
operator depends on asymptotic conditions; for a discussion see 
Refs.~\cite{Coester:1965zz,Schierholz:1968uck,Namyslowski:1977pr}.} 

The effective interaction operator in Eq.~(\ref{scattering_wf}) 
conserves the total LF momentum of the states.
Using translational invariance, we can write the matrix element 
in the numerator of Eq.~(\ref{scattering_wf}) in the form
\be
\langle p, p_p; h, p_h | \; \hat{T} \; |p, p_{p1}; h, p_{h1} \rangle
\; &=& \; (2\pi)^3 \; \delta (p_p^+ + p_h^+ - p_{p1}^+ - p_{h1}^+)
\; \delta^{(2)} (\bm{p}_{pT} + \bm{p}_{hT} - \bm{p}_{p1T} - \bm{p}_{h1T})
\nonumber
\\[1ex] 
&\times & \; T (p_p, p_h; p_{p1}, p_{h1}) ,
\ee
where no assumption is made about the LF energies of the initial and final states. 
The on-shell part of the 
scattering term in Eq.~(\ref{scattering_wf}), in which the total final LF energy 
is equal to initial one, is obtained by retaining the pole term of the energy 
denominator,
\be
\lefteqn{
\left. \frac{\langle p, p_p; h, p_h | \; \hat{T} \; |p, p_{p1}; h, p_{h1} \rangle}
{\frac{1}{2}(p_p^- + p_h^- - p_{p1}^- - p_{h1}^- - i0)} \right|_{\rm on-shell}}
&& \nonumber 
\\[2ex]
&=&
i \, (2\pi) \, \delta (p_p^- + p_h^- - p_{p1}^- - p_{h1}^- ) \;
\langle p, p_p; h, p_h | \; \hat{T} \; |p, p_{p1}; h, p_{h1} \rangle
\\[2ex]
&=&
i \, (2\pi)^4 \, \delta (p_p^- + p_h^- - p_{p1}^- - p_{h1}^- ) \;
\delta (p_p^+ + p_h^+ - p_{p1}^+ - p_{h1}^+)
\; \delta^{(2)} (\bm{p}_{pT} + \bm{p}_{hT} - \bm{p}_{p1T} - \bm{p}_{h1T}) \; T
\\[2ex]
&=& i \, (2\pi)^4 \, \delta^{(4)} (p_p + p_h - p_{p1} - p_{h1} ) \; \frac{T}{2} .
\label{scattering_wf_onshell}
\ee
Here $T$ is the on-shell matrix element of effective interaction, which coincides
with the invariant amplitude of the physical $ph \rightarrow ph$ 
scattering process,
\be
T(p_p, p_h; p_{p1}, p_{h1}) &=& \mathcal{M}(s_{ph}, t_{ph}) ,
\label{invariant_amplitude}
\\[2ex]
s_{ph} \; \equiv \;  (p_{p1} + p_{h1})^2 &=& (p_p + p_{h})^2, \\[1ex]
\label{invariant_amplitude_s}
t_{ph} \; \equiv \;  (p_{p1} - p_p)^2 &=& (p_{h1} - p_{h})^2 .
\label{invariant_amplitude_t2}
\ee
Equations~(\ref{scattering_wf})--(\ref{invariant_amplitude_t2}) allow us to express
the FSI matrix element in terms of the physical $ph \rightarrow ph$ amplitude
within our scheme of approximations. The factor 1/2 in Eq.~(\ref{scattering_wf_onshell})
accounts for the fact that the interaction in the matrix element is present only 
from $x^+ = 0$ to $\infty$ (i.e, ``half the time''), while in the scattering process
it is present from $x^+ = -\infty$ to $\infty$. We note that the same factor 1/2 is 
obtained in an equivalent calculation of the FSI effect in the current matrix element
using invariant perturbation theory, where it appears from the Cutkosky rules for the 
on-shell part of the Feynman diagram.

We can now derive from Eq.~(\ref{current_fsi_general}) the explicit expressions for 
the transition current to $X_1$ states. The $d \rightarrow pn$ matrix element 
in Eq.~(\ref{current_fsi_general}) is expressed in terms of the deuteron LF wave 
function Eq.~(\ref{lfwf_def}). The FSI matrix element is substituted by
Eq.~(\ref{scattering_wf}). The non-interaction term on the R.H.S.\ results in 
an expression of the same form as the IA result, Eq.~(\ref{current_ia_2}). 
The interaction term is expressed in terms of the on-shell scattering amplitude
using Eqs.~(\ref{scattering_wf_onshell}) and (\ref{invariant_amplitude}). 
Altogether the transition current to $X_1$ states, Eq.~(\ref{current_fsi_general}), 
can be written as the sum of an IA and an FSI term,
\be
\langle p X_1 | \hat J^\mu (0) | d \rangle \; &\equiv& \;
\langle p h X' | \hat J^\mu (0) | d \rangle \; = \; 
\langle \ldots \rangle \, \textrm{[IA]}
\; + \; \langle \ldots \rangle \, \textrm{[FSI]}
\label{current_ia_fsi}
\\[2ex]
\langle p h X' | \hat J^\mu (0) | d \rangle \, \textrm{[IA]} \; &=& \;
\frac{p_d^+}{p_n^+} \;
\langle h, p_h; X'| \hat J^\mu (0) | n, p_n \rangle \; 
(2\pi)^{3/2} \, \Psi_d (\alpha_p , \bm{p}_{pT})
\nonumber
\\[2ex]
&& \; [p_n^+ = p_d^+ - p_p^+, \; \bm{p}_{nT} = - \bm{p}_{pT}] ,
\label{current_ia_x1}
\\[2ex]
\langle p h X' | \hat J^\mu (0) | d \rangle \, \textrm{[FSI]} &=& \int [d p_{p1}] 
\; \frac{p_d^+}{p_n^+} \;
\langle h, p_{h1} ; X' | \hat J^\mu (0) | n, p_n \rangle \; 
(2\pi)^{3/2} \, \Psi_d (\alpha_{p1} , \bm{p}_{p1T})
\nonumber \\
&\times & \;
\frac{2\pi}{p_{h1}^+} \; \delta (p_p^- + p_h^- - p_{p1}^- - p_{h1}^-) \; 
\frac{i}{2} \, {\mathcal M}(s_{ph}, t_{ph})
\nonumber
\\[2ex]
&& \; [p_n^+ = p_d^+ - p_{p1}^+, \; \bm{p}_{nT} = - \bm{p}_{p1T};
\nonumber \\
&& \; p_{h1}^+ = p_h^+ + p_p^+ - p_{p1}^+, \;
\bm{p}_{h1T} = \bm{p}_{hT} + \bm{p}_{pT} - \bm{p}_{p1T}] ,
\label{current_fsi}
\ee
where $s_{ph}$ and $t_{ph}$ are given by Eqs.~(\ref{invariant_amplitude_s}) and
(\ref{invariant_amplitude_t2}).
\subsection{Distorted spectral function}
The deuteron tensor for tagged DIS in the presence of FSI is obtained as the product 
of the current matrix element and its complex conjugate, summed over all
final states $X$, Eq.~(\ref{w_mu_nu_current}). In accordance with the distinction
between final states with zero and one slow hadron, $X_0$ and $X_1$,
we now write this sum as
\be
W^{\mu\nu}_d \; &=& \; \sum_{X_0} \langle d | \hat{J}^\mu(0) | p X_0 \rangle
\; \langle p X_0  | \hat{J}^\nu(0) | d \rangle
\nonumber \\[1ex]
&+& \; \sum_{X_1} \langle d | \hat{J}^\mu(0) | p X_1 \rangle
\; \langle p X_1  | \hat{J}^\nu(0) | d \rangle ,
\label{tensor_x0_x1}
\ee
and substitute the expressions Eq.~(\ref{current_x0}) and 
Eqs.~(\ref{current_ia_fsi})--(\ref{current_fsi}) for the different
current matrix elements.
It is easy to see that in Eq.~(\ref{tensor_x0_x1}) the sum over states $X_0$ 
(for which the current is always of IA form), and the part of the sum over states 
$X_1$ involving the
IA term of the currents (in which no FSI takes place), 
reproduce the original IA result for the deuteron tensor,
Eq.~(\ref{tensor_ia}),
\be
&& \; \sum_{X_0} \langle d | \hat{J}^\mu(0) | p X_0 \rangle [\textrm{IA}]
\; \langle p X_0  | \hat{J}^\nu(0) | d \rangle [\textrm{IA}]
\nonumber \\[1ex]
&+& \; \sum_{X_1} \langle d | \hat{J}^\mu(0) | p X_1 \rangle [\textrm{IA}]
\; \langle p X_1  | \hat{J}^\nu(0) | d \rangle [\textrm{IA}]
\nonumber \\[1ex]
&=& \; \sum_{X} \langle d | \hat{J}^\mu(0) | p X \rangle [\textrm{IA}]
\; \langle p X  | \hat{J}^\nu(0) | d \rangle [\textrm{IA}]
\;\; = \;\; W_d^{\mu\nu} [\textrm{IA}] .
\label{w_mu_nu_ia_x0_x1}
\ee
Here we have used that the sum over states $X_0$ and $X_1$ exhausts the full
set of inclusive final states $X$, Eq.~(\ref{x_x0_x1}).

Corrections to the IA tensor arise from the FSI terms in the currents in
the sum over states $X_1$ in Eq.~(\ref{tensor_x0_x1}). These corrections
come in two types: (a) the products of the FSI term of one current and the IA 
term of the other (linear FSI); (b) the product of the FSI terms from both 
currents (quadratic FSI).

Consider the linear FSI correction arising from the product of the FSI current
Eq.~(\ref{current_fsi}) and the complex conjugate IA current Eq.~(\ref{current_ia_x1}).
Because of the momentum integral in the FSI current, the momentum of the active
neutron in the FSI current, $p_n^+$ and $\bm{p}_{nT}$, is generally different from 
that in the IA current. The corresponding neutron current matrix elements can therefore 
not generally be combined to form the neutron tensor (which is diagonal in the 
neutron momentum), as was done for the IA in Eq.~(\ref{tensor_ia}).
An important simplification arises from the fact that the characteristic momenta in 
deuteron wave function are much smaller than in nucleon current matrix element.
The latter is therefore not affected by the small shift of the active neutron momentum
caused by the FSI integral and can be evaluated at the nominal active neutron momentum
defined by the IA.

A similar simplification can be made regarding the slow hadron momentum in the
neutron current matrix element. Under the FSI integral the slow hadron produced
by the nucleon current has momentum $p_{h1}^+$ and $\bm{p}_{h1T}$, which differs
from the momentum it has in the IA, $p_{h}^+$ and $\bm{p}_{hT}$, by the momentum 
transfer through the rescattering process. Assuming that this momentum transfer
is much smaller than the typical slow hadron momentum, we can evaluate the
nucleon current at the nominal slow hadron momentum defined by the IA.
Together, the two approximations imply that the nucleon current matrix elements 
are evaluated at the \textit{same nucleon and slow hadron momenta} 
in both the FSI and IA, so that their product can be replaced by the
nucleon tensor.

With these simplifications we can write the FSI term of the current matrix element,
Eq.~(\ref{current_fsi}), in the form
\be
\langle p h X' | \hat J^\mu (0) | d \rangle \, \textrm{[FSI]} 
\; &=& \; \frac{p_d^+}{p_n^+} \;
\langle h, p_h; X' | \hat J^\mu (0) | n, p_n \rangle \; 
(2\pi)^{3/2} \; i I_d ,
\nonumber
\\[1ex]
&& \; [p_n^+ = p_d^+ - p_p^+, \; \bm{p}_{nT} = -\bm{p}_{pT}], 
\nonumber \\
\label{current_fsi_shifted}
\\[2ex]
I_d \equiv I_d (\alpha_p, \bm{p}_{pT}, \alpha_h, \bm{p}_{hT})
\; &\equiv & \; \int [d p_{p1}] \;
\; \frac{2\pi}{p_{h1}^+} \; \delta (p_p^- + p_h^- - p_{p1}^- - p_{h1}^-) \; 
\frac{1}{2} \; \Psi_d (\alpha_{p1} , \bm{p}_{p1T}) \;
{\mathcal M}(s_{ph}, t_{ph}) 
\nonumber
\\[1ex]
&& \; [p_{h1}^+ = p_h^+ + p_p^+ - p_{p1}^+, \;
\bm{p}_{h1T} = \bm{p}_{hT} + \bm{p}_{pT} - \bm{p}_{p1T}] .
\label{rescattering_integral}
\ee
The function $I_d$ represents the integral of the deuteron LF wave function and
the rescattering amplitude over the phase space available for the rescattering process,
defined by the LF momenta of the final--state particles. Note that we have extracted 
the factor of $i$ from the rescattering integral and 
exhibit it explicitly in Eq.~(\ref{current_fsi_shifted}).
The complete deuteron tensor emerging from the IA and FSI current matrix elements,
Eqs.~(\ref{current_ia_fsi}) and (\ref{current_fsi_shifted}), 
including the pure IA contribution Eq.~(\ref{w_mu_nu_ia_x0_x1}), is then obtained as
\be
W_d^{\mu\nu} \; &\equiv& \; W^{\mu\nu}_d (p_d, q, p_p) \;\; = \;\; 
W_d^{\mu\nu} \textrm{[IA]} \; + \; W^{\mu\nu}_d\textrm{[FSI]}
\; + \; W^{\mu\nu}_d\textrm{[FSI$^2$]} ,
\\[2ex]
W_d^{\mu\nu} \textrm{[IA]} \; &=& \; (2\pi)^3 
\; \left( \frac{p_d^+}{p_n^+} \right)^2 \; |\Psi_d|^2
\; W_n^{\mu\nu} (p_n, \widetilde{q}) ,
\label{w_mu_nu_ia_alt}
\\[1ex]
W_d^{\mu\nu} \textrm{[FSI]} \; &=& \; (2\pi)^3 \; \left( \frac{p_d^+}{p_n^+} \right)^2 \;
\sum_h \;
\int_{\rm phas} [dp_h] \; (-2) \, \textrm{Im} \, [\Psi_d I_d ] 
\; W^{\mu\nu}_{n, h} (p_n, \widetilde{q}, p_h) ,
\label{w_mu_nu_fsi}
\\[1ex]
W_d^{\mu\nu} \textrm{[FSI$^2$]} \; &=& \; (2\pi)^3 
\; \left( \frac{p_d^+}{p_n^+} \right)^2 \; 
\sum_h \;
\int_{\rm phas} [dp_h] \; | I_d |^2 \; W^{\mu\nu}_{n, h} (p_n, \widetilde{q}, p_h)
\label{w_mu_nu_fsi2}
\\[1ex]
\nonumber
&& \; [p_n^+ = p_d^+ - p_p^+, \; \bm{p}_{nT} = 0].
\ee
The linear and quadratic FSI terms include integration over the phase space of the 
unobserved slow hadron $h$; the physical limits of the phase space are denoted by 
``phas'' and will be specified below. In addition, they include the summation over 
the relevant slow hadron species $h$. From Eqs.~(\ref{w_mu_nu_ia_alt})--(\ref{w_mu_nu_fsi2}) 
expressions for the tagged deuteron structure functions can be obtained
by performing suitable projections (see Appendix~\ref{app:projection}). 
The operations are the same as in the IA calculation
in Sec.~\ref{subsec:spectral}. When applying the projections,
the conditional neutron tensor produces the conditional neutron structure functions,
\be
W_{n, h}^{\mu\nu} (p_n, \widetilde{q}, p_h) 
\; &\rightarrow& \; F_{2n, h}(\widetilde{x}, Q^2; \zeta_h, p_{hT}) 
\; = \; F_{2n} (\widetilde{x}, Q^2)  D_h(\widetilde{x}, Q^2; \zeta_h, p_{hT}) 
\;\;\; \textrm{etc}.
\nonumber
\\[2ex]
&& \; [\widetilde{x} = x/(2 - \alpha_p)] .
\ee
The tagged deuteron structure functions can be expressed in a form analogous to the IA, 
Eq.~(\ref{spectral_impulse}), in terms of a ``distorted'' spectral function,
\be
F_{2d} (x, Q^2; \alpha_p, p_{pT}) \; &=& \;
S_d(\alpha_p, \bm{p}_{pT}) [\textrm{dist}] \; F_{2n} \left(\widetilde{x}, Q^2 \right) ,
\\[3ex]
S_d [\textrm{dist}] 
\; &\equiv& \; S_d\textrm{[IA]} \; + \; S_d\textrm{[FSI]} \; + \; S_d\textrm{[FSI$^2$]} .
\label{s_d_distorted}
\ee
The explicit expressions of the terms are
\be
S_d\textrm{[IA]} \; &=& \; \frac{|\Psi_d|^2}{2 - \alpha_p} ,
\label{s_d_ia}
\\[3ex]
S_d\textrm{[FSI]} \; &=& \; \frac{1}{2 - \alpha_p} \; 
\sum_h \;
\int_{\rm phas} [dp_h] \; 
(-2) \, \textrm{Im} \, [\Psi_d I_d ] \; D_h ,
\label{s_d_fsi}
\\[3ex]
S_d\textrm{[FSI$^2$]} \; &=& \; \frac{1}{2 - \alpha_p} \;
\sum_h \;
\int_{\rm phas} [dp_h] \; |I_d |^2 \; D_h .
\label{s_d_fsi2}
\ee

It remains to determine the kinematic limits of the phase space integral over the 
final-state slow hadron LF momentum in Eqs.~(\ref{s_d_fsi}) and (\ref{s_d_fsi2}),
or Eqs.~(\ref{w_mu_nu_fsi}) and (\ref{w_mu_nu_fsi2}). We work in a collinear frame 
(cf.~Sec.~\ref{subsec:collinear}) and parametrize the LF plus momentum of the 
slow hadron and the recoil nucleon as fractions of $p_d^+/2$ 
[cf.~Eq.~(\ref{rescattering_lf})],
\beq
p_h^+ \; = \; \frac{\alpha_h p_d^+}{2},  
\hspace{2em}
p_p^+ \; = \; \frac{\alpha_p p_d^+}{2} .
\hspace{2em}
\eeq
On general grounds the plus momentum fractions of the slow hadron and the recoil nucleon 
in the final state of DIS on the deuteron are bounded by [cf.\ 
Sec.~\ref{subsec:momentum_distribution} and Eq.~(\ref{zeta_limit_nucleon}); we 
approximate $\xi_d \approx x_d$ and set $x_d = x/2$, cf.\ Eq.~(\ref{x_def})]
\beq
0 \; < \; \alpha_h + \alpha_p \; < \; 2 (1 - x_d) \; = \; 2 - x .
\label{alpha_limit}
\eeq
For given $\alpha_p$, the phase space integral over the slow hadron momentum is 
therefore restricted to
\beq
\alpha_h \; < \; 2 - \alpha_p - x .
\label{alpha_h_limit}
\eeq
The LF momentum fraction of the slow hadron with respect to the active nucleon is
\beq
\zeta_h \;\; = \;\; \frac{p_h^+}{p_d^+ - p_p^+} \;\; = \;\;
\frac{\alpha_h}{2 - \alpha_p} .
\eeq
From Eq.~(\ref{alpha_h_limit}) it follows that
\beq
\zeta_h \; = \; \frac{\alpha_h}{2 - \alpha_p} \; < \; 1 - \frac{x}{2 - \alpha_p}
\; = \; 1 - \widetilde{x} ,
\eeq
as it should be for the DIS final state on the nucleon, cf.\ Eq.~(\ref{zeta_limit_nucleon}).
Thus $\zeta_h$ is has the correct kinematic limits within our scheme of approximations.

The expressions of Eqs.~(\ref{w_mu_nu_ia_alt})--(\ref{w_mu_nu_fsi2}) and (\ref{s_d_ia})--(\ref{s_d_fsi2})
were derived neglecting spin degrees of freedom (deuteron as S--wave bound state, nucleon spins
averaged over). They can easily be generalized to account for spin, by including the summation over
the nucleon spins and the S-- and D--wave components of the deuteron wave function. In this case an 
interference contribution to the FSI appears only if the spectator protons in the final state of the 
IA and FSI amplitudes have the same spin projection. This can happen either if they had the same spin 
in the inital deuteron wave function (S--wave or D--wave) and the proton spin was preserved in the 
rescattering process, or if they had different spins (S--D--wave interference) and the proton spin was 
flipped during the rescattering process (more complex trajectories, in which the spin of the active 
neutron in the two amplitudes is different, are also possible). Such S--D--wave interference effects 
are small at rest frame momenta $|\bm{p}_p| < 200 \, \textrm{MeV}$, where the deuteron's nonrelativistic
momentum density is dominated by the S--wave (see Fig.~\ref{fig:nonrel}).
\subsection{Factorized approximation}
In the rescattering integral Eq.~(\ref{rescattering_integral}) and the spectral function 
Eqs.~(\ref{s_d_ia})--(\ref{s_d_fsi2}), the deuteron wave function is evaluated at physical 
nucleon momenta, away from the singularities. Furthermore, since we neglect polarization
degrees of freedom, the wave function does not contain any complex factors associated
with spin dependence (cf.~Sec.~\ref{subsec:rotationally}). The deuteron wave function 
can therefore be taken as explicitly real. The expression in the integrand of Eq.~(\ref{s_d_fsi}) 
can thus be simplified as
\beq
-2 \, \textrm{Im} \, [\Psi_d I_d ]
\;\; \rightarrow \;\; -2 \, \Psi_d  \, \textrm{Im} \, [I_d ] .
\eeq
The phase of the rescattering integral Eq.~(\ref{rescattering_integral}) is 
determined by the phase of the rescattering amplitude,
\beq
\left. \begin{array}{l} 
\textrm{Re} \, I_d 
\\[2ex]
\textrm{Im} \, I_d 
\end{array} \right\}
\;\; = \;\; \int [d p_{p1}] \;
\; \frac{2\pi}{p_{h1}^+} \; \delta (p_p^- + p_h^- - p_{p1}^- - p_{h1}^-) \; 
\frac{1}{2} \; \Psi_d 
\left\{ \begin{array}{l}
\textrm{Re} \, {\mathcal M}(s_{ph}, t_{ph}) 
\\[2ex]
\textrm{Im} \, {\mathcal M}(s_{ph}, t_{ph}) 
\end{array} \right\} ,
\label{rescattering_integral_real}
\eeq
where the momentum assignments are the same as in Eq.~(\ref{rescattering_integral}).

The evaluation of the rescattering integral of Eq.~(\ref{rescattering_integral})
or Eq.~(\ref{rescattering_integral_real}) is 
described in Appendix~\ref{app:integral}. There we solve the constraint of
the LF-energy-conserving delta function and determine the kinematic
limits of the intermediate spectator nucleon LF variables $\zeta_{h1}$ and $\bm{p}_{h1T}$.
We also show that the integral can be represented in a manifestly relativistically 
invariant form, which is useful for comparing the LF formulation of FSI with the 
relativistically invariant formulation (virtual nucleon formulation) used in quasi--elastic
high--energy scattering on the deuteron.

In the rescattering integral the deuteron wave function is convoluted with the proton--hadron
($ph$) elastic scattering amplitude. At the recoil momenta considered here
($p_p \lesssim 200 \, \textrm{MeV}$ in the deuteron rest frame), the integral is 
dominated by average proton momenta in the deuteron wave function. 
In this region the momentum dependence of the deuteron wave function is much steeper than 
that of the $ph$ amplitude, because the deuteron size is much larger than the range of the 
$hN$ interaction,
\beq
\textrm{deuteron size} \;\; \gg \;\; \textrm{range of hadron--nucleon interaction} .
\label{hierarchy}
\eeq
This circumstance allows us to neglect the dependence of the $ph$ scattering amplitude 
on the initial nucleon momentum $p_{p1}$ in the rescattering integral (we do, however, retain
the dependence of the scattering amplitude on the final $p_p$). The invariant momentum
transfer in the $ph$ scattering amplitude becomes
\beq
t_{ph} \;\; \rightarrow \;\; (p_d/2 - p_p)^2 \;\; = \;\; t' \; + \; O(\epsilon_d M_N) ,
\eeq
and we replace the elastic amplitude in Eq.~(\ref{rescattering_integral_real}) by
\beq
{\mathcal M}(s_{ph}, t_{ph}) \;\; \rightarrow \;\; {\mathcal M}(s_{ph}, t') .
\eeq
The rescattering integral then factorizes as
\be
\left. \begin{array}{l} 
\textrm{Re} \, I_d 
\\[2ex]
\textrm{Im} \, I_d 
\end{array} \right\}
\; &=& \; \frac{1}{2} \; \Phi_d \; \times \; 
\left\{ \begin{array}{l}
\textrm{Re} \, {\mathcal M}(s_{ph}, t')
\\[2ex]
\textrm{Im} \, {\mathcal M}(s_{ph}, t')
\end{array} 
\right\} ,
\label{rescattering_integral_factorized}
\nonumber 
\\[3ex]
\Phi_d \; &\equiv& \;
\int [d p_{p1}] \;
\; \frac{2\pi}{p_{h1}^+} \; \delta (p_p^- + p_h^- - p_{p1}^- - p_{h1}^-) \; 
\; \Psi_d (\alpha_{p1} , \bm{p}_{p1T}) .
\label{phi_d_def}
\ee
The function $\Phi_d$ represents the integral of the deuteron wave function over
the phase space defined by the rescattering process, but is independent of the 
rescattering amplitude. It can be computed in closed form and permits a very
efficient evaluation of the rescattering integral. In this approximation
the parts of the distorted spectral function, Eqs.~(\ref{s_d_ia})--(\ref{s_d_fsi2}),
become
\be
S_d\textrm{[IA]} &=& \frac{|\Psi_d|^2}{2 - \alpha_p} ,
\label{s_d_ia_fact}
\\[3ex]
S_d\textrm{[FSI]} &=& \frac{\Psi_d}{2 - \alpha_p} \; 
\sum_h \;
\int_{\rm phas} [dp_h] \; 
\Phi_d \; [- \textrm{Im} \, {\mathcal M}] \; D_h ,
\label{s_d_fsi_fact}
\\[3ex]
S_d\textrm{[FSI$^2$]} &=& \frac{1}{2 - \alpha_p} \; 
\sum_h \;
\int_{\rm phas} [dp_h] \; \Phi_d ^2 \; \frac{1}{4}
[ (\textrm{Re} \, {\mathcal M})^2 +  (\textrm{Im} \, {\mathcal M})^2] 
\; D_h .
\label{s_d_fsi2_fact}
\ee
 
\begin{figure}
\includegraphics[width=.6\textwidth]{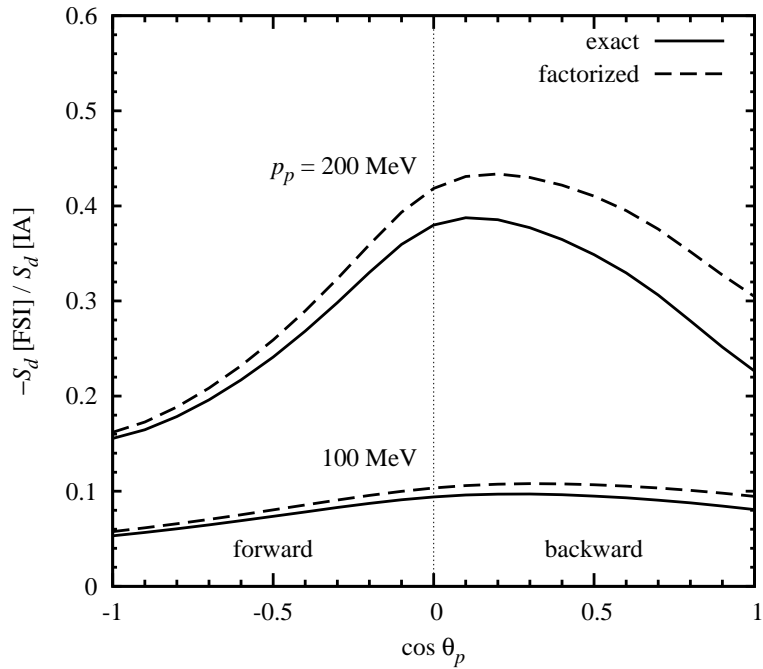}
\caption[]{Comparison of exact result and the factorized approximation for the distorted
linear FSI term of the distorted deuteron spectral function, 
Eqs.~(\ref{s_d_fsi}) and (\ref{s_d_fsi_fact}). The plot shows the ratio 
$-S_d[\textrm{FSI}] / S_d[\textrm{IA}]$ as a function of the cosine of the recoil 
momentum angle in the deuteron rest frame, $\cos\theta_p$, for two values of the 
recoil momentum modulus $p_p \equiv |\bm{p}_p|$, as indicated on the plot. 
Model parameters are described in the text.}
\label{fig:sd3_fact}
\end{figure}

Figure~\ref{fig:sd3_fact} shows the numerical results for the FSI term in the distorted spectral 
function obtained with the exact integral Eq.~(\ref{s_d_fsi}) and the factorized approximation
Eq.~(\ref{s_d_fsi_fact}), in typical kinematics (the choice of parameters is explained
below). One sees that the factorized formula provides an excellent approximation to the 
exact integral in our kinematics.
%
%
\subsection{Positivity properties}
Some comments are in order regarding the sign of the FSI correction and the positivity 
of the spectral function. The imaginary part of the elastic rescattering amplitude is 
related to the total proton--hadron cross section at the given energy by the optical 
theorem (cf.~Appendix~\ref{app:rescattering_amplitude} for the case that the hadron 
is a nucleon, $h = p, n$), and therefore satisfies
\beq
\textrm{Im} \, {\mathcal M} (s_{ph}, \, t_{ph} = 0) \; > \; 0.
\eeq
As a result, the contribution to the spectral function that arises from the interference of 
FSI and IA amplitudes and is linear in the FSI amplitude, Eq.~(\ref{s_d_fsi}), 
is explicitly negative
\beq
S_d\textrm{[FSI]} \; < \; 0 .
\label{sd_fsi_negative}
\eeq
In contrast, the contribution that arises from the square of the FSI amplitude,
Eq.~(\ref{s_d_fsi2}), is explicitly positive, 
\beq
S_d\textrm{[FSI$^2$]} \; > \; 0 .
\label{sd_fsi2_positive}
\eeq
These findings have a simple interpretation in terms of conventional quantum--mechanical
scattering theory. The linear term in the FSI amplitude represents the loss of flux
due to absorption of the outgoing hadron--nucleon wave at a given value of the final
nucleon momentum. The quadratic term represents the gain in cross section due to
scattering of the outgoing wave into a configuration with the given value of the
final nucleon momentum. In the language of wave optics, the two effects can be
referred to as ``absorption'' and ``refraction.'' One expects that absorption is the 
dominant effect at low recoil momenta, while refraction becomes dominant at
large recoil momenta. This expectation is borne out by the numerical results described
below.

The total distorted spectral function must be positive, 
\beq
S_d\textrm{[dist]} \; > \; 0 .
\label{sd_dist_positive}
\eeq
as it represents the physical
cross section for tagged DIS on the deuteron. In our scheme this is ensured by the 
fact that the hadronic tensor (i.e., the cross section) is calculated as the square of 
the current matrix element with the outgoing distorted wave. The further approximations
made in the distorted spectral function do not change this basic property. Because
the linear term in the FSI amplitude is negative, Eq.~(\ref{sd_fsi_negative}),
both the linear and quadratic terms are needed to ensure positivity of the overall
spectral function. This is again demonstrated  by the numerical results.
\subsection{Recoil momentum dependence}
We now evaluate the distorted spectral function numerically and study the magnitude 
of the distortion and its kinematic dependence on the recoil proton momentum.
The parameters of the slow-hadron distribution and the rescattering amplitude are described 
in Sec.~\ref{subsec:implications} and Appendix~\ref{app:rescattering_amplitude}. Throughout we 
consider a value of $x \sim \xi \sim 0.1$, which is so small that it does not significantly 
restrict the slow hadron momentum, so that the integration can be carried out over the full
range $0 < \zeta_h < 1$ [cf.~Eq.~(\ref{zeta_limit_nucleon}) and Figs.~\ref{fig:slow} and
\ref{fig:slow_pmin}]; the calculations can easily be extended to larger values of $x$.

The spectral function can be studied as a function of any of the recoil momentum variables 
described in Sec.~\ref{subsec:recoil}. The most transparent representation is obtained using 
as independent variables the modulus and angle of the recoil momentum in the deuteron 
rest frame (which is a special collinear frame, cf.~Sec.~\ref{subsec:recoil}),
\beq
p_p \;\; \equiv \;\; |\bm{p}_p(\text{RF})|,
\hspace{2em}
\cos \theta_p \;\; = \frac{p_p^z(\text{RF})}{|\bm{p}_p (\text{RF})|} .
\eeq
Their relation to the variables $t'$ and $\alpha_p$ is given by 
Eqs.~(\ref{t_prime_from_p_p_approx}) and (\ref{alpha_from_cos_theta}).

%
%
\begin{figure}
\includegraphics[width=.6\textwidth]{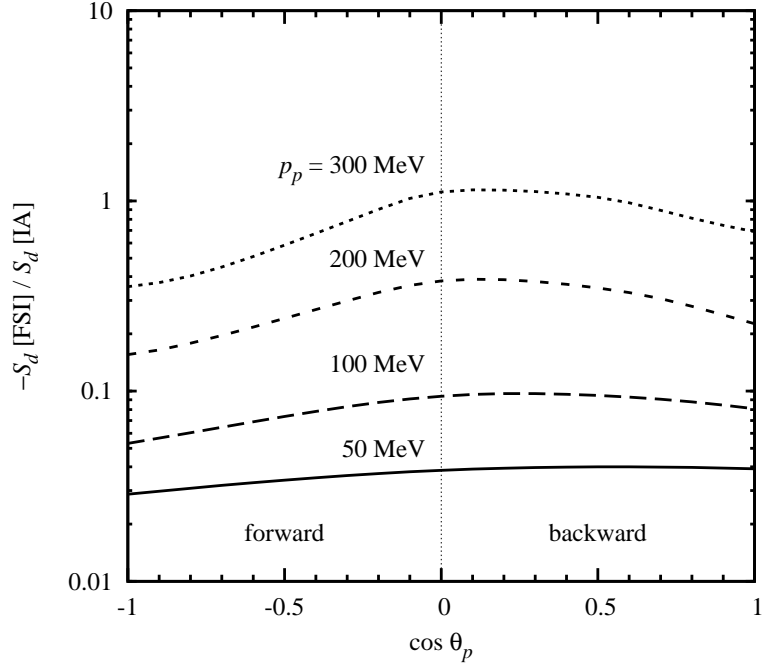}
\caption[]{The ratio of the FSI and IA deuteron spectral functions, 
$-S_d[\textrm{FSI}] / S_d[\textrm{IA}]$, Eqs.~(\ref{s_d_ia}) and (\ref{s_d_fsi}),
as a function of the cosine of the recoil momentum angle in the deuteron rest frame,
$\cos\theta_p$, for several values of the recoil momentum modulus $p_p \equiv |\bm{p}_p|$,
as indicated on the plot. Note that the ratio shown here includes only the term linear 
in the FSI amplitude, not the quadratic one. Model parameters are described in the text.}
\label{fig:sd_fsi}
\end{figure}
Figure~\ref{fig:sd_fsi} shows the ratio of the linear FSI term in the distorted 
spectral function, $S_d[{\rm FSI}]$, Eq.~(\ref{s_d_fsi}), to the IA term,
$S_d[{\rm IA}]$, Eq.~(\ref{s_d_ia}), as a function of $\cos\theta_p$, for several 
values of $p_p$. This ratio describes the relative correction to the IA arising from 
the linear FSI term. Note that $S_d[{\rm FSI}] < 0$, and the ratio is plotted with a minus 
sign in order to display it on a logarithmic scale. The following features are
apparent:
\begin{itemize}
\item The correction from the linear FSI term increases in magnitude with the recoil momentum, 
from a few percent at $p_p \sim 50\, \textrm{MeV}$ to $O(1)$ at
$\sim 200\, \textrm{MeV}$.
\item The correction is isotropic at low $p_T$ and becomes peaked at $\cos\theta_p \sim 0$ 
at larger momenta, corresponding to the nucleon emerging at approximately $\theta_p \sim \pi/2 = 90\deg$
in the deuteron rest frame. 
\end{itemize}
%
%
\begin{figure}
\begin{tabular}{c}
\includegraphics[width=.56\textwidth]{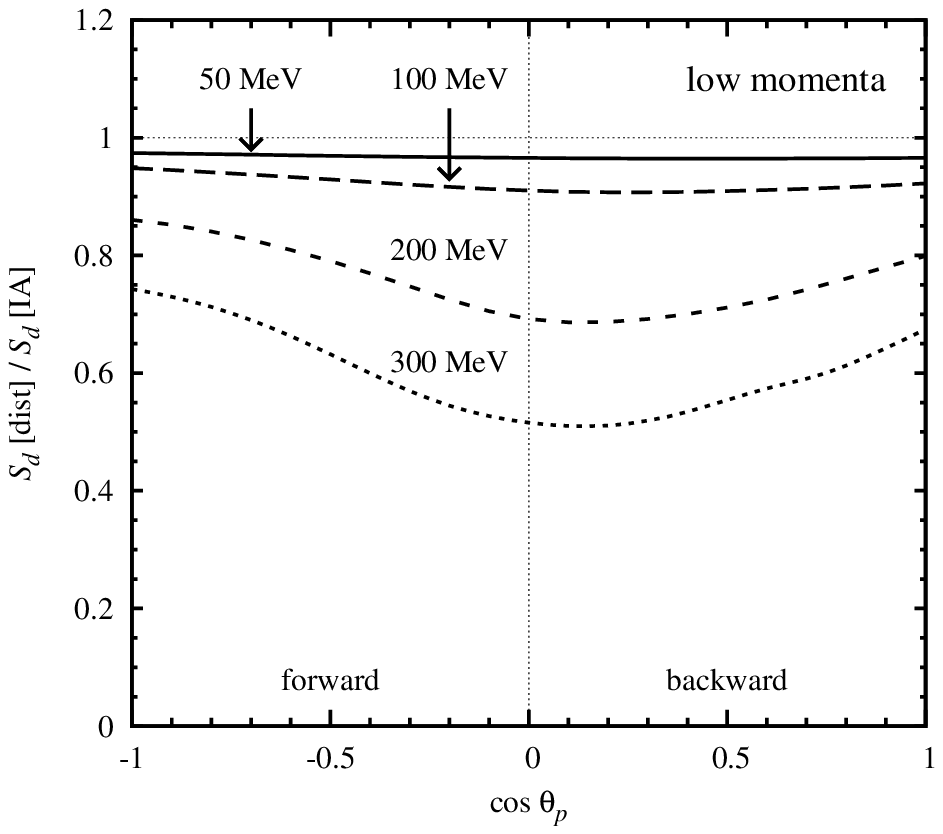}
\\[-2ex]
\includegraphics[width=.55\textwidth]{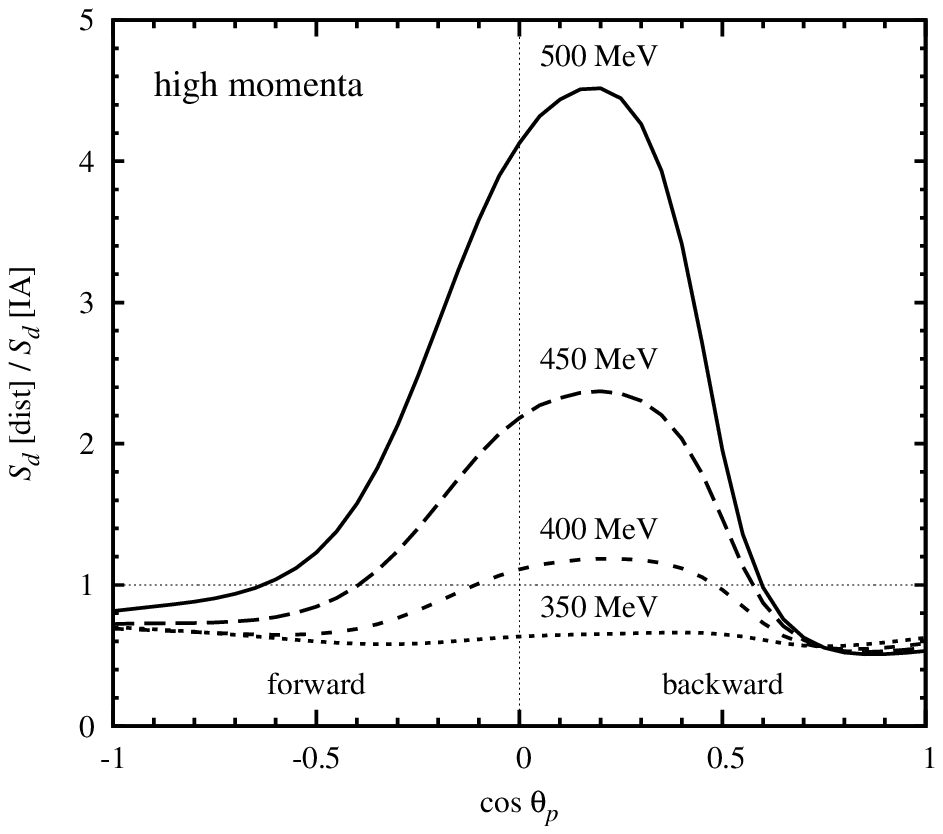}
\end{tabular}
\caption[]{The ratio of the distorted spectral function to the IA spectral function, 
$S_d[\textrm{dist}] / S_d[\textrm{IA}]$, Eqs.~(\ref{s_d_distorted})--(\ref{s_d_fsi2}),
as a function of the cosine of the recoil momentum angle in the deuteron rest frame,
$\cos\theta_p$, for several values of the recoil momentum modulus $p_p \equiv |\bm{p}_p|$.
The upper and lower plot show low and high values of the recoil momentum, as indicated 
on the plots. The distorted spectral function includes the IA, FSI and FSI$^2$ terms.
Model parameters are described in the text.}
\label{fig:sd3_rel}
\end{figure}
Figure~\ref{fig:sd3_rel} shows the ratio of the entire distorted spectral function, 
$S_d[\textrm{dist}]$, including the IA, FSI and FSI$^2$ terms, 
Eqs.~(\ref{s_d_distorted})--(\ref{s_d_fsi2}),
to the IA spectral function $S_d[\textrm{IA}]$. In other words, it shows the factor 
by which the IA spectral function is modified by the entire FSI effect. 
The following features are apparent:
\begin{itemize}
\item At low recoil momenta $p_p \lesssim 300\, \textrm{MeV}$ the linear FSI term
dominates and the FSI effect is mainly absorptive, reducing the spectral function 
relative to the IA. In this region the distorted spectral function has a minimum
at $\cos\theta_p \sim 0$.
\item At higher recoil momenta $p_p \gtrsim 300\, \textrm{MeV}$ the FSI$^2$ term
becomes dominant at forward and sideways angles, $\cos\theta_p < 0.7$, resulting 
in a large positive correction relative to the IA.
The distorted spectral function now shows a maximum at $\cos\theta_p \sim 0$. 
The transition between the low- and high-momentum regimes is rather sudden.
\item At backward angles $\cos\theta_p > 0.7$ the FSI$^2$ term is suppressed, so that the
FSI remains absorptive even at large recoil momenta. The spectral function in this
region shows little variation with the recoil momentum at $p_p \gtrsim 300\, \textrm{MeV}$.
Overall this results in a forward--backward asymmetry of the spectral function at large
momenta.
\item The distorted spectral function is positive for all recoil momenta $\bm{p}_p$, 
as required on general grounds, cf.\ Eq.~(\ref{sd_dist_positive}).
\end{itemize}
The observed dependencies are naturally explained by considering the kinematics of the 
scattering process in the rest frame. At low recoil momenta the main rescattering effect 
is always at $\theta_p \sim 90\deg$, because the only way in which the forward--moving 
DIS hadron with momenta $\sim 1\, \textrm{GeV}$ could transfer a momentum of the order
$p_p \sim 100\, \textrm{MeV}$ to the spectator proton is by pushing it sideways. 
At larger recoil momenta $p_p \gtrsim 300\, \textrm{MeV}$ is becomes increasingly 
possible for the forward--going DIS hadron to push the spectator forward, resulting
in an enhancement of the spectral function in the forward region. In contrast, the backward
region is protected from this effect, as it is kinematically impossible for the 
forward-going DIS hadron to push the spectator backwards.

The results shown in Fig.~\ref{fig:sd_fsi} and \ref{fig:sd3_rel} are close to those 
obtained in Ref.~\cite{Frankfurt:1994kt}  for the distorted spectral function of quasi-elastic 
deuteron breakup $d(e,e'p)n$ at intermediate energies $\sim \textrm{few GeV}$ in the 
Glauber approximation. This is natural, as our distorted spectral function also describes
quasi-elastic breakup, if the tagged DIS structure function of the nucleon is replaced by 
its elastic structure function (i.e., by the square of the nucleon elastic form factor) 
at $\sim \textrm{few GeV}$. 
Note that the present calculation includes only the S--wave of the deuteron bound state;
if the D--wave were included the results of Fig.~\ref{fig:sd_fsi} and \ref{fig:sd3_rel} 
at low recoil momenta $p_p < 200 \, \textrm{MeV}$ would be practically unchanged, and
at higher recoil momenta the pattern would still be the same \cite{Frankfurt:1994kt}.

%
%
\begin{figure}
\includegraphics[width=.55\textwidth]{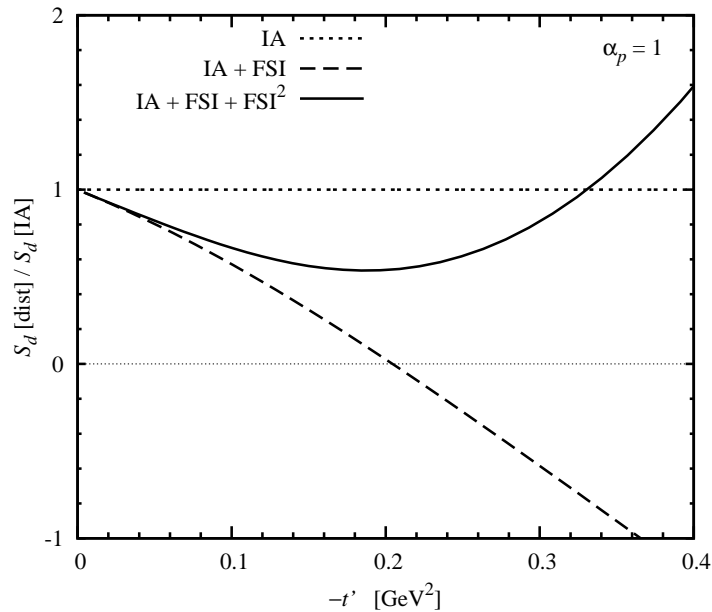}
\caption[]{The ratio of the distorted spectral function to the IA spectral function, 
$S_d[\textrm{dist}] / S_d[\textrm{IA}]$, Eqs.~(\ref{s_d_distorted})--(\ref{s_d_fsi2}),
as a function of $-t'$, for a fixed value $\alpha_p = 1$. The plot shows separately
the IA, IA + FSI, and IA + FSI + FSI$^2$ results.}
\label{fig:sd_tdep}
\end{figure}
Figure~\ref{fig:sd_tdep} shows the distorted spectral function as a function of
the invariant momentum transfer $t'$ and the recoil LF fraction $\alpha_p$, as
used in neutron structure extraction and on-shell extrapolation.
The plot again shows the ratio $S_d[\textrm{dist}] / S_d[\textrm{IA}]$, 
Eqs.~(\ref{s_d_distorted})--(\ref{s_d_fsi2}), and gives separately the results
for the IA, the sum IA + FSI, and the sum IA + FSI + FSI$^2$ (total).
One sees that 
\begin{itemize}
\item For $|t'| \lesssim 0.1 \, \textrm{GeV}^2$ the correction arises mainly from the
linear FSI term and is negative.
\item For $|t'| \gtrsim 0.2 \, \textrm{GeV}^2$ the FSI$^2$ term dominates and
causes a steep rise of the spectral function.
\item The distorted spectral function is again positive for all $t'$.
\end{itemize}
\subsection{Analytic properties}
We must also investigate the effect of FSI on the analytic properties of the
spectral function in $t'$. The IA current matrix element contains the nucleon pole 
of the deuteron wave function at $t' = 0$, which causes the IA spectral function to 
behave as $\sim R/(t')^2$ in the limit $t' \rightarrow 0$
(see Sec.~\ref{subsec:analytic_ia}). It is easy to see that the FSI contribution
to the current matrix element is non-singular in the limit $t' \rightarrow 0$.
This follows from the fact that the rescattering integral $I_d$,
Eq.~(\ref{rescattering_integral}) [or $\Phi_d$ in the factorized 
approximation, Eq.~(\ref{phi_d_def})] is a smooth function of the recoil
momentum in the physical region $|\bm{p}_p(\textrm{RF})| > 0$ (or $t' < t'_0$), 
remains regular at $|\bm{p}_p(\textrm{RF})| = 0$, and can thus be continued to the 
unpysical point $|\bm{p}_p(\textrm{RF})|^2 = t'_0 = 
-M_N \epsilon_d + \epsilon^2/4$ (or $t' = 0$) without encountering singularities
of the deuteron wave function. In the invariant formulation using Feynman graphs,
it follows from the fact that the nucleon pole arises from the nucleon tree graph, 
while the loop graphs describing FSI can at most modify the subleading behavior.
A formal proof of this ``no-loop theorem'' was 
given in Ref.~\cite{Sargsian:2005rm}. 

\begin{figure}
\includegraphics[width=.57\textwidth]{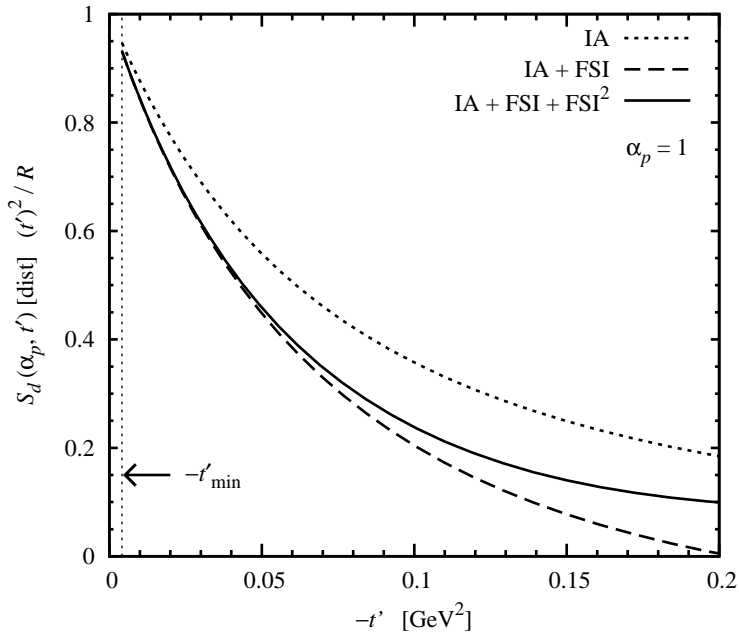}
\caption[]{The distorted spectral function $S_d[\textrm{dist}]$, 
Eqs.~(\ref{s_d_distorted})--(\ref{s_d_fsi2}), with the pole factor 
$R/(t')^2$ extracted, as function of $t'$, for $\alpha_p = 1$. 
(cf.\ Fig.~\ref{fig:spectral_ia}). The plot shows 
separately the IA, IA + FSI, and IA + FSI + FSI$^2$ results.}
\label{fig:sd_tdep_rescal}
\end{figure}

Figure~\ref{fig:sd_tdep_rescal} shows the distorted deuteron spectral function with the
pole factor removed. The plot shows separately the IA, the sum IA + FSI, and 
the sum IA + FSI + FSI$^2$ (total). One sees that
\begin{itemize}
\item The FSI correction vanishes when approaching the pole.
\item The FSI$^2$ correction vanishes even faster than the FSI one.
\item The magnitude of the FSI correction is $\sim 1/2$ of the IA at 
$|t'| \sim 0.1\, \textrm{GeV}^2$, decreasing approximately linearly with $|t'|$.
\end{itemize}
%
%

The fact that FSI does not modify the nucleon pole singularity of the IA
spectral function is of central importance for the extraction of neutron structure
from tagged DIS on the deuteron with proton tagging. It implies that FSI can
be eliminated in a model-independent manner through the on-shell extrapolation
procedure described in Sec.~\ref{subsec:analytic_ia}. FSI modifies the measured
tagged structure at $t' < t'_0$ but drops out when performing the 
extrapolation to $t' = 0$ \cite{Sargsian:2005rm}.
\subsection{Sum rules and unitarity}
Important physical requirements of the deuteron spectral function are the nucleon 
number and LF momentum sum rules, Eqs.~(\ref{spectral_number_sumrule}) and 
(\ref{spectral_momentum_sumrule}). They express the fact that the initial state
consists of two nucleons and does not involve non-nucleonic degrees of freedom.
The interactions summarized by the deuteron wave function distribute the LF momentum
among the two nucleons but do not change the baryon number or the overall LF momentum 
of the system. The IA result for the spectral function satisfies both sum rules,
and we would like them to be satisfied in the presence of FSI as well. We now want 
to discuss how this is accomplished within our model of FSI.

The nucleon number sum rule Eq.~(\ref{spectral_number_sumrule}) demands that the
integral of the distorted spectral function over the recoil proton momentum
($\alpha_p, \bm{p}_{pT}$) satisfy
\be
&& \; \int \frac{d\alpha_p}{\alpha_p} \int d^2 p_{pT} \; S_d (\alpha_p, \bm{p}_{pT})
[\textrm{dist}]
\nonumber
\\[1ex]
&=& \;
2 (2\pi)^3 \int d\Gamma_p \; S_d (\alpha_p, \bm{p}_{pT})
[\textrm{dist}]
\;\; = \;\; 1.
\label{number_sum_rule_dist}
\ee
Since the IA spectral function alone already satisfies the sum rule, 
Eq.~(\ref{number_sum_rule_dist}) requires that the integral over the total FSI 
correction (linear and quadratic) be zero,
\beq
\int d\Gamma_p \; S_d (\alpha_p, \bm{p}_{pT})
[\textrm{FSI} + \textrm{FSI}^2]
\;\; = \;\; 0.
\label{number_sum_rule_fsi}
\eeq
Equations~(\ref{number_sum_rule_dist}) viz.\ (\ref{number_sum_rule_fsi}) are realized
in our formulation if the scattering process between the slow hadron and the spectator 
proton is elastic, i.e., if it only redistributes the momentum between the particles 
but preserves the overall flux. This is achieved if the operator converting the
plane-wave state into the scattering state [the time evolution operator 
of Eq.~(\ref{time_evolution_operator})] is unitary in the two-particle space of the slow 
hadron and spectator nucleon,
\beq
\hat U^\dagger (\infty, 0) \;  \hat U (\infty, 0) \;\; = \;\; 1.
\label{unitarity_U}
\eeq
In general the operator $\hat U (\infty, 0)$ would have matrix elements between states of 
different LF energy, so that the unitarity condition Eq.~(\ref{unitarity_U}) would involve 
summation over two-particle intermediate states of arbitrary LF energy. In the on-shell 
approximation of Eq.~(\ref{scattering_wf_onshell}) we effectively restrict the operator 
to have only energy-conserving matrix elements. In this approximation Eq.~(\ref{unitarity_U})
is realized if the effective interaction satisfies the condition
(here $p_{h1}, p_{h2}$ and $p_{p1}, p_{p2}$ denote arbitrary on-shell hadron and 
nucleon 4--momenta satisfying $p_{h2} + p_{p2} = p_{h1} + p_{p1}$)
\be
&& {\textstyle\frac{1}{2}}
T(p_{p2}, p_{h2}; p_{p1}, p_{h1}) - {\textstyle\frac{1}{2}} 
T^\ast (p_{p2}, p_{h2}; p_{p1}, p_{h1})
\nonumber \\
&=& i \int d\Gamma_{h} \int d\Gamma_p \; 
(2\pi)^4 \delta^{(4)} (p_p + p_h - p_{p1} - p_{h1})
\nonumber \\
&\times& {\textstyle\frac{1}{4}} \; T^\ast (p_{p2}, p_{h2}; p_p, p_h)
\; T(p_p, p_h; p_{p1}, p_{h1}) .
\label{unitarity_T}
\ee
Equation~(\ref{unitarity_T}) has the form of the standard unitarity relation for the
$T$--matrix, but with $T$ replaced by $T/2$, corresponding to the fact that in
Eqs.~(\ref{scattering_wf}), (\ref{time_evolution_operator}), and (\ref{scattering_wf_onshell}) 
the $T$ matrix appears with a factor $1/2$ relative to the standard definition
of the $S$--matrix. If the interaction is chosen such as to satisfy Eq.~(\ref{unitarity_T}),
one can show that the linear and quadratic FSI term in the distorted spectral 
function, Eqs.~(\ref{s_d_fsi}) and (\ref{s_d_fsi2}), indeed 
obey Eq.~(\ref{number_sum_rule_fsi}). The proof involves converting the momentum 
integrals to a form such that Eq.~(\ref{unitarity_T}) can be applied; 
we shall not present the details here.

The unitarity condition Eq.~(\ref{unitarity_T}) has to be understood within our
scheme of approximations based on the hierarchy Eq.~(\ref{hierarchy}). Our FSI 
calculation describes correction to the IA spectral function for recoil momenta
of the order of the inverse deuteron size and are meaningful in this parametric 
region only. The unitarity condition Eq.~(\ref{unitarity_T}) involves momenta 
of the order of the inverse range of the nucleon-nucleon interaction, which 
lie outside the region where we consider the rescattering process. We can therefore 
assume that unitarity is realized by contributions from parametrically large 
recoil momenta, where we can not --- and do not need to --- model the rescattering amplitude. 
In other words, we suppose that any change of flux in the low-momentum region in which 
we are interested will be compensated by a change in the high-momentum region which
we cannot control. In this spirit we have parametrized the rescattering amplitude
at low momenta without explicitly implementing the unitarity condition 
Eq.~(\ref{unitarity_T}); see Appendix~\ref{app:rescattering_amplitude}.

The LF momentum sum rule Eq.~(\ref{spectral_momentum_sumrule}) follows from the
nucleon number sum rule Eq.~(\ref{spectral_number_sumrule}) if the function
$(2 - \alpha_p) S_d (\alpha_p, p_{pT})$ is symmetric under 
$\alpha_p \rightarrow 2 - \alpha_p$, which amounts to interchange of the LF momenta 
of the active nucleon and the spectator. The IA result embodies this symmetry exactly 
thanks to the symmetry of the deuteron LF wave function. The FSI correction does
not satisfy it exactly, as the rescattering integral $I_d$ in Eqs.~(\ref{s_d_fsi}) 
and (\ref{s_d_fsi2}) is generally not symmetric under $\alpha_p \rightarrow 2 - \alpha_p$.
However, the symmetry of the spectral function is still achieved within the parametric 
approximation based on Eq.~(\ref{hierarchy}), as the variation of $I_d$ in $\alpha_p$ 
around $\alpha_p = 1$ is much slower than that of $\psi_d$, so that $I_d$ can effectively
be regarded as a constant for the purpose of the reflection symmetry, and the symmetry
of Eqs.~(\ref{s_d_fsi}) is again brought about by that of the deuteron wave function.
In this sense also the momentum sum rule of the spectral function is preserved by the FSI 
within our scheme of approximations. Note that our physical picture of FSI applies only 
in a limited range of $x$, so that it is not possible to test the momentum sum rule
for the deuteron structure function, Eq.~(\ref{sumrule_f2}), within our model.
\section{Neutron structure extraction}
\label{sec:neutron}
Our findings regarding the momentum and angular dependence of FSI have implications
for the extraction of neutron structure functions from deuteron DIS data with 
proton tagging. A full assessment of the strategy requires an estimate of the 
uncertainties of the tagged structure function measurements and should be made 
with realistic pseudodata. Nevertheless, some general conclusions can be drawn 
already at this level.

The preferred method for extracting the free neutron structure function is the
on-shell extrapolation in $t'$ at fixed $\alpha_p$ (see Sec.~\ref{subsec:analytic_ia}).
The procedure eliminates modifications due to nuclear binding as well as FSI.
The accuracy of the extrapolation depends on several factors: (a) the uncertainties 
of the tagged structure function data; (b) the smoothness of the $t'$ dependence
of the spectral function after removing the pole factor, which is determined by the
FSI; (c) the distance between the physical region and the on-shell point, which depends 
on the recoil fraction $\alpha_p$.

If accurate measurements of the tagged structure functions can be made
down to rest-frame recoil momenta $p_p(\textrm{RF}) \sim$ few 10 MeV, one
may perform the on-shell extrapolation in $t'$ at LF fractions $\alpha_p \approx 1$,
where one can come closest to the pole in $t'$ (see Fig.~\ref{fig:alphar_tprime}).
In this situation our model predicts that the FSI corrections have smooth 
$t'$ dependence, and a magnitude of $<$ 10\% of the IA at the lowest $t'$ values,
so that they are reliably eliminated by the extrapolation procedure.
Since there are no singularities in $t'$ between $t'_0$ and 0 the 
extrapolation can be performed using a polynomial fit to the $t'$ dependence of 
the tagged structure function data \cite{Sargsian:2005rm}. 

If accurate measurements of the tagged structure functions are only possible
at larger recoil momenta $p_p(\textrm{RF}) \sim$ 100--200 MeV, one may instead
focus on the backward region $\alpha_p > 1$ (or $\cos\theta_p > 0$), where our
model predicts that FSI are relatively small; see Figs.~\ref{fig:sd3_rel} and
\ref{fig:sd_tdep}. FSI are also suppressed in the forward region, $\alpha_p > 1$, 
but the predictions there are more model-dependent. In this case there is a 
trade-off between coming as close as possible to the pole (which favors $\alpha_p = 1$)
and minimizing FSI (which favors $\alpha_p = 1$). While our model predicts that
the $t'$ dependence is smooth even in the presence of FSI, the distances from the
pole are such that the magnitude of the correction is substantial ($\sim$ 50\% of 
the IA at $t' =$ 0.1 GeV$^2$; see Fig.~\ref{fig:sd_tdep_rescal}). In this situation 
one may no longer rely on polynomial extrapolation but fit the data with a more 
complex parametrization of the spectral function based on the expected functional 
form of the FSI correction.
\section{Summary and outlook}
\label{sec:summary}
In this work we have presented a theoretical framework for the analysis of DIS
on the deuteron with spectator nucleon tagging.
Nuclear and nucleonic structure are separated using the apparatus of LF quantum 
mechanics appropriate for high-energy scattering processes. The IA determines the
basic dependence of the tagged cross section on the recoil momentum and its analytic
properties (nucleon pole). In the region of intermediate $x$ (roughly $0.1 < x < 0.5$)
FSI arise mainly from the interaction of the spectator with slow hadrons produced in 
the fragmentation of the active nucleon (rest frame momenta $|\bm{p}_h| \sim$ 1 GeV). 
We have described this effect in a schematic
model using empirical slow-hadron distributions (protons, neutrons) and the
nucleon-nucleon scattering amplitude. Our treatment is based on a hierarchy
of dynamical scales (deuteron size $\gg$ range of rescattering) and gives rise to
a self-consistent physical picture. The main conclusions regarding FSI in tagged DIS
can be summarized as follows:
\begin{itemize}
\item \textit{Absorption and refraction.}
The rescattering between the slow DIS hadrons and the spectator nucleon
involves absorptive interactions (linear in the imaginary part of amplitude,
dominant at recoil momenta $|\bm{p}|_p <$ 200 MeV) and refractive interactions 
(quadratic in real and imaginary parts, dominant at higher momenta). The net
effect is to reduce the flux of spectator nucleons at low momenta and
increase it at higher momenta.
\item \textit{Angular dependence.}
The FSI effect on the recoil momentum distribution is approximately isotropic 
in the deuteron rest frame at momenta $|\bm{p}_p| <$ 100 MeV. At higher momenta it
becomes angle-dependent, with the strongest effect occurring in the direction
approximately perpendicular to the $\bm{q}$--vector (sideways direction).
\item \textit{Analyticity in $t'$.}
The FSI correction to the IA spectral function vanishes at the nucleon pole
$t' \rightarrow 0$ (relative to the IA) and exhibits a smooth dependence on $t'$ 
up to $|t'| \sim$ 0.1 GeV$^2$. It can be eliminated through on-shell 
extrapolation $t' \rightarrow 0$.
\item \textit{Relative magnitude.}
The FSI correction reduces the IA spectral function by $\sim 50\%$ at 
$|t'| \sim$ 0.1 GeV$^2$. The effect becomes proportionately smaller as $|t'|$ decreases.
\end{itemize}

Our results show that the extraction of free neutron structure through on-shell extrapolation
are possible if accurate measurements of the recoil momentum dependence can be performed
in the region $|t'| \ll$ 0.1 GeV$^2$ (or $|\bm{p}_p| \ll$ 200 MeV). The analytic structure of 
the FSI correction and its moderate size indicate that the nucleon pole residue can be
extracted reliably even in the presence of experimental errors. We emphasize that the 
extrapolation eliminates not only Fermi motion but also nuclear binding effects, as the phase 
space for interactions vanishes at the on-shell point. Measurements of tagged DIS over
a wide kinematic range will become possible at a future EIC with suitable forward detectors. 
Simulations of neutron structure extraction through on-shell extrapolation using the IA 
cross section model suggest that the procedure is feasible under 
realistic conditions \cite{Cosyn:2014zfa,Guzey:2014jva}.
The dominant systematic uncertainty in the tagged structure function results from the 
uncertainty in the transverse recoil momentum $\bm{p}_{pT}$, as caused by the finite
detector resolution and the intrinsic momentum spread of the deuteron beam \cite{LD1506}.
These simulations can now be updated to include FSI effects in the cross section model;
results will be reported elsewhere.

Tagged DIS on the deuteron has also been proposed as a tool to explore the dynamical origin 
of the nuclear modification of the nucleon's partonic structure. The idea is that the 
observed recoil momentum effectively controls the spatial size of the $pn$ configuration
in the deuteron, which makes is possible to study nuclear modifications of the nucleon 
structure functions in configurations of defined size (``tagged EMC effect'')
\cite{Frankfurt:1988nt,Melnitchouk:1996vp,Atti:2010yf,Hen:2014vua}. The main 
challenge in such measurements is to separate initial-state modifications of the partonic 
structure from FSI effects. Our model provides an a-priori estimate of the size
of the FSI effect and can be used to assess the sensitivity of such measurements to a 
putative nuclear modification of nucleon structure. In particular, the results of
Fig.~\ref{fig:sd3_rel} show that in the backward region $\cos\theta_p > 0.7$ the FSI
effect is practically independent of the modulus of the recoil momentum for values 
$|\bm{p}_p| \gtrsim$ 300 MeV. An observed variation of the tagged structure function with 
$|\bm{p}_p|$ in this region could therefore be ascribed to initial-state modifications. 
The formulation of a practical procedure for tagged EMC effect studies at the EIC based on
these findings should be the object of future work.

In applications to neutron structure and the EMC effect one aims to eliminate or minimize the 
FSI effects in tagged DIS. The same measurements could be used to study the FSI as an object 
in iteslf, by going to kinematics where the effects are maximal (recoil angles 
$-0.2 < \cos\theta_p < 0.4$; see Fig.~\ref{fig:sd3_rel}) and verifying their strong kinematic 
dependence. Such measurements on the deuteron would help to understand better the pattern of 
nuclear breakup in DIS on heavier nuclei (e.g., slow neutron rates and angular distributions), 
which in turn would assist other studies of hard processes in nuclei 
(centrality dependence, hadronization and jets in nuclei) \cite{Boer:2011fh}.

In the present study of FSI in tagged DIS we considered the case of unpolarized 
electron and deuteron and made several simplifying assumptions about deuteron structure,
the DIS hadron spectrum, and the nature of the rescattering process. The treatment 
could be refined in several aspects while remaining within the same physical picture:
\begin{itemize}
\item \textit{Spin degrees of freedom and polarization.} The spin degrees of freedom
of the deuteron and the nucleons could be incorporated in the LF quantum-mechanical
description without essential difficulties. A connection
between the deuteron's $NN$ LF wave function and the non-relativistic wave function
along the lines of Sec.~\ref{subsec:rotationally} and Eq.~(\ref{nonrel}) can be derived
including spin; it involves relativistic spin rotations describing the transition from 
three-dimensional spin to LF helicities. The deuteron now includes S-- and D--wave states 
($L =$ 0 and 2), giving rise to a rich structure. Expressions for the helicity representation 
of the deuteron $NN$ LF wave function are given in Ref.~\cite{Frankfurt:1983qs}, and the tagged 
DIS structure functions in the IA for the polarized deuteron can be obtained by substituting 
the unpolarized deuteron LF momentum density in Eq.~(\ref{spectral_density_deuteron}) by the 
corresponding polarized density given there. A detailed treatment of polarized deuteron LF 
structure and double-polarized tagged DIS will be presented in Ref.~\cite{Cosyn17}.
When describing FSI in polarized tagged DIS one must account also for (a) the dependence
of the slow hadron distribution on the spin of the active nucleon (there are presently 
few data on target fragmentation in polarized nucleon DIS); (b) the spin dependence
of the rescattering amplitude; (c) contributions to the cross section resulting from the
interference of amplitudes with different initial spin states (S--D wave interference). 
The correlation between the nucleon spin and its
momentum in the deuteron, combined with a spin dependence of the slow hadron distribution
and the rescattering in FSI, could give rise to pronounced spin-orbit effects in tagged DIS
on the polarized deuteron. Electron polarization would bring into play the nucleon spin 
structure functions and enable spin-spin and further spin-orbit effects in tagged DIS.
\item \textit{Inelastic rescattering.} In the present calculation we implement FSI through 
elastic rescattering of slow hadrons (protons, neutrons) on the spectator. This scheme allows 
us to describe FSI in the hadronic tensor at the probabilistic level (because the same 
hadrons appear in the current matrix element with and without FSI) and preserve the nucleon 
number sum rule through elastic unitarity. It is clear that certain inelastic channels
are open at the momenta $|\bm{p}_h| \sim$ 1 GeV considered here and can have sizable cross 
section, for example, production of $\Delta$ isobars in nucleon-nucleon collisions.
Including such channels in the FSI calculation is possible in theory but very difficult
in practice. It requires a coupled-channel formalism, in which one considers all stable 
hadrons appearing in the final state (in the example of $\Delta$ excitation, this would
be two nucleons and one pion) and implements all possible interactions between them.
One would also need to know the amplitudes for the ``direct'' production of these hadrons through
fragmentation of the active nucleon, which interfere with those of ``indirect'' production
in the rescattering process. It is not obvious whether these amplitudes could be extracted
from the DIS hadron multiplicities without further modeling.
\item \textit{Rescattering of pions.} We have focused here on FSI induced by the 
rescattering of slow protons and neutrons in the DIS final state, as these are the
dominant hadrons at $x_{\rm F} < -0.3$ and have large cross sections for rescattering
on the spectator nucleon. FSI could also 
arise from the rescattering of pions, whose multiplicity rises strongly at $x_{\rm F} > -0.3$.
This effect could be calculated with the same formalism as used here (the formulas
for the rescattering integral in Appendix~\ref{app:integral} 
are given for a general slow hadron mass). The pion--nucleon
amplitude at pion momenta $|\bm{p}_h| \sim$ 1 GeV is well constrained by data.
One interesting aspect of pions is that they can emerge in the backward direction
of the DIS process, i.e., opposite to the $\bm{q}$--vector 
[see Sec.~\ref{subsec:momentum_distribution} and Eq.~(\ref{pz_from_zeta})], and 
therefore push the spectator in the backward direction. It would be worth investigating
pion--induced FSI in a separate study.
\end{itemize}

The physical picture of FSI developed here refers to the kinematic region of intermediate 
$0.1 \lesssim x \lesssim 0.5$. In this region the slow hadron distributions are approximately 
independent of $x$, while diffractive hadron production is not yet important. 
Tagged DIS experiments can of course be performed also at larger or
smaller values of $x$, with various scientific objectives. It is worthwhile summarizing 
what changes in the physical picture of FSI are expected in these regions.
\begin{itemize}
\item \textit{FSI and diffraction at small $x$.} At $x \ll 0.1$ diffractive DIS becomes a 
distinctive source of slow nucleons in the target fragmentation region. The $x_{\rm F}$ spectra
of protons in DIS on the proton measured at HERA show a diffractive peak near $x_{\rm F} = -1$
with an integrated multiplicity of $\sim 0.1$; see Ref.~\cite{Wolf:2009jm} for a review. 
Physically this effect is explained by a 
color-singlet exchange between the electromagnetic current and the nucleon, such that
the DIS process leaves the nucleon intact and recoiling with a momentum $\sim$ few 100 MeV. 
If such diffractive production happens in tagged DIS on the deuteron, there is a significant 
probability for the diffractive nucleon and the spectator to recombine and form the deuteron, 
as they have the same spin-isospin quantum numbers and similar momenta as the original 
proton-neutron pair in the deuteron wave function. In measurements of tagged DIS at
small $x$ one selects the channel where this recombination does not happen and a
proton-neutron scattering state is produced instead of the deuteron. In this situation
it is essential that the wave function of the scattering state is constructed such that 
it is orthogonal to the deuteron, i.e., that it is obtained as the solution of the dynamical 
equation with the same effective interaction as gives the deuteron bound state, 
cf. Eqs.~(\ref{time_evolution_operator}) and (\ref{scattering_wf}). It also requires
that off-shell energies are allowed in the rescattering process. A detailed treatment
of FSI in tagged DIS at small $x$ will be presented in a forthcoming article
\cite{Guzey17}.
\item
\textit{FSI and in tagged DIS at large $x$.} In DIS on the nucleon at $x \gtrsim 0.5$ 
the distribution of hadrons in the target fragmentation region differs substantially 
from that at lower $x$. The reason is that the hadron LC fraction is
kinematically restricted to $\zeta_h < 1 - \xi \approx 1 - x$, such that only
small values of $\zeta_h$ are allowed at $x \rightarrow 1$. Physically speaking
the DIS process almost ``empties'' the nucleon of LF momentum, and the produced hadrons
have to share the small rest. These hadrons therefore have large momenta in the 
target rest frame, cf.~Eq.~(\ref{pz_from_zeta}) and Figs.~\ref{fig:slow} and \ref{fig:slow_pmin}, 
and their interactions with the spectator are suppressed by the formation time. Our picture
therefore suggests that FSI may be suppressed in tagged DIS at large $x$. However, since 
$F_{2n} \ll F_{2p}$ at $x \rightarrow 1$, even small FSI would have a large relative effect
on the extracted neutron structure function. The $x \rightarrow 1$ limit of tagged DIS
therefore requires a dedicated study.
\end{itemize}
%
%
\appendix
\section{Deuteron wave function}
\label{app:deuteron}
In this appendix we describe a simple two--pole parametrization of the non-relativistic 
deuteron wave function (S--wave only, no D--wave), which has the correct analytic properties 
at small momenta and provides an excellent approximation to realistic wave functions over 
the range of momenta considered here. It is of the form (see e.g.\ Ref.~\cite{Tiburzi:2000je}) 
\be
\widetilde\Psi_d (\bm{k})_{\rm two-pole} \; &=& \; \frac{1}{\sqrt{c}}
\left( \frac{1}{|\bm{k}|^2 + a^2} - \frac{1}{|\bm{k}|^2 + b^2} \right) ,
\label{twopole}
\\[2ex]
\int d^3 k \; |\widetilde\Psi_d (\bm{k})_{\rm two-pole}|^2 \; &=& \; 1,
\hspace{2em} c \; = \; \frac{\pi^2 (a - b)^2}{ab (a + b)} ,
\ee
where the parameter $a^2$ is determined by the position of the nucleon pole
[cf.~Eq.~(\ref{a2_def})],
\be
a^2 &=& M_N \epsilon_d - \frac{\epsilon_d^2}{4}
\;\; = \;\; \frac{t'_0}{2} ,
\label{a2_def_alt}
\ee
and $b^2$ is determined empirically from the average deuteron size. The numerical values are
\beq
a \; = \; 0.0456\, \textrm{GeV} \hspace{2em}
\textrm{[from Eq.~(\ref{a2_def_alt})]},
\hspace{4em}
b \; = \; 0.2719 \, \textrm{GeV} \hspace{2em}
\textrm{(empirically).}
\eeq
%
%
\begin{figure}
\begin{tabular}{l}
\includegraphics[width=.5\textwidth]{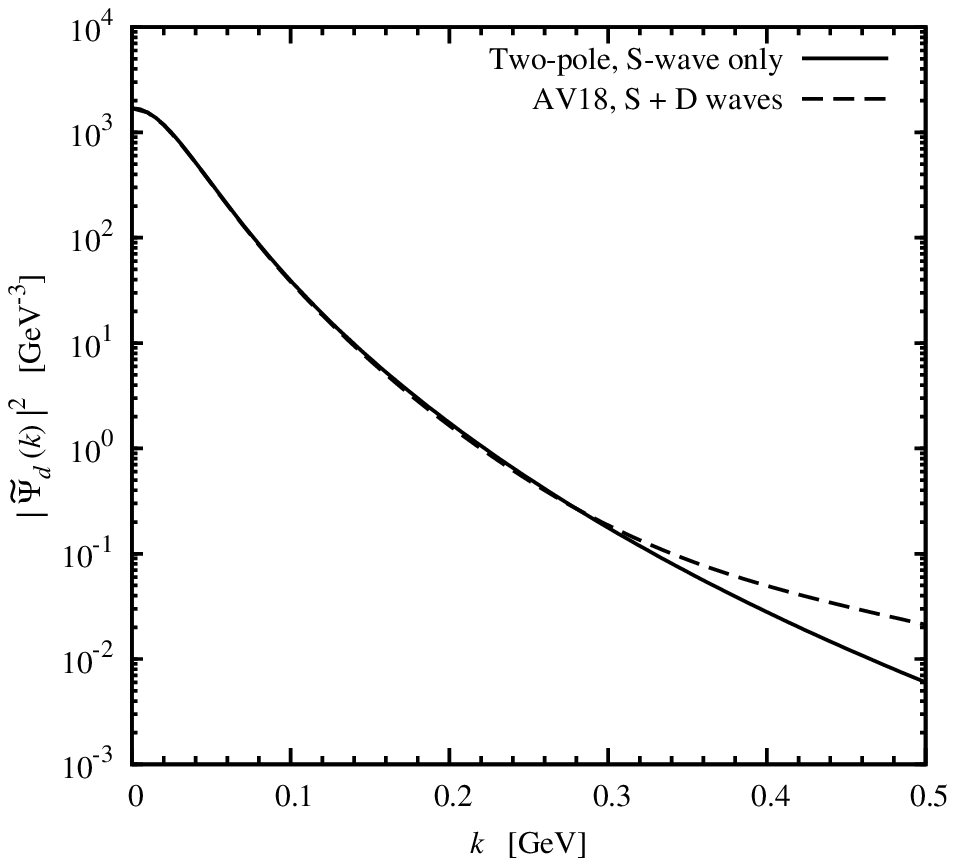}
\\[-3ex]
(a)
\\[4ex]
\includegraphics[width=.5\textwidth]{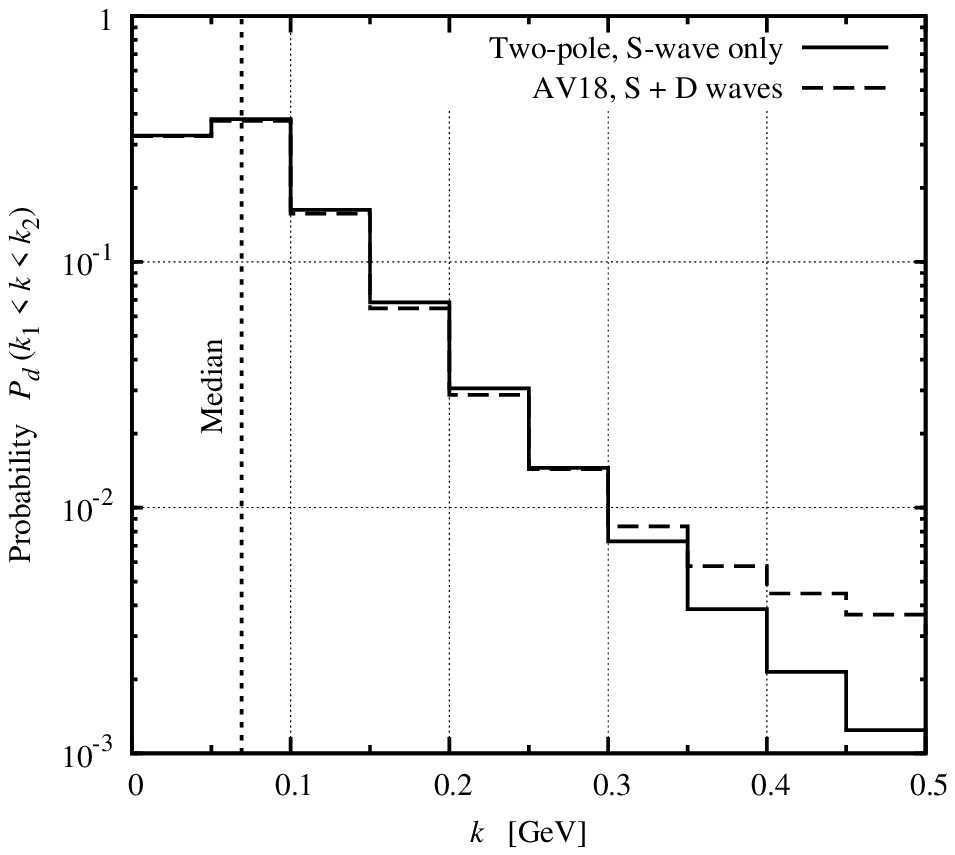}
\\[-3ex]
(b)
\end{tabular}
\caption[]{(a) Momentum density of the non-relativistic deuteron wave function,
$|\widetilde\Psi_d (\bm{k})|^2$, as a function of $k \equiv |\bm{k}|$.
Solid line: Two--pole parametrization, Eq.~(\ref{twopole}) (S--wave only).
Dashed line: Wave function obtained with the AV18 NN potential (S + D waves)
\cite{Wiringa:1994wb}. (b) Probability to find nucleon with momentum $k_1 < k < k_2$,
Eq.~(\ref{dmom_int}), for $k_{1,2}$ in steps of 50 MeV. The constant values shown at $k_1 < k < k_2$ 
give the value of $P_d(k_1 < k < k_2)$ for that range. The median momentum is indicated by a
vertical line.}
\label{fig:nonrel}
\end{figure}
Figure~\ref{fig:nonrel}a compares the momentum density $|\widetilde\Psi_d (\bm{k})|^2$
obtained with the two--pole parametrization, Eq.~(\ref{twopole}), with that of the
deuteron wave function obtained with the AV18 NN potential \cite{Wiringa:1994wb}. 
One sees that the two--pole form provides a completely 
adequate description of the momentum density over a wide range of momenta 
$|\bm{k}| < 0.3\, \textrm{GeV}$. The discrepancy at larger momenta is due to
the fact that the two--pole form contains only the S--wave, while the
AV18 wave function includes the D--wave components, which
becomes dominant at larger momenta.

Figure~\ref{fig:nonrel}b shows the integral of the deuteron momentum density over finite 
intervals $k_1 < |\bm{k}| < k_2$, corresponding to the probability to find a nucleon with momentum 
in that range, 
\beq
P_d (k_1 < |\bm{k}| < k_2) \; = \; 
\int d^3 k \; \theta(k_1 < |\bm{k}| < k_2) \; |\widetilde\Psi_d (\bm{k})|^2 
\; = \; 4\pi \int_{k_1}^{k_2} dk \; k^2 \; |\widetilde\Psi_d (\bm{k})|^2 .
\label{dmom_int}
\eeq
The result with the two-pole wave function is again compared with AV18. The histogram gives 
an intuitive picture of the momentum distribution of nucleons in the deuteron and enables 
simple estimates of the contribution of different momentum regions to observables. 
The median of the momentum distribution is 68 MeV for the two-pole wave function
(69 MeV for AV18). Note that the median nucleon momentum in the deuteron is considerably larger 
than the ``binding momentum'' defined as $\sqrt{M_N \epsilon_d} = 45 \, \textrm{MeV}$;
the different values illustrate the presence of multiple dynamical scales in the 
deuteron wave function.
\section{Projection formulas}
\label{app:projection}
In this appendix we derive the explicit expressions of the proton-tagged deuteron structure 
functions in terms of the deuteron LF momentum density and the inclusive neutron structure 
functions, Eqs.~(\ref{IA_F2_wf}) and (\ref{IA_FL_wf}), starting from the 
``master formula'' for the scattering tensors in the collinear frame, 
Eq.~(\ref{w_impulse}). The same derivation can be used with the 
distorted spectral function in the presence of FSI.
We write Eq.~(\ref{w_impulse}) in schematic form as
\beq
W_d^{\mu\nu} (p_d, q, p_p) \;\; = \;\; [...] \; W_n^{\mu\nu} (p_n, \widetilde{q}),
\hspace{3em}
[...] \;\; \equiv \;\;
[2 (2\pi)^3] \, \frac{2 \; |\Psi_d (\alpha_p, \bm{p}_{pT})|^2}{(2 - \alpha_p)^2},
\label{master_short}
\eeq
and substitute the deuteron tensor parametrized by Eq.~(\ref{W_decomposition}),
and the neutron tensor parametrized by Eq.~(\ref{nucleon_tensor_1}). 
We consider the deuteron tensor averaged over the transverse direction of the 
recoil momentum, in which only the structures with $F_{Ld}$ and $F_{Td}$ are 
present. Equations for the structure functions can be obtained by taking specific 
components of the tensor equation Eq.~(\ref{master_short}) in the collinear 
frame (see Sec.~\ref{subsec:collinear}). 
From the longitudinal component $\mu\nu = ++$ we obtain
\be
F_{Ld} 
\; &=& \;
[...] \;\; \frac{L_d^2}{(L_d^+)^2} \;
 \left\{ -\frac{(\widetilde q^+)^2}{\widetilde q^2} \; F_{Ln}
\; + \; 
\left[ \frac{(\widetilde L_n^+)^2}{\widetilde L_n^2}
+ \frac{(\widetilde q^+)^2}{\widetilde q^2}
\right] \; F_{Tn} \right\}
\\[2ex]
&=& \;
[...] \;\; \frac{L_d^2}{(L_d^+)^2} \;
 \left[ \frac{(\widetilde q^+)^2}{\widetilde q^2} \; (F_{Tn} - F_{Ln})
\; + \; \frac{(\widetilde L_n^+)^2}{\widetilde L_n^2} \; F_{Tn} \right] .
\label{projected_1}
\ee
From the transverse components $\mu\nu = ij, (i, j = x, y)$ we obtain
\be
F_{Td} - F_{Ld} \; &=& \; [...] \; \left[
F_{Tn} - F_{Ln} \; + \; 
\frac{|\widetilde{\bm{L}}_{nT}|^2}{2 \widetilde L_n^2} F_{Tn} \right] .
\label{projected_2}
\ee
Here we have used that, after averaging over the direction of the transverse
recoil momentum, rotational symmetry allows us to replace 
\beq
\widetilde L_{nT}^i \widetilde L_{nT}^j \; \rightarrow \; \delta^{ij} 
|\widetilde{\bm{L}}_{nT}|^2/2. 
\eeq
Explicit expressions for the structure functions are obtained from 
Eqs.~(\ref{projected_1}) and (\ref{projected_2}) by substituting
the specific expressions for the vector components in the collinear frame
(see Sec.~\ref{subsec:collinear}). 
The deuteron vector $L_d$ is given by Eq.~(\ref{L_deuteron}), and
the nucleon vector components in the IA are
\beq
\widetilde q^+ \; = \; q^+, \hspace{2em}
\widetilde{\bm{L}}_{nT} \; = \; -\bm{p}_{pT}.
\eeq
The exact expressions for the structure functions are complicated and not instructive.
We quote only the expressions in the scaling limit $Q^2 \gg M_N^2, |\bm{p}_{pT}^2|$, where
\beq
-\frac{L_d^2 (\widetilde q^+)^2}{(L_d^+)^2 \widetilde q^2}
\; \rightarrow \; 1,
\hspace{3em}
\frac{L_d^2 (\widetilde L_n^+)^2}{(L_d^+)^2 \widetilde L_n^2}
\; \rightarrow \; 1,
\hspace{3em}
\frac{|\widetilde{\bm{L}}_{nT}|^2}{2 \widetilde L_n^2} 
\; = \; O\left(\frac{|\bm{p}_{pT}|^2}{Q^2}\right) .
\eeq
In this limit the $L$ and $T$ deuteron structure functions are obtained as
\be
F_{Ld} \; &=& \; \textrm{[...]} \; F_{Ln},
\\[1ex]
F_{Td} \; &=& \; \textrm{[...]} \; F_{Tn} .
\ee
Reverting to the long form of Eq.~(\ref{master_short}) and writing the
arguments of the structure functions, this is
\be
F_{Ld}(x, Q^2) 
\; &=& \; [2 (2\pi)^3] \, 
\frac{2 \; |\Psi_d (\alpha_p, \bm{p}_{pT})|^2}{(2 - \alpha_p)^2}
\; F_{Ln} (\widetilde x, Q^2),
\\[1ex]
F_{Td}(x, Q^2) 
\; &=& \; [2 (2\pi)^3] \, 
\frac{2 \; |\Psi_d (\alpha_p, \bm{p}_{pT})|^2}{(2 - \alpha_p)^2}
\; F_{Tn} (\widetilde x, Q^2),
\ee
The corresponding formula for the deuteron structure function $F_{2d} = x_d F_{Td}$, 
Eq.~(\ref{FT_from_F2}), is
\be
F_{2d}(x, Q^2) 
\; &=& \; [2 (2\pi)^3] \, 
\frac{|\Psi_d (\alpha_p, \bm{p}_{pT})|^2}{2 - \alpha_p}
\; F_{2n} (\widetilde x, Q^2),
\ee
where we have used that $x_d = x/2, \; \widetilde x = x/(2 - \alpha_p)$, and
$F_{2n}(\widetilde x, Q^2) = \widetilde x F_{Tn}(\widetilde x, Q^2)$.
\section{Elastic scattering amplitude}
\label{app:rescattering_amplitude}
In this appendix we give an empirical parametrization of the nucleon--nucleon
elastic scattering amplitude at small angles and incident momenta $p \lesssim 1\, \textrm{GeV}$
(in the rest frame of the target nucleon), for use in our calculation of FSI in tagged 
DIS on the deuteron.

Measurements of nucleon--nucleon elastic and total cross sections at incident
momenta $p \sim 1\, \textrm{GeV}$ have been performed in several experiments; 
see Ref.~\cite{LechanoineLeLuc:1993be} for a review of the data.
Neutron--proton scattering measures directly the strong--interaction cross section; 
in proton--proton scattering one also has to account for electromagnetic 
interactions (Coulomb scattering) \cite{Silverman:1988df,Dobrovolsky:1983wz}. 
For both channels ($np, pp$) the differential strong--interaction cross section for elastic 
scattering at forward angles can be parametrized as
\beq
\frac{d\sigma_{\rm el}}{dt}
\;\; = \;\; |f(0)|^2 \; e^{bt} ,
\label{dsigma_dt_elastic}
\eeq
where $f(t)$ is a complex amplitude and $b$ the exponential slope. The amplitude is
of the form
\beq
f(t) \;\; = \;\; A(t) \; + \; \textrm{spin-dependent amplitudes} ,
\eeq
where the central term $A(t)$ is non-zero at $t = 0$ and can be expressed as
\beq
A(0) \;\; = \;\; \textrm{Im}\, A(0) \, (i + \rho_0), \hspace{2em} \rho_0 \;\; 
\equiv \;\; \textrm{Re}\, A(0) / \textrm{Im}\, A(0) .
\eeq
The imaginary part at $t = 0$ is related to the nucleon--nucleon total cross 
section by the optical theorem
\beq
[\textrm{Im}\, A(0)]^2 \;\; = \;\; \frac{\sigma_{\rm tot}^2}{16\pi} .
\eeq
The contribution of spin--dependent amplitudes at $t = 0$ can be described by the 
parameter
\beq
\beta_0 \; = \; |\textrm{spin-dependent amplitudes at $t = 0$}|^2 / [\textrm{Im}\, A(0)]^2 .
\eeq
The differential cross section Eq.~(\ref{dsigma_dt_elastic}) can then be represented as
\beq
\frac{d\sigma_{\rm el}}{dt}
\;\; = \;\; \frac{\sigma_{\rm tot}^2}{16\pi} \; (1 + \rho_0^2 + \beta_0) \; e^{bt} .
\label{dsigma_dt_elastic_optical}
\eeq
Experimental values of the parameters $\sigma_{\rm tot}, \rho_0, \beta_0,$ and $b$ at
several energies are summarized in Table~\ref{tab:elastic}.
%
%
\begin{table}
\begin{tabular}{|r|r|r|r|r|r|r|}
\hline
& $p$ [GeV] & $T$ [GeV] & 
$\sigma_{\rm tot}$ [mb] & $\rho_0$ & $\beta_0$ & 
$b \; [\textrm{GeV}^{-2}]$
\\
\hline
$np$ & 1.26 & 0.633 & 36.1      & $-0.253$ & $0.181 \pm 0.074$ & 5.09 $\pm 0.51$ \\
     & 1.68 & 0.985 & $\sim 40$ & $-0.414$ & $<0.01          $ & 5.35 $\pm 0.49$ \\
\hline
$pp$ & 1.28 & 0.648 & 41.1      & $ 0.202$ & $0.096 \pm 0.030$ & 3.56 $\pm 1.09$ \\
     & 1.69 & 0.992 & 47.5      & $-0.178$ & $0.025 \pm 0.012$ & 6.24 $\pm 0.37$ \\
\hline
\end{tabular}
\caption{Parameters of the small--angle elastic scattering amplitude measured in 
$np$ \cite{Silverman:1988df} and $pp$ \cite{Dobrovolsky:1983wz} 
scattering experiments. Here $p$ is the momentum of the incident nucleon
in the target rest frame (lab momentum), and $T = \sqrt{p^2 + M_N^2} - M_N$ is the
incident kinetic energy (lab energy).}
\label{tab:elastic}
\end{table}

In terms of the invariant amplitude Eq.~(\ref{invariant_amplitude}) the differential
cross section for nucleon-nucleon elastic scattering is expressed as
\beq
\frac{d\sigma_{\rm el}}{dt}
\;\; = \;\; \frac{|{\mathcal M}(s, t)|^2}{64\pi s \, p_{\rm cm}^2}
\label{dsigma_dt_invariant}
\eeq
where $s$ is the squared CM energy and $p_{\rm cm} = \sqrt{s/4 - M_N^2}$ is the CM momentum.
Comparing Eq.~(\ref{dsigma_dt_invariant}) with the empirical formula 
Eq.~(\ref{dsigma_dt_elastic_optical}), we parametrize the invariant amplitude as
\beq
\left.
\begin{array}{rclrcl}
|{\mathcal M}(s, 0)|^2 &=& 4 s p_{\rm cm}^2 \sigma_{\rm tot}^2 
(1 + \rho_0^2 + \beta_0) &
\hspace{2em}
|{\mathcal M}(s, t)| &=&  |{\mathcal M}(s, 0)| \; e^{bt/2} ,
\\[3ex]
\textrm{Im}\, {\mathcal M}(s, t) &=& 
\displaystyle \frac{|{\mathcal M}(s, t)|}{\sqrt{1 + \rho_0^2}}, &
\textrm{Re}\, {\mathcal M}(s, t) &=& 
\displaystyle \frac{\rho_0 \, |{\mathcal M}(s, t)|}{\sqrt{1 + \rho_0^2}}
\end{array}
\hspace{2em}
\right\} .
\label{m_parametrization}
\eeq
These formulas apply at fixed $s$, and the parameters $(\sigma_{\rm tot}, \rho_0, \beta_0, b)$ 
generally depend on $s$. Equation~(\ref{m_parametrization})
can be adapted to the cases of $np$ and $pp$ scattering by choosing appropriate
parameters (cf.\ Table~\ref{tab:elastic}) 
and provide a sufficient description of the nucleon--nucleon 
elastic amplitude for our purposes.
For a simple parametrization of the average $np$ and $pp$ amplitude we take the average
of the parameter values at the lower energy of Table~\ref{tab:elastic} ,
\beq
\sigma_{\rm tot} \; = \; 39 \, \textrm{mb} ,
\hspace{2em}
\rho_0 \; = \; -0.03,
\hspace{2em}
\beta_0 \; = \; 0.14,
\hspace{2em}
b \; = \; 4.3 \, \textrm{GeV}^{-2},
\label{amplitude_parameters_fixed}
\eeq
For a more realistic parametrization one may use the energy--dependent parameters quoted
in Ref.~\cite{LechanoineLeLuc:1993be}. We note that the Re/Im ratio of the amplitude, $\rho_0$, 
is poorly constrained by experimental data and relies on theoretical calculations. 
\section{Rescattering integral}
\label{app:integral}
In this appendix we evaluate the rescattering integral determining the FSI 
correction to the current matrix element, Eq.~(\ref{rescattering_integral}).
It is usesful to consider a phase space integral for hadron--proton scattering 
of the general form
\be
I (p_h, p_p, \ldots)
\; &\equiv & \; \int [d p_{p1}] \; \frac{2\pi}{p_{h1}^+} \; 
\delta (p_{h1}^- + p_{p1}^- - p_{h}^- - p_p^-) \; F(p_{p1}, p_h, p_p, ...) ,
\label{rescat_general}
\\[1ex]
p_{h1}^+ \; &=& \; p_h^+ + p_p^+ - p_{p1}^+, 
\label{cons_plus}
\\[1ex]
\bm{p}_{h1T} \; &=& \; \bm{p}_{hT} + \bm{p}_{pT} - \bm{p}_{p1T} ,
\label{cons_transv}
\\[1ex]
p_{h1}^- \; &=& \; (|\bm{p}_{h1T}|^2 + M_h^2 )/p_{h1}^+ . 
\ee
Here $p_{h1}$ and $p_{p1}$ are the hadron and proton 4--momenta in the initial state,
and $p_{h}$ and $p_p$ are the 4--momenta in the final state, with 
$p_{h1}^2 = p_h^2 = M_h^2, \; p_{p1}^2 = p_p^2 = M_N^2$. 
LF plus and transverse momenta are conserved
in the scattering process, cf.\ Eqs.~(\ref{cons_plus}) and (\ref{cons_transv}),
and the integration is over the initial--state phase space 
defined by the LF minus momentum-- (or energy--) conserving delta function.
The integral is regarded as a function of the final--state momenta. The integrand
depends on the initial and final hadron and proton 4--momenta, as well as on other 
4--momenta, e.g.\ the momenta associated with the current matrix element of the 
high--energy scattering process. 

The integral Eq.~(\ref{rescat_general}) can be converted to a manifestly relativistically 
invariant form. Using the conditions of momentum conservation, Eqs.~(\ref{cons_plus}) 
and (\ref{cons_transv}), one easily shows
\beq
\frac{1}{p_{h1}^+} \, \delta (p_{h1}^- + p_{p1}^- - p_{h}^- - p_p^-) 
\;\; = \;\; \delta [p_{h1}^+ (p_{h1}^- + p_{p1}^- - p_{h}^- - p_p^-)] 
\;\; = \;\; \delta [(p_{h} + p_p - p_{p1})^2 - M_h^2] ,
\eeq
so that
\be
I (p_h, p_p, \ldots)
&=& \int [d p_{p1}] \; 2\pi \, \delta [(p_{h} + p_p - p_{p1})^2 - M_h^2] 
\; F(p_{p1}, p_h, p_p, ...)
\label{rescat_invariant}
\\[1ex]
&=& \int \frac{d^4p_{p1}}{(2\pi)^4} \; 2\pi \, \delta (p_{p1}^2 - M_N^2) 
\; 2\pi \, \delta [(p_{h} + p_p - p_{p1})^2 - M_h^2] 
\; F(p_{p1}, p_h, p_p, ...) .
\label{rescat_invariant_cut}
\ee
The last integral has the form of a cut Feynman integral, where the hadron and proton
propagators are replaced by mass--shell delta functions, as appears in the invariant
formulation of nuclear rescattering processes (virtual nucleon formulation).

It is convenient to evaluate the rescattering integral in the invariant form 
Eq.~(\ref{rescat_invariant}). We choose a collinear frame (cf.~Sec.~\ref{subsec:collinear}) 
and describe the 4--momenta by their LF components, with the plus components given
as multiple of $p_d^+/2$,
\beq
\left.
\begin{array}{rcllrcl}
p_h^+ &=& \displaystyle \frac{\alpha_h p_d^+}{2}, \hspace{2em} & 
\bm{p}_{hT}, \hspace{2em} &
p_h^- &=& \displaystyle \frac{2 ( |\bm{p}_{hT}|^2 + M_h^2)}{\alpha_h p_d^+}, 
\\[3ex]
p_p^+ &=& \displaystyle \frac{\alpha_p p_d^+}{2}, &
\bm{p}_{pT}, &
p_p^- &=& \displaystyle \frac{2 (|\bm{p}_{pT}|^2 + M_N^2)}{\alpha_p p_d^+}, 
\\[3ex]
p_{p1}^+ &=& \displaystyle \frac{\alpha_{p1} p_d^+}{2}, &
\bm{p}_{p1T}, &
p_{p1}^- &=& \displaystyle \frac{2 (|\bm{p}_{p1T}|^2 + M_N^2)}{\alpha_{p1} p_d^+}
\end{array}
\hspace{2em}
\right\} .
\label{rescattering_lf}
\eeq
We introduce the total plus momentum fraction and transverse momentum of the hadron--proton
system,
\beq
\alpha \; \equiv \; \alpha_h + \alpha_p, 
\hspace{3em}
\bm{P}_T \; \equiv \; \bm{p}_{hT} + \bm{p}_{pT} ,
\eeq
and the invariant mass of the hadron--proton system,
\be
s \; &\equiv& \; s_{hp} \;\; \equiv \;\; (p_h + p_p)^2 \; = \; \alpha \left( 
\frac{|\bm{p}_{hT}|^2 + M_h^2}{\alpha_h} 
\; + \; 
\frac{|\bm{p}_{pT}|^2 + M_N^2}{\alpha_p} 
\right) \; - \; \bm{P}_T^2 ,
\ee
such that
\be
\frac{p_d^+ (p_h^- + p_p^-)}{2}
\; &=& \; \frac{|\bm{p}_{hT}|^2 + M_h^2}{\alpha_h} 
\; + \; 
\frac{|\bm{p}_{pT}|^2 + M_N^2}{\alpha_p} \;\; = \;\; \frac{\bm{P}_T^2 + s}{\alpha} .
\ee
The argument of the delta function in Eq.~(\ref{rescat_invariant}) can now be expressed as
\beq
(p_h + p_p - p_{p1})^2 - M_h^2 \;\; = \;\;
\frac{\alpha}{\alpha_{p1}} \left[ R^2 
\; - \; \left(\bm{p}_{p1T} - \frac{\alpha_{p1}}{\alpha} \bm{P}_T \right)^2 \right] ,
\label{delta_argument}
\eeq
where
\beq
R^2 \;\; \equiv \;\;
 - \frac{\alpha_{p1}^2}{\alpha^2} s \; + \; 
\frac{\alpha_{p1}}{\alpha} \, (s + M_N^2 - M_h^2) \; - \; M_N^2 .
\label{R2_def}
\eeq
From Eq.~(\ref{delta_argument}) it follows that the argument can reach zero only
if $R^2 > 0$. According to Eq.~(\ref{R2_def}) this is the case if
\beq
x_{\rm min} \; < \; \frac{\alpha_{p1}}{\alpha} \; < \; x_{\rm max} ,
\label{condition_alpha}
\eeq
where
\beq
x_{\rm min, \; max} \;\; \equiv \;\; 
\frac{s + M_N^2 - M_h^2 \mp {\textstyle \sqrt{(s + M_N^2 - M_h^2)^2 - 4 s M_N^2}}}{2s}
\;\; = \;\;
\frac{{\textstyle\sqrt{p_{\rm cm}^2 + M_N^2}} \mp p_{\rm cm}}{\sqrt{s}} .
\label{condition_p_cm}
\eeq
In the last step we have introduced the center--of--mass momentum of the hadron--nucleon 
system, $p_{\rm cm}$,
\beq
(s + M_N^2 - M_h^2)^2 - 4 s M_N^2 \;\; = \;\; 4 s p_{\rm cm}^2 .
\eeq
The ratio $\alpha_{p1}/\alpha$ in Eq.~(\ref{condition_alpha}) represents the fraction 
of the LF plus momentum of the hadron--nucleon system carried by the initial nucleon.
The condition Eq.~(\ref{condition_p_cm}) has a simple
physical meaning. In the CM frame the LF plus momentum fraction is given by the ratio 
of the nucleon plus momentum, $\sqrt{p_{\rm cm}^2 + M_N^2} 
+ p_{\rm cm}^z$, to the mass of the system, $\sqrt{s}$, 
and the minimum and maximum values correspond to the situation that
the nucleon momentum is opposite to, or along, the $z$--axis, $p_{\rm cm}^z = \mp p_{\rm cm}$.
The bounds satisfy $0 < x_{\rm min, \; max} < 1$, and the limiting values for
small and large energies are
\beq
\left.
\begin{array}{lclll}
x_{\rm min, \; max} &\rightarrow& 1/2, \hspace{2em} & s \rightarrow (M_h + M_N)^2, 
\hspace{2em} & \textrm{or} \hspace{2em} p_{\rm cm} \rightarrow 0, 
\\[2ex]
x_{\rm min, \; max} &\rightarrow& 0, 1, & s \rightarrow \infty, \hspace{2em} & 
\textrm{or} \hspace{2em} p_{\rm cm} \rightarrow \infty  
\end{array}
\hspace{2em}
\right\} .
\eeq
In sum, the phase space for the initial nucleon momentum $p_{p1}$ in the rescattering
integral Eq.~(\ref{rescat_invariant}) is defined by the conditions that the LF fraction 
lie in the interval Eq.~(\ref{condition_alpha}), and that the transverse momentum 
lie on the (shifted) circle corresponding to zero value of Eq.~(\ref{delta_argument}),
\beq
\left| \bm{p}_{p1T} - \frac{\alpha_{p1}}{\alpha} \bm{P}_T \right| \;\; = \;\; R .
\eeq
It remains to express the phase space element in the LF momentum variables and
account for the Jacobian factors. Using Eqs.~(\ref{phase_space_lf}) and
Eq.~(\ref{delta_argument}) we obtain
\be
I \; &=& \; \int [d^4 p_{p1}] \; 2\pi \, \delta [(p_{h} + p_p - p_{p1})^2 - M_h^2] ...
\\[2ex]
&=& \; \frac{1}{8 \pi^2 \alpha} \int_{\alpha x_{\rm min}}^{\alpha x_{\rm max}}
d\alpha_{p1} \int d^2 p_{p1T} \; \delta \left[ 
\left(\bm{p}_{p1T} - \frac{\alpha_{p1}}{\alpha} \bm{P}_T \right)^2 \; - \; R^2 \right]
...
\ee
Finally, introducing the shifted transverse momentum as integration variable,
\beq
\bm{p}_{p1T} \;\; = \;\; \frac{\alpha_{p1}}{\alpha} \bm{P}_T \; + \; \bm{l}_T ,
\eeq
and using that 
\beq
\delta (|\bm{l}|^2 - R^2) \;\; = \;\; \frac{\delta (|\bm{l}| - R)}{2 R} , 
\eeq
we obtain
\beq
I \;\; = \;\; \frac{1}{16 \pi^2 \alpha} \int_{\alpha x_{\rm min}}^{\alpha x_{\rm max}}
d\alpha_{p1} \int_0^{2\pi} d\phi_l \; F[|\bm{l}_T| = R] .
\label{rescattering_integral_formula}
\eeq
Equation~(\ref{rescattering_integral_formula}) represents a practical formula for the
evaluation of the phase space integral. As a test we compute the phase volume,
i.e., the integral of unity, $F = 1$, and obtain
\beq
\frac{1}{16 \pi^2 \alpha} \int_{\alpha x_{\rm min}}^{\alpha x_{\rm max}}
d\alpha_{p1} \int_0^{2\pi} d\phi_l \;\; = \;\; \frac{x_{\rm max} - x_{\rm min}}{8\pi}
\;\; = \;\; \frac{p_{\rm cm}}{4\pi \sqrt{s}} ,
\eeq
which agrees with the standard phase volume obtained by evaluating the invariant integral 
Eq.~(\ref{rescat_invariant_cut}) in the CM frame.
\section*{Acknowledgments}
This work was motivated by Jefferson Lab's Laboratory-Directed R\&D project LD1403/LD1506
``Physics potential of polarized light ions with EIC@JLab'' (see Ref.~\cite{LD1506}).
We are indebted to W.~Cosyn, V.~Guzey, D.~Higinbotham, Ch.~Hyde,
K.~Park, P.~Nadel-Turonski, and M.~Sargsian, for numerous discussions
of theoretical and experimental aspects of nuclear DIS with spectator
tagging. We thank R.~Schiavilla for helpful communications regarding
scattering theory and FSI in deuteron breakup.

This material is based upon work supported by the U.S.~Department of Energy, Office of Science, 
Office of Nuclear Physics, under contract DE-AC05-06OR23177.
The research of M.S.\ was supported by the U.S.~Department of Energy, Office of Science, 
Office of Nuclear Physics, under Award No. DE-FG02-93ER40771.
\end{document}